\ifdefined\pdfoutput
\pdfoutput=1
\fi
\documentclass[english,12pt]{article}
\usepackage[T1]{fontenc}
\usepackage[utf8]{inputenc}
\usepackage[english]{babel}
\usepackage{graphicx}
\usepackage[width=\textwidth]{caption}
\usepackage[letterpaper,margin=1.25in]{geometry}
\usepackage{setspace}
\usepackage{array,float,booktabs}
\usepackage{lmodern}
\usepackage{mathtools}
\allowdisplaybreaks
\usepackage{amsfonts,amsmath,amsthm,amssymb,bm}
\usepackage{enumitem}
\usepackage{natbib,color}
\usepackage[pdfstartview=FitH,bookmarks=true,bookmarksopen=true,breaklinks=true,colorlinks=true,linkcolor=blue,citecolor=blue,filecolor=blue,urlcolor=black]{hyperref}
\usepackage{tikz}

\numberwithin{equation}{section}
\newtheorem{theorem}{Theorem}[section]
\newtheorem{lemma}[theorem]{Lemma}
\newtheorem{proposition}[theorem]{Proposition}
\newtheorem{corollary}[theorem]{Corollary}
\theoremstyle{definition}
\newtheorem{definition}{Definition}

{\theoremstyle{plain}\newtheorem{assumption}{Assumption}}

\newcommand{\G}{\mathbb{G}}
\newcommand{\Pn}{\mathbb{P}_n}
\newcommand{\E}{\operatorname{E}}
\newcommand{\R}{\mathbb{R}}
\newcommand{\Ltwo}{L^2_0(Q_0)}
\newcommand{\cl}{\operatorname{cl}}
\newcommand{\Sph}{\mathcal{S}}
\newcommand{\cE}{\mathcal{E}}
\newcommand{\cD}{\mathcal{D}}
\newcommand{\cF}{\mathcal{F}}
\newcommand{\cP}{\mathcal{P}}
\newcommand{\dto}{\rightsquigarrow}
\newcommand{\ip}[2]{\langle #1,#2\rangle}
\newcommand{\indep}{\mathrel{\perp\mkern-10mu\perp}}
\newcommand{\plus}[1]{\bigl(#1\bigr)_+}
\newcommand{\orb}{\mathrm{orb}}
\newcommand{\conp}{\stackrel{p}{\longrightarrow}}
\newcommand{\conas}{\stackrel{a.s.}{\longrightarrow}}
\newcommand{\defeq}{\coloneqq}

\makeatletter
\newcommand{\inpart}[2]{\phantomsection\protected@edef\@currentlabel{(#2)}\label{#1}(#2)}
\makeatother
\newcommand{\floatnotes}[1]{\par\smallskip{\footnotesize\textit{Notes}: #1\par}}

\title{Bootstrap Likelihood Ratio Tests: Quotient Geometry and Conditional Inference\thanks{The author is grateful to Jean-Marie Dufour for many discussions, guidance and references on this topic over the years, which led to this work. Part of this research was conducted during the author's 2025--2026 sabbatical at the Department of Economics, University of Toronto, whose hospitality is gratefully acknowledged.}}
\author{Purevdorj Tuvaandorj\thanks{York University, \texttt{tpujee@yorku.ca}.}}
\date{November 2026}

\begin{document}
\setcitestyle{round}
\setlength{\bibsep}{0pt plus 0.3ex}
\renewcommand{\bibfont}{\setstretch{1.22}}
\maketitle
\begin{abstract}
\noindent Likelihood ratio tests are often bootstrapped when parameters are not identified under the null hypothesis. We formulate the hypothesis on the quotient of the parameter space by observational equivalence. The parametric bootstrap then depends on the restricted estimator only through its equivalence class, which converges, as Redner (1981) showed, even when the estimator does not. When the null model is regular, the bootstrap is consistent if the alternative splits off the nuisance directions and, in a Gaussian limit experiment, exactly when the limiting statistic is independent of the nuisance score. Otherwise, for a given statistic, the only asymptotically similar critical value based on the restricted fit is, under regularity conditions, its null quantile given that fit, computed by a conditional bootstrap. Applied to an averaged likelihood ratio, this gives the weighted-average-power optimal similar test. Under independent homoskedastic Gaussian errors, the conditional test is exact in regressions with a nonlinear term.

\end{abstract}

\section{Introduction}\label{sec:intro}

Likelihood ratio (LR) tests are often bootstrapped when a parameter is not identified under the null hypothesis, as in tests of the number of unobserved types, of a threshold or of an additional hidden unit. The limit of the LR statistic is then typically nonstandard and nonpivotal, and the parametric bootstrap from the model fitted under the null is the natural source of critical values. Its validity has been established case by case. A general argument is challenging because the unidentified components of the restricted estimator have no probability limit, so there is no point around which to localize the bootstrap.

\citet{Redner1981} showed that when the parameter is not identified, the maximum likelihood estimator, though it may fail to converge to any point of the parameter space, is consistent for the true equivalence class in the quotient of the parameter space by observational equivalence, where a point is a distribution: estimation is well posed on the quotient. \citet{FengMcCulloch1996} connected this consistency to the bootstrap LR test of \citet{McLachlan1987} for the number of mixture components, and \citet{MoreiraPorterSuarez2009} used its analogue for the linear instrumental-variables (IV) reduced form to prove the validity of a residual bootstrap of the score statistic. We ask whether the parametric bootstrap of the LR statistic is likewise well posed when the null hypothesis itself is formulated on the quotient, as a set of distributions. The bootstrap distribution then depends on the restricted estimator only through its equivalence class (Lemma~\ref{lem:quotient}). At a regular point of the null (Definition~\ref{def:nullpar}) and under Assumption~\ref{ass:null}, the null model is locally asymptotically normal in an identified finite-dimensional parameter and the restricted fit is root-$n$ consistent for it, so the bootstrap samples from a regular null model, although the LR statistic has a nonstandard limit.

We give a sufficient condition for the bootstrap to be consistent, a necessary and sufficient condition, and repairs for when it fails. First, if the local directions of the full model split as the nuisance tangent space plus a cone of departures, the LR limit is the same along every root-$n$ drift of the null and the restricted-fit bootstrap is consistent (Theorems~\ref{thm:drift} and~\ref{thm:boot}). The argument expands the likelihood only at the true distribution, and the limit may be the supremum of a Gaussian process. Splitting holds in tests of instrument relevance, of an additional hidden unit and, where the known expansions apply, of the number of types (Sections~\ref{sec:setup} and~\ref{sec:examples}); in each, the alternative adds a component to the null model (a first stage, a hidden unit, a type) whose possible departures do not depend on the values of the null parameters. This is the bootstrap counterpart of Redner's result. 

Second, at a regular point of the null and in a Gaussian limit experiment, bootstrap consistency is equivalent to independence of the limiting statistic and the nuisance score, which splitting implies (Theorem~\ref{thm:diagnostic}). Independence can fail when a nuisance parameter governs how well the alternative is identified, as the first stage does for a hypothesis on the structural coefficient in the linear IV model.

Third, the ordinary bootstrap counts the nuisance noise twice, once through the fit and once through the draw (Lemma~\ref{lem:fittedshift}). The \emph{conditional bootstrap} removes the second copy by conditioning on the bootstrap fit returning to the observed fit. Because the nuisance score is complete and sufficient for the local position of the null, the conditional test is similar whatever the alternative, and, by Neyman structure, for the given statistic its critical value is the only similar one computed from the restricted fit, when the conditional distribution is continuous and strictly increasing there (Theorems~\ref{thm:condlimit} and~\ref{thm:fitcond}, Corollary~\ref{cor:unique}). Applied to an averaged likelihood ratio, it gives the weighted-average-power optimal similar test (Theorem~\ref{thm:condWAP}). In the IV model, it is the conditional LR test of \citet{Moreira2003}, for which \citet{MoreiraPorterSuarez2004} propose two conditional bootstraps, one fixing the first-stage statistic at its observed value and one weighting the draws by a kernel as in \citet{BoothHallWood1992}, and \citet{DavidsonMacKinnon2008} propose another.

The other repair, \emph{orbit completion}, enlarges the alternative until the statistic no longer depends on the nuisance score. In a sample, it has an LR form, against an enlarged model, and a score form, which needs no such model and estimates the nuisance parameters under the null, as in the $C(\alpha)$ statistic of \citet{Neyman1959}; in the IV model, these are the \citet{AndersonRubin1949} statistic and its score form (Theorem~\ref{thm:orbit} and Proposition~\ref{prop:orbitsample}).

In the endogenous binary treatment model of \citet{HanMcCloskey2019}, the conditional LR test is asymptotically similar uniformly over bounded root-$n$ drifts of the take-up effect from zero and, in their design, substantially more powerful than their robust Wald test. Under independent homoskedastic Gaussian errors, the conditional test is exact for regressions with a nonlinear term whose coefficient may be small, such as yield-curve, break-date and threshold models. In the break-date experiment of \citet{ElliottMullerWatson2015}, the best similar test nearly attains their bound on weighted average power over all valid tests. In the yield curves of \citet{DieboldLi2006}, the exact sets for the decay of a second \citet{Svensson1994} curvature term are the whole admissible range in most months without significant evidence of that term, whereas $\chi^2_1$ and Wald sets are short and undercover in simulations calibrated to the data.

\paragraph*{Related literature.}
Limit distributions of LR statistics under loss of identifiability \citep{Davies1977,Davies1987,Andrews1993,AndrewsPloberger1994,Hansen1996,DacunhaCastelleGassiat1999,Gassiat2002,LiuShao2003,AzaisGassiatMercadier2009,AndrewsCheng2012} are inputs to our analysis. Conditioning on a statistic that is sufficient for the nuisance parameter under the null is the classical route to similar tests \citep[Chapter~4]{LehmannRomano2005}, and 
\citet{Moreira2003} made powerful use of it under weak identification. Extensions include, among others, optimal invariant tests \citep{AndrewsMoreiraStock2006}, moment-condition models \citep{Kleibergen2005,AndrewsMikusheva2016}, nearly integrated regressors \citep{JanssonMoreira2006}, non-Gaussian errors \citep{CattaneoCrumpJansson2012} and permutation tests \citep{Tuvaandorj2025}. We apply the same principle to the parametric bootstrap of the LR statistic in general likelihood models, including statistics with nonstandard limits.

On the bootstrap, \citet{GuKoenkerVolgushev2018} establish its consistency for LR tests of homogeneity in mixtures and \citet{HigginsJochmans2024} develop it for fixed-effect models. For tests of the rank of a matrix, it has been studied for minimum-distance statistics \citep{PortierDelyon2014,ChenFang2019} and, with restricted estimates, for the LR test of the cointegration rank \citep{CavaliereRahbekTaylor2012}. \citet{DiCiccioYoung2008} study the conditional properties of the unconditional parametric bootstrap in exponential families, and \citet{CavaliereGeorgiev2020} show that inference can remain valid when the bootstrap distribution has a random limit and give conditions for the validity of the bootstrap in conditional inference. We derive the conditioning that makes the bootstrap test similar.

\citet{Kaji2021} develops a theory of weak identification in semiparametric models through an underlying regular parameter and uses it to improve estimators; our questions are the validity of the bootstrap and the construction of critical values. Near a regular point, the null parameter $\eta$ is a local form of the minimal sufficient parameter of \citet{Barankin1960} and \citet{Picci1977} (Section~\ref{sec:setup}), as is the local reduced-form parameter of \citet{ChenChristensenTamer2018}, who use it to build confidence sets for identified sets.

Section~\ref{sec:setup} formulates the hypothesis on the quotient and develops the local geometry. Section~\ref{sec:main} gives the criteria for bootstrap validity, and Section~\ref{sec:examples} examples in which the ordinary bootstrap works. Section~\ref{sec:fail} explains its failure and develops the conditional bootstrap and orbit completion. Section~\ref{sec:apps} presents applications to an endogenous treatment and to regressions with a nonlinear term. Proofs are in the Appendix and the Supplemental Appendix.

Sections~\ref{sec:setup}--\ref{sec:fail} assume i.i.d.\ observations, extended to fixed covariates in Section~\ref{sec:suff}. The exact conditional test of Section~\ref{sec:nonlin} rests only on a Gaussian linear null model, so it allows fixed regressors.

\paragraph*{Notation.}
For vectors and matrices, $\|\cdot\|$ is the Euclidean norm, $\|A\|\defeq\operatorname{tr}(A'A)^{1/2}$, and $x_+\defeq\max(x,0)$. A cone is a set closed under multiplication by nonnegative scalars. In an inner-product space, $\cl A$ is the closure of $A$ and $A+B\defeq\{a+b:a\in A,\ b\in B\}$ the Minkowski sum; $d(x,A)\defeq\inf_{a\in A}\|x-a\|$; $A\perp B$ means that $\ip ab=0$ for all $a\in A$ and $b\in B$, and then $A+B$ is written $A\oplus B$; $A^\perp$ is the orthogonal complement of $A$; $B\ominus A\defeq B\cap A^\perp$ for linear subspaces $A\subseteq B$; and $\Sph(C)\defeq\{h/\|h\|:h\in C,\ h\neq0\}$ for a cone $C$. For distributions $\mu,\nu$ on a Euclidean space, the bounded-Lipschitz distance $d_{\mathrm{BL}}(\mu,\nu)\defeq\sup_g|\int g\,d\mu-\int g\,d\nu|$, over $g$ with $|g|\le1$ and $|g(x)-g(y)|\le\|x-y\|$, metrizes weak convergence, and the Hellinger distance between distributions $P$ and $Q$ is $d_{\mathrm{H}}(P,Q)\defeq\bigl(\int(\sqrt{dP}-\sqrt{dQ})^2\bigr)^{1/2}$. Vectors are columns; $(a',b')'$ stacks the vectors $a$ and $b$, $[A\ \,B]$ places matrices side by side, and $(X,Y)$ is a pair of random elements. We write $\dto$ for convergence in distribution, $X\indep Y$ for independence, w.p.a.1 for ``with probability approaching one'', and CLT and CMT for the central limit and continuous-mapping theorems; $\Phi$ is the standard normal distribution function; functions of random variables are Borel measurable.

\section{Quotient formulation}\label{sec:setup}

Let $X_1,\dots,X_n$ be i.i.d.\ with distribution in a model $\cP\defeq\{P_\theta:\theta\in\Theta\}$ dominated by a $\sigma$-finite measure, with densities $p_\theta$, chosen equal for equal distributions, and log-likelihood $L_n(\theta)\defeq\sum_{i=1}^n\log p_\theta(X_i)$. Define $\theta_1\sim\theta_2$ if $P_{\theta_1}=P_{\theta_2}$, write $[\theta]$ for the equivalence class and $\Theta/{\sim}$ for the quotient. A null hypothesis is a set of equivalence classes. Let $\Theta_0\subset\Theta$ be the union of these classes, so that the hypothesis is the set $\Theta_0/{\sim}$ of points of the quotient, and
\[
H_0:\ [\theta]\in\Theta_0/{\sim},\qquad
\mathrm{LR}_n=2\Bigl[\sup_{\theta\in\Theta}L_n(\theta)-\sup_{\theta\in\Theta_0}L_n(\theta)\Bigr].
\]
Since $L_n(\theta)$ depends on $\theta$ only through $P_\theta$, $\mathrm{LR}_n$ depends only on the sets $\cP$ and $\cP_0\defeq\{P_\theta:\theta\in\Theta_0\}$. The map $[\theta]\mapsto P_\theta$ is a bijection from $\Theta/{\sim}$ onto $\cP$, so a point of the quotient is a distribution, and the quotient map $\theta\mapsto[\theta]$ is the minimal sufficient parameter of \citet{Barankin1960} and \citet{Picci1977}. In a test of the number of mixture components of an identifiable family, a point of the null is a mixing distribution with at most $m_0$ atoms. A restriction on a parameter that is not identified everywhere, such as $\beta=\beta_0$ in the IV example below, is read as the set of classes that contain a parameter satisfying it: $H_0$ states that the distribution is compatible with the restriction, and $\mathrm{LR}_n$ is unchanged because $L_n(\theta)$ is constant on classes.

The restricted estimator is $\tilde\theta\in\arg\max_{\theta\in\Theta_0}L_n(\theta)$. The parametric bootstrap draws $X_1^*,\dots,X_n^*$ i.i.d.\ from $P_{\tilde\theta}$ and computes $\mathrm{LR}_n^*$ by the same formula. Write $\mathcal L^*$ for the conditional distribution given the data. The bootstrap is \emph{consistent} for a distribution $\mathcal L_0$ if $d_{\mathrm{BL}}(\mathcal L^*(\mathrm{LR}_n^*),\mathcal L_0)\conp 0$.

\begin{lemma}[Quotient reduction]\label{lem:quotient}
	The conditional distribution of $\mathrm{LR}_n^*$ given the data is the distribution of $\mathrm{LR}_n$ under $P_{\tilde\theta}^n$. It depends on $\tilde\theta$ only through $[\tilde\theta]$, and so not on the choice among observationally equivalent maximizers.
\end{lemma}
\noindent The lemma holds because $\mathrm{LR}_n^*$ is the same function of the bootstrap sample as $\mathrm{LR}_n$ is of the data, and every element of $[\tilde\theta]$ gives the same $P_{\tilde\theta}$. The bootstrap is thus centered at the fitted null distribution, not at a point of $\Theta$.

\paragraph*{Example: linear IV.}
Let $(y_i,x_i,z_i')'$ be i.i.d.\ with $z_i\in\R^k$, $k\ge2$, $\Sigma_{zz}\defeq\E[z_iz_i']$ positive definite, $y=x\beta+u$, $x=z'\pi+v$ and $(u,v)'\,\vert\, z\sim \mathcal N(0,\Omega)$. The reduced form is $r\defeq(y,x)'\,\vert\, z\sim\mathcal N(\Pi'z,\Sigma)$ with $\Pi=\pi[\beta\ \,1]$, so, with the distribution of $z_i$ fixed, a point of the quotient is a pair $(\Pi,\Sigma)$, the closure of the set of coefficients $\Pi$ is $\mathcal V_1\defeq\{\Pi\in\R^{k\times2}:\operatorname{rank}\Pi\le1\}$, and at $\pi=0$ the class of $\theta=(\beta,\pi',(\operatorname{vech}\Omega)')'$ contains every $\beta$. Under instrument irrelevance, $H_0:\pi=0$, the restricted estimate of $\beta$ is arbitrary, but the bootstrap draws from $\mathcal N(0,n^{-1}\sum_ir_ir_i')$ given $z$ (Lemma~\ref{lem:quotient}). At $\pi=0$, the structural hypothesis $\beta=\beta_0$ and its alternative give the same distributions. As a set of distributions, the hypothesis is $\gamma=\beta_0\pi$ for the reduced-form coefficient $\gamma\defeq\beta\pi$ of $y$, and after replacing $y$ by $y-x\beta_0$, it is $H_0:\gamma=0$.

\subsection{Local geometry}\label{sec:assumptions}
The results require two conditions at the true null distribution: the null model is regular there (Assumption~\ref{ass:null}), and the LR statistic is asymptotically a supremum over efficient scores (Assumption~\ref{ass:expansion}). An expansion of the likelihood with product structure (Assumption~\ref{ass:directional}) implies the second (Proposition~\ref{prop:split}), and Proposition~\ref{prop:model} gives conditions on the model itself for Assumptions~\ref{ass:null} and~\ref{ass:directional}.

Suppose that $H_0$ holds, and write $Q_0\in\cP_0$ for the true distribution, so that $Q_0=P_{\theta_0}$ for every $\theta_0$ in the true equivalence class; the letter $Q$ is reserved for null distributions. Write $\Ltwo$ for the mean-zero functions in $L^2(Q_0)$ with inner product $\ip fg\defeq\E_{Q_0}[fg]$ and norm $\|f\|\defeq\ip ff^{1/2}$. Norms, closures, orthogonality and complements are taken in $L^2(Q_0)$ or, for vectors, in Euclidean space. Functions are fixed versions, for which the decompositions below hold pointwise and on which $\G_n$, defined next, is linear; the closure of a class of functions consists of pointwise limits along $L^2(Q_0)$-convergent sequences, which keeps a Donsker class Donsker \citep[Section~2.10]{vanderVaartWellner1996}. Write $\Pn$ for the empirical measure and $\G_n\defeq n^{1/2}(\Pn-Q_0)$, so that $\G_nf=n^{-1/2}\sum_{i=1}^n(f(X_i)-\E_{Q_0}[f])$.

\begin{definition}[Regular point of the null]\label{def:nullpar}
	$Q_0\in\cP_0$ is a \emph{regular point} of the null model if there are a neighborhood $\mathcal U$ of $Q_0$ in $\cP_0$ for $d_{\mathrm{H}}$, an open set $\mathcal H\subset\R^d$ and a one-to-one map $\eta\mapsto Q_\eta$ from $\mathcal H$ onto $\mathcal U$, with $Q_{\eta_0}=Q_0$, that is differentiable in quadratic mean (DQM) at $\eta_0$ with score $s_0\in\Ltwo^{\,d}$ and nonsingular information $I_0\defeq\E_{Q_0}[s_0s_0']$, and such that $\eta_k\to\eta_0$ whenever $d_{\mathrm{H}}(Q_{\eta_k},Q_0)\to0$.
\end{definition}

\noindent Near a regular point, the null model is an identified family, indexed by $\eta$ and DQM at $\eta_0$; nothing is required of the full model. The parameter $\theta$ of the full model may be unidentified, but each null distribution near $Q_0$, that is, each class $[\theta]\subset\Theta_0$ near the true one, has a unique $\eta$, and we write $\overline{L}_n(\eta)\defeq L_n(\theta)$ for any $\theta\in\Theta_0$ with $P_\theta=Q_\eta$; the value does not depend on the choice of $\theta$. In the tests for a threshold and for an additional hidden unit, $\eta$ omits the threshold location or the index coefficients of the inactive unit. In a test of the number of types, a null distribution with exactly $m_0$ distinct interior atoms and positive weights is a regular point when finite mixtures of the family are identified and the information for the atoms and weights is nonsingular: near it, the atoms and weights, ordered to remove relabeling, then index the null one-to-one. A null distribution with fewer atoms is a singular point of the null model, outside our results.

Let $L\defeq\{c's_0:c\in\R^d\}$ be the nuisance tangent space, $\Pi_L$ the orthogonal projection onto $L$ in $L^2(Q_0)$ and $\Pi^\perp$ the projection onto its orthogonal complement. Because $\mathcal H$ is open and null distributions close to $Q_0$ have parameters close to $\eta_0$, the tangent set of the null model at $Q_0$ is all of $L$. When $P_{\tilde\theta}\in\mathcal U$, let $\tilde\eta$ be the $\eta$ with $Q_\eta=P_{\tilde\theta}$, the parameter of the restricted fit; as $\mathcal U\subset\cP_0$, it maximizes $\overline{L}_n$ over $\mathcal H$.

\begin{assumption}[Regular identified null]\label{ass:null}
	$Q_0$ is a regular point of the null model, and:
	\begin{enumerate}[label=(A\arabic*)]
		\setcounter{enumi}{-1}
		\item\label{A0} $Q_0^n(P_{\tilde\theta}\in\mathcal U)\to1$.
		\item\label{A1} Under $Q_0$, $n^{1/2}(\tilde\eta-\eta_0)=O_p(1)$ and, with $\Delta_n\defeq\G_ns_0$, $\sup_{\theta\in\Theta_0}L_n(\theta)-\overline{L}_n(\eta_0)=\tfrac12\Delta_n'I_0^{-1}\Delta_n+o_p(1)$.
	\end{enumerate}
\end{assumption}
\noindent \ref{A0} follows from consistency of the restricted fit in the quotient topology \citep{Redner1981} when $\mathcal U$ is also a neighborhood of $Q_0$ in that topology. \ref{A1} holds under the usual conditions for maximum likelihood \citep[Chapter~5 and Theorem~16.7]{vanderVaart1998}.

\paragraph*{Example (continued).}
Under irrelevance, the null model is $r\,\vert\, z\sim\mathcal N(0,\Sigma)$, indexed by $\eta=\operatorname{vech}\Sigma$; under $\gamma=0$, it is the Gaussian regression $r\,\vert\, z\sim\mathcal N((0,z'\pi)',\Sigma)$, indexed by $\eta=(\pi',(\operatorname{vech}\Sigma)')'$, regular also at $\pi=0$, where $\beta$ is not identified. Assumption~\ref{ass:null} holds in both, and $L$ is spanned by the scores of $\Sigma$, centered quadratic forms in $r-\Pi_0'z$, and, under $\gamma=0$, the first-stage scores $(z'c)\,e_2'\Sigma_0^{-1}(r-\Pi_0'z)$, $c\in\R^k$, where $(\Pi_0,\Sigma_0)$ is the true reduced form and $e_1,e_2$ are the unit vectors of $\R^2$.

\begin{assumption}[Efficient-score representation]\label{ass:expansion}
	There is a set $\cE\subset\Ltwo$ of unit-norm functions orthogonal to $L$ such that $\cF\defeq\cE\cup\{s_{0,1},\dots,s_{0,d}\}$ is $Q_0$-Donsker and, under $Q_0$,
	\begin{equation}\label{eq:efficientrep}
		\mathrm{LR}_n=\sup_{e\in\cE}\plus{\G_n e}^2+o_p(1).
	\end{equation}
\end{assumption}

\noindent The functions in $\cE$ are normalized efficient scores: scores of the full model with their projection on the nuisance tangent space removed, scaled to unit norm. In the hidden-unit and IV examples, \eqref{eq:efficientrep} is verified directly. In mixture-type models it follows from a directional expansion \citep{DacunhaCastelleGassiat1999,Gassiat2002,LiuShao2003,AzaisGassiatMercadier2009} and the splitting result below.

\begin{assumption}[Directional expansion with product structure]\label{ass:directional}
	There are a cone $K_0\subset\Ltwo$ with $K_0\not\subset L$ and a set $\cD$ with $\Sph(L+K_0)\subset\cD\subset\cl\,\Sph(L+K_0)$ such that, under $Q_0$,
	\begin{equation}\label{eq:directional}
		\sup_{\theta\in\Theta}L_n(\theta)-\overline{L}_n(\eta_0)=\tfrac12\sup_{h\in\cD}\plus{\G_n h}^2+o_p(1),
	\end{equation}
	and, with $\cE\defeq\cl\,\Sph(\Pi^\perp K_0)$, the class $\cE\cup\{s_{0,1},\dots,s_{0,d}\}$ is $Q_0$-Donsker.
\end{assumption}

\noindent Here, $\cD$ is the set of normalized local directions of the full model at $Q_0$, the unit-norm derivatives of its smooth paths through $Q_0$ and their limits (made precise in \ref{M3} below), and $K_0$ is the cone of \emph{departures}, the local directions that the full model adds to the tangent space $L$ of the null. That the two combine freely, $\Sph(L+K_0)\subset\cD$, is a \emph{product structure}, as when the full model is parametrized near the null by a product of a neighborhood of $\eta_0$ and a set of departures.

\begin{proposition}[Splitting]\label{prop:split}
	Assumptions~\ref{ass:null} and~\ref{ass:directional} imply Assumption~\ref{ass:expansion} with $\cE=\cl\,\Sph(\Pi^\perp K_0)$; more precisely, under $Q_0$,
	\[
	\sup_{\theta\in\Theta}L_n(\theta)-\overline{L}_n(\eta_0)=\tfrac12\Delta_n'I_0^{-1}\Delta_n+\tfrac12\sup_{e\in\cE}\plus{\G_ne}^2+o_p(1).
	\]
\end{proposition}

\noindent The nuisance term cancels against \ref{A1} in $\mathrm{LR}_n$. Departures may approach the nuisance space, as when a new type merges with an existing one; the closure in $\cE$ then adds the limits of their normalized efficient parts.

For $P\in\cP$ with $P\ll Q_0$, let $\chi_P\defeq\E_{Q_0}[(dP/dQ_0-1)^2]^{1/2}$ and, if $0<\chi_P<\infty$, $s_P\defeq(dP/dQ_0-1)/\chi_P$, the direction in which $P$ departs from $Q_0$.

\begin{proposition}[Conditions on the model]\label{prop:model}
	Let $Q_0$ be a regular point of the null model at which, as $\eta\to\eta_0$, $Q_\eta\ll Q_0$ and $\E_{Q_0}[(dQ_\eta/dQ_0-1-(\eta-\eta_0)'s_0)^2]=o(\|\eta-\eta_0\|^2)$. Suppose that, for some $\varepsilon>0$ and a cone $K_0\subset\Ltwo$ with $K_0\not\subset L$:
	\begin{enumerate}[label=(M\arabic*)]
		\item\label{M1} for every $\varepsilon'>0$, w.p.a.1, $L_n(\theta)<\overline{L}_n(\eta_0)$ for every $\theta\in\Theta$ with $d_{\mathrm{H}}(P_\theta,Q_0)>\varepsilon'$;
		\item\label{M2} every $P\in\cP$ with $0<d_{\mathrm{H}}(P,Q_0)\le\varepsilon$ has $P\ll Q_0$ and $\chi_P<\infty$, and their $s_P$ form a class with a finite bracketing integral in $L^2(Q_0)$;
		\item\label{M3} every limit point of $s_P$ as $\chi_P\to0$ lies in $\cl\,\Sph(L+K_0)$; every $d\in\Sph(L+K_0)$ is the derivative of a path in $\cP$, that is, $\E_{Q_0}[(dP_u/dQ_0-1-ud)^2]=o(u^2)$ as $u\downarrow0$ for some $P_u\in\cP$; and $\cl\,\Sph(\Pi^\perp K_0)$ is $Q_0$-Donsker.
	\end{enumerate}
	Then, Assumptions~\ref{ass:null} and~\ref{ass:directional} hold, with $\cD=\cl\,\Sph(L+K_0)$.
\end{proposition}

\noindent Conditions \ref{M1}--\ref{M3} restrict the model, not the LR statistic. \ref{M1} holds under Wald's conditions on the quotient \citep{Redner1981}. \ref{M2} and \ref{M3} are the conditions of \citet{Gassiat2002} for a simple null, imposed only on a Hellinger ball, as in \citet{LiuShao2003}, so that Gaussian scale parameters are allowed; the proof (Section~\ref{app:auxgeom}) applies her argument to the full and the null model. The product structure is in \ref{M3}: derivatives of paths fill $L+K_0$. In parametric models, \ref{M2} and \ref{M3} can be checked by expanding $dP_\theta/dQ_0-1$ to first order in the distance-and-direction parametrization of \citet{ChoWhite2018}; in tests of the number of types, the first-order term vanishes in some directions, and locally conic expansions of second order are needed \citep{DacunhaCastelleGassiat1999}.

\paragraph*{Example (continued).}
Under instrument irrelevance, the score along the path $\Pi=uH$, $H\in\R^{k\times2}$, is $z'H\Sigma_0^{-1}r$, which is linear in $r$, whereas the scores of $\Sigma$ are centered quadratic forms in $r$. Since $r$ is centered Gaussian given $z$, their product is odd in $r$ and has mean zero, so the two sets are orthogonal. As $\Pi$ and $\Sigma$ vary independently, the local directions form $L+K_0$ with $K_0\defeq\{z'H\Sigma_0^{-1}r:H\in\mathcal V_1\}\perp L$, and normalizing gives $\cE=\{(a'\Sigma_{zz}^{-1/2}z)(b'\Sigma_0^{-1/2}r):\|a\|=\|b\|=1\}$. Thus, $\G_ne=a'\Xi_nb$ with $\Xi_n\defeq n^{-1/2}\sum_i\Sigma_{zz}^{-1/2}z_ir_i'\Sigma_0^{-1/2}$, and $\mathrm{LR}_n=\sigma_1(\Xi_n)^2+o_p(1)$, with $\sigma_1\ge\sigma_2$ the singular values; the limit is the largest eigenvalue of a $\mathrm{Wishart}_2(k,I_2)$ matrix, that of the reduced-rank regression statistic of \citet{Anderson1951}.

\section{Bootstrap validity}\label{sec:main}

\subsection{Sufficient conditions}\label{sec:suff}

Let $\G$ be a $Q_0$-Brownian bridge on $\cF$, a tight centered Gaussian process with $\operatorname{Cov}(\G f,\G g)=\ip fg$, and define
\[
\Lambda\defeq\sup_{e\in\cE}\plus{\G e}^2,\qquad F(x)\defeq\Pr(\Lambda\le x).
\]

\begin{theorem}[Null limit and drift invariance]\label{thm:drift}
Under Assumptions~\ref{ass:null} and~\ref{ass:expansion}, let $a_n\to a\in\R^d$ and $\eta_n\defeq\eta_0+a_n/n^{1/2}$. Under $Q_{\eta_n}^n$, \inpart{thm:drift:a}{a}~$\G_n\dto\G$ in $\ell^\infty(\cE)$, with $\G_n$ still centered at $Q_0$; and \inpart{thm:drift:b}{b}~$\mathrm{LR}_n\dto\Lambda$. In particular, $\mathrm{LR}_n\dto\Lambda$ under $Q_0^n$.
\end{theorem}

\noindent By Le Cam's third lemma, the drift shifts $\G_nf$ by $\ip f{a's_0}$. On the nuisance space this shift enters $\mathrm{LR}_n$ only through $\Delta_n'I_0^{-1}\Delta_n$, which has canceled; on $\cE$ it is zero because $\cE\perp L$. The argument uses both linearity of $L$ and orthogonality of $\cE$ to $L$: at a boundary point of the null parameter space the null tangent set is in general a cone that is not linear, and the bootstrap can fail \citep{Andrews2000}.

\begin{theorem}[Bootstrap validity]\label{thm:boot}
Under Assumptions~\ref{ass:null} and~\ref{ass:expansion}, the restricted-fit bootstrap is consistent for $\mathcal L(\Lambda)$. If also, for a level $\alpha\in(0,1)$, $F$ is continuous and strictly increasing at $c_\alpha\defeq F^{-1}(1-\alpha)$, and $c_n^*(\alpha)$ is the $(1-\alpha)$ quantile of $\mathcal L^*(\mathrm{LR}_n^*)$, then $c_n^*(\alpha)\conp c_\alpha$ and $Q_0^n(\mathrm{LR}_n>c_n^*(\alpha))\to\alpha$.
\end{theorem}

\noindent The continuity condition is mild: the distribution of the square of the positive part of the supremum of a centered Gaussian process is continuous except possibly at the left end of its support \citep{Tsirelson1975}. In regressions, one holds the covariates fixed and draws $y_i^*\,\vert\, x_i$ from the fitted conditional distribution; the proofs carry over conditionally on almost every covariate sequence when the assumptions hold in their triangular-array forms, for the empirical process and for local asymptotic normality \citep[Section~2.11]{vanderVaartWellner1996}.

\paragraph*{Example (continued).}
For instrument irrelevance, Assumptions~\ref{ass:null} and~\ref{ass:expansion} hold (Section~\ref{sec:assumptions} and Proposition~\ref{prop:iv}), so the limit of $\mathrm{LR}_n$ is the same along every root-$n$ drift of $\Sigma$ and the restricted-fit bootstrap is consistent (Theorems~\ref{thm:drift} and~\ref{thm:boot}). It is even exact, because given $z$ the null distribution of $\mathrm{LR}_n$ is free of $\Sigma$.

\subsection{Necessary and sufficient conditions}\label{sec:exact}

Two properties of the problem at $Q_0$ were used above:
\begin{enumerate}[label=(G\arabic*),leftmargin=*]
\item\label{G1} \emph{The null tangent set is a linear space $L$} (Definition~\ref{def:nullpar}).
\item\label{G2} \emph{The alternative splits off the nuisance}: the local directions of the full model form $L+K_0$ (Assumption~\ref{ass:directional}).
\end{enumerate}

\noindent Without \ref{G2}, the limit of $\mathrm{LR}_n$ can vary with the local position of the null, and the bootstrap, which draws from the fitted null, reproduces the limit at the fitted position, which may differ from that at the true one. To make this precise, let $\mathcal K$ be a closed set of local null parameters containing $0$, and for $h\in\mathcal K$ let $\theta_n(h)\in\Theta_0$ be a null sequence with $\theta_n(0)=\theta_0$; in the regular case, $\mathcal K=\R^d$ and $P_{\theta_n(h)}=Q_{\eta_0+I_0^{-1/2}h/n^{1/2}}$. Suppose that, w.p.a.1, $P_{\tilde\theta}=P_{\theta_n(\hat H_n)}$ for a random $\hat H_n\in\mathcal K$, the \emph{local position} of the restricted fit; in the regular case, $\hat H_n=I_0^{1/2}n^{1/2}(\tilde\eta-\eta_0)$.

\begin{proposition}[Consistency and local constancy]\label{prop:dichotomy}
Suppose that $\mathrm{LR}_n\dto\Lambda(h)$ under $P^n_{\theta_n(h_n)}$ for every $h\in\mathcal K$ and every sequence $h_n\to h$ in $\mathcal K$, write $\mu(h)\defeq\mathcal L(\Lambda(h))$, and let $\hat H_n$ be tight. Then, \inpart{prop:dichotomy:a}{a}~$h\mapsto\mu(h)$ is continuous for $d_{\mathrm{BL}}$ and $d_{\mathrm{BL}}(\mathcal L^*(\mathrm{LR}_n^*),\mu(\hat H_n))\conp0$; \inpart{prop:dichotomy:b}{b}~if $\hat H_n\dto H$, the bootstrap is consistent for $\mu(0)$ if and only if $\mu(H)=\mu(0)$ almost surely.
\end{proposition}

\noindent This is the local-limit logic of \citet[Theorem~2.3]{Beran1997}; \citet[Theorem~1]{DrtonWilliams2011} argue similarly to express the bootstrap test's asymptotic size in the limit experiment. Theorem~\ref{thm:boot} is the case of constant $\mu$; in the failures below, the bootstrap reproduces the random $\mu(H)$ instead of $\mu(0)$. Inconsistency does not always distort the size: $\mu(h)$ can vary with $h$ while its $(1-\alpha)$ quantile stays fixed.

\paragraph*{Gaussian limit experiments.}
We now characterize $\mu(h)$, the limit law of $\mathrm{LR}_n$ under $\theta_n(h)$, when the local directions of the null and full models lie in a finite-dimensional space $V\defeq\{c's_V:c\in\R^m\}$, for some $s_V\in\Ltwo^m$ with nonsingular $J_V\defeq\E_{Q_0}[s_Vs_V']$: $\mu(h)$ is then the distribution of an LR statistic in a Gaussian shift experiment. We identify $c's_V\in V$ with $h\defeq J_V^{1/2}c\in\R^m$, which has the same norm, and let $Z_n\defeq J_V^{-1/2}\G_ns_V$, the standardized score, so that $\G_n(c's_V)=h'Z_n$.

The closures of the local directions of the null and full models are then cones $C_0\subseteq C_1$ in $\R^m$. If $P_u\in\cP_0$ is a path with derivative $c's_V$ as in~\ref{M3}, Le Cam's third lemma gives $Z_n\dto\mathcal N(h,I_m)$ under $P^n_{n^{-1/2}}$. For a nonempty closed set $C$, let
\begin{equation*}
\Psi_C(z)\defeq\sup_{v\in C}(2z'v-\|v\|^2)=\sup_{v\in C}(\|z\|^2-\|z-v\|^2)=\|z\|^2-d(z,C)^2,
\end{equation*}
the LR statistic given $z$ for $\mathcal N(0,I_m)$ against $\mathcal N(v,I_m)$, $v\in C$. For a cone, maximizing over $t\ge0$ with $v=tu$, $u\in\Sph(C)$, gives $\Psi_C(z)=\sup_{u\in\Sph(C)}\plus{u'z}^2$, with $\sup\emptyset=0$. This is the form of the directional expansion behind Proposition~\ref{prop:model} (Lemma~\ref{lem:direxp}), which, applied to both models under its conditions, gives $\mathrm{LR}_n=T(Z_n)+o_p(1)$, where
\begin{equation}\label{eq:coneLRdiagnostic}
T(z)\defeq\Psi_{C_1}(z)-\Psi_{C_0}(z)=d(z,C_0)^2-d(z,C_1)^2
\end{equation}
is the LR statistic for $h\in C_0$ against $h\in C_1$ given an observation from $\mathcal N(h,I_m)$. Hence, with $Z\sim\mathcal N(0,I_m)$, $\mu(h)=\mathcal L(T(Z+h))$. When $C_0=L$, we take $s_V=(s_0',s_\perp')'$ with the components of $s_\perp$ orthogonal to $L$, so that $J_V$ is block diagonal with blocks $I_0$ and $\E_{Q_0}[s_\perp s_\perp']$ and $L$ is represented by $\R^d\times\{0\}$, and write $z=(z_L',z_\perp')'$ with $z_L\in\R^d$, and $f(z_L,z_\perp)$ for $f(z)$ when $f$ is a function of $z$. Then $Z_{n,L}=I_0^{-1/2}\Delta_n$ is the standardized nuisance score, $Z_{n,\perp}=\E_{Q_0}[s_\perp s_\perp']^{-1/2}\G_ns_\perp$ is the standardized efficient score, and $T(z)=\|z_\perp\|^2-d(z,C_1)^2$, because $\Psi_L(z)=\|z_L\|^2$. Under Assumption~\ref{ass:directional} with $L+K_0\subset V$, $C_1=\cl(L+K_0)$ and the expansion is Proposition~\ref{prop:split}, a case that \citet[Example~2]{DrtonWilliams2011} note in the limit experiment.

\begin{theorem}[Gaussian-shift criterion]\label{thm:diagnostic}
Let $C_0\subseteq C_1$ be nonempty closed cones in $\R^m$, $Z\sim \mathcal N(0,I_m)$ and $T$ as in~\eqref{eq:coneLRdiagnostic}. Suppose the conditions of Proposition~\ref{prop:dichotomy} hold with $\mathcal K=C_0$ and $\mu(h)=\mathcal L(T(Z+h))$, and $\hat H_n\dto H$, supported in $C_0$.
\begin{enumerate}[label=(\alph*)]
\item\label{thm:diagnostic:a} The bootstrap is consistent for $\mathcal L(T(Z))$ if and only if $T(Z+h)$ has the distribution of $T(Z)$ for all $h\in\operatorname{supp}H$. If the local restricted fit converges to a Euclidean projection of $Z$ onto $C_0$, then $\operatorname{supp}H=C_0$.
\item\label{thm:diagnostic:b} If $C_0=L$ is a linear subspace and $\operatorname{supp}H=L$, consistency is equivalent to
\begin{equation}\label{eq:diagnosticindependence}
T(Z)\indep Z_L,\qquad Z_L\defeq\Pi_L Z.
\end{equation}
\item\label{thm:diagnostic:c} If $C_0=L$, these are equivalent:
$T(z+u)=T(z)$ for all $z\in\R^m$ and $u\in L$; $C_1+L=C_1$; $C_1=L+(C_1\cap L^\perp)$. Each implies~\eqref{eq:diagnosticindependence} and holds if $C_1$ is convex.
\end{enumerate}
\end{theorem}

\begin{figure}[t]
\begin{center}
\caption{Alternative cones in the Gaussian limit experiment}\label{fig:cones}
\begin{tikzpicture}[scale=0.85,>=stealth]
\begin{scope}
\fill[gray!22] (-2.7,0) rectangle (2.7,2.2);
\draw[very thick] (-2.7,0) -- (2.7,0) node[right] {$L$};
\draw[->,thick] (0,0) -- (0.9,1.6) node[right] {$K_0$};
\node[anchor=west] at (-2.6,1.75) {$C_1=L+K_0$};
\coordinate (z) at (-1.6,-0.95);
\coordinate (zu) at (1.3,-0.95);
\draw[dashed,thick] (z) -- (-1.6,0);
\draw[dashed,thick] (zu) -- (1.3,0);
\fill (z) circle (1.6pt) node[left] {$z$};
\fill (zu) circle (1.6pt) node[right] {$z+u$};
\draw[->] (-1.45,-1.35) -- (1.15,-1.35) node[midway,below] {\small $u\in L$};
\node at (0,-2.25) {\small Irrelevance: $T(z+u)=T(z)$};
\end{scope}
\begin{scope}[xshift=7.6cm]
\fill[gray!22] (0,0) -- (2.7,1.89) -- (2.7,-1.89) -- cycle;
\fill[gray!22] (0,0) -- (-2.7,1.89) -- (-2.7,-1.89) -- cycle;
\draw[thin] (-2.7,-1.89) -- (2.7,1.89);
\draw[thin] (-2.7,1.89) -- (2.7,-1.89);
\draw[very thick] (-2.7,0) -- (2.7,0) node[right] {$L$};
\node at (-1.95,0.55) {$C_1$};
\coordinate (z) at (0.2,1.25);
\coordinate (zu) at (2.2,1.25);
\draw[dashed,thick] (z) -- (0.722,0.505);
\fill (z) circle (1.6pt) node[above left=-1pt] {$z$};
\fill (zu) circle (1.6pt) node[right] {$z+u$};
\draw[->] (0.35,1.95) -- (2.05,1.95) node[midway,above] {\small $u\in L$};
\node at (0,-2.25) {\small Structural, $\pi=0$: $T(z+u)\neq T(z)$};
\end{scope}
\end{tikzpicture}
\end{center}
\floatnotes{Shaded regions are $C_1$, $L$ is the nuisance tangent space and $T(z)=d(z,L)^2-d(z,C_1)^2$; dashed segments are distances to $C_1$. Left: translation along $L$ maps $C_1=L+K_0$ onto itself, so the distribution of $T$ does not depend on the nuisance position, as in the IV test of irrelevance. Right: a fan of lines through the origin containing $L$, as for the structural IV null at $\pi=0$ and in the endogenous-treatment model; translation along $L$ changes $T$.}
\end{figure}
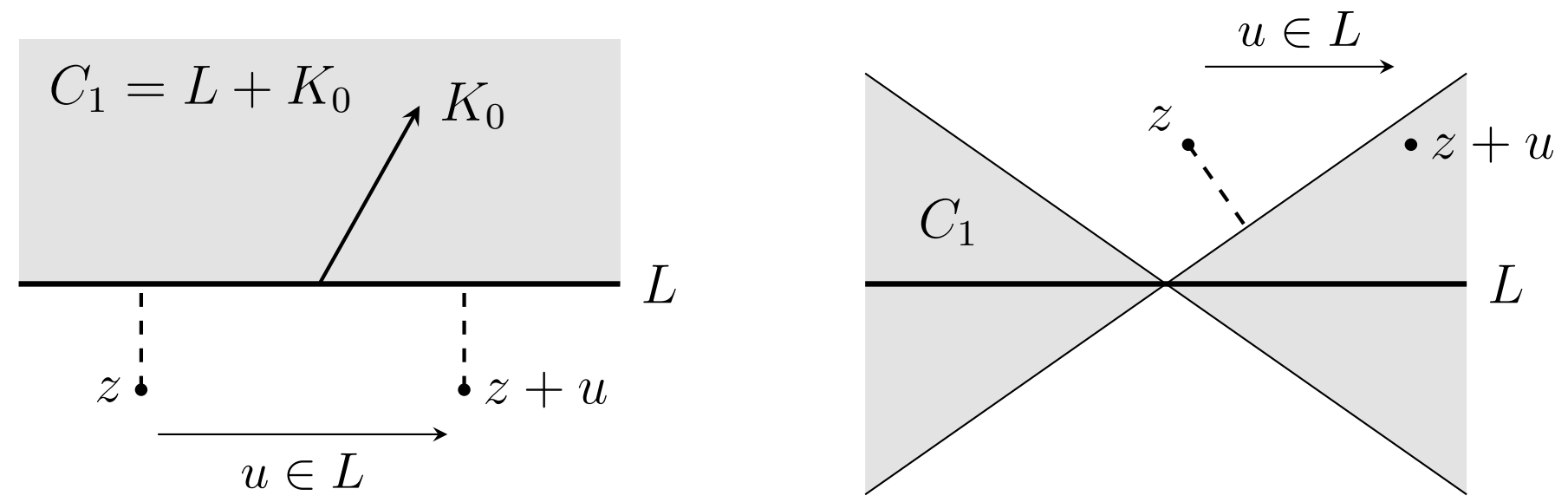

\noindent Part~\ref{thm:diagnostic:c} gives a geometric sufficient condition for~\eqref{eq:diagnosticindependence}, translation invariance of $T$ along $L$ (Figure~\ref{fig:cones}); we do not know whether it is necessary for every closed cone $C_1$.

\paragraph*{Example (continued).}
In the limit experiment of the irrelevance test, $z_\perp=\operatorname{vec}\xi$, with $\xi\in\R^{k\times2}$ the limit of $\Xi_n$, $C_1=L\oplus\operatorname{vec}\mathcal V_1$, and $T(z)=\|\xi\|^2-d(\xi,\mathcal V_1)^2=\sigma_1(\xi)^2$ does not involve $z_L$, as in part~\ref{thm:diagnostic:c} and the left panel of Figure~\ref{fig:cones}. For $H_0:\gamma=0$, let $b_t\defeq\Sigma_0^{-1/2}e_2/\|\Sigma_0^{-1/2}e_2\|$ and $b_s\defeq\Sigma_0^{1/2}e_1/\|\Sigma_0^{1/2}e_1\|$, unit vectors of $\R^2$ that are orthogonal because $e_2'\Sigma_0^{-1/2}\Sigma_0^{1/2}e_1=0$, and let $s\defeq\xi b_s$ and $t\defeq\xi b_t$ in $\R^k$, so that $[s\ \,t]=\xi[b_s\ \,b_t]$ has the singular values of $\xi$. The first-stage scores, now in $L$, are represented by $t$, and the efficient score of $\gamma$ by $s$. At $\pi=0$, $C_1$ consists of the scores of $\Sigma$ plus the matrices $[s\ \,t]$ of rank at most one, so $C_1\supset L$ but $C_1+L\ne C_1$, and
\[
T(z)=\|s\|^2-\sigma_2([s\ \,t])^2=\sigma_1([s\ \,t])^2-\|t\|^2
\]
depends on $t$, as in the right panel of Figure~\ref{fig:cones}. The theorem applies with $C_0=L$: $\mathrm{LR}_n\dto T(Z+h)$ along root-$n$ drifts $h\in L$ of the null, and the local position of the restricted fit converges to $Z_L$, so that $\operatorname{supp}H=L$ (part~\ref{thm:diagnostic:a} and Proposition~\ref{prop:iv2}). Thus, \ref{G2} fails, the distribution of $T(Z+h)$ varies with $h$ (Proposition~\ref{prop:iv2}), and by part~\ref{thm:diagnostic:b} the bootstrap is inconsistent. At a fixed $\pi_0\ne0$, where $\beta$ is identified, the alternative adds only the direction $[\pi_0\ \,0]$ to $L$, and $T(Z)\sim\chi^2_1$ whatever $Z_L$.

\paragraph*{Plug-in bootstraps.}
A plug-in bootstrap resamples from the unrestricted fit, so its draws depend on $Z_n$ in every direction. It fails for nondifferentiable functions of asymptotically normal estimators, such as the absolute value \citetext{\citealp{FangSantos2019}; see also \citealp{HiranoPorter2012}}. The restricted-fit draws $T(\hat H_n+Z^*)$ depend on the data only through $\hat H_n\in L$, so, even for such functions, invariance in distribution of $T$ along $L$ suffices, as in Theorem~\ref{thm:boot}.

\section{Where the ordinary bootstrap works}\label{sec:examples}

At the regular null points below, tests of the number of types and of an additional hidden unit satisfy \ref{G1} and \ref{G2}. In each, the likelihood expansion is assumed or follows from known results, and Proposition~\ref{prop:split} and Theorems~\ref{thm:drift} and~\ref{thm:boot} explain why the bootstrap works, including cases where its validity was proved directly.

\paragraph*{Unobserved types.}
Consider testing the number of unobserved types, $m_0$ against $m_0+1$ components of a finite mixture. At a null mixture with distinct interior components and positive weights, the null is regular under the conditions of Section~\ref{sec:assumptions}, and the directions in which an additional type moves the density, a new component with small weight or the split of an existing one, form a cone that adds to the tangent space $L$ of the $m_0$-type model, as in \ref{G2}. The limit depends only on the parts of these directions orthogonal to $L$, so root-$n$ changes in the null mixture leave it unchanged. Where the expansion~\eqref{eq:directional} holds \citep{DacunhaCastelleGassiat1999,Gassiat2002,LiuShao2003,AzaisGassiatMercadier2009}, the bootstrap from the restricted $m_0$-type fit is therefore consistent, as \citet{GuKoenkerVolgushev2018} prove for ${m_0=1}$ and \citet{KasaharaShimotsu2019} for a penalized statistic in normal mixtures.

\paragraph*{One more hidden unit.}\label{sec:nn}

Let $y$ be binary, $x\in\R^{d_x}$, $\tilde x\defeq(1,x')'$ and $\ell(t)\defeq1/(1+e^{-t})$. A hidden unit, the building block of a neural network, is a nonlinear function $\psi(x;\gamma)$ of the covariates, typically $\varphi(\gamma'\tilde x)$ for a known $\varphi$, with $\gamma$ the intercept and slopes of an index. Adding one to a logit gives
\[
\Pr(y=1\,\vert\, x)=\ell(\beta'\tilde x+a\,\psi(x;\gamma)),\qquad \theta\defeq(\beta',a,\gamma')',
\]
with $\beta\in B\subset\R^{d_x+1}$ compact, $a\in[-A,A]$ for some $A>0$ and $\gamma$ in a compact set $\Gamma$. Under $H_0:a=0$, the unit is inactive, $\gamma$ is not identified, and the null model is the logit model indexed by $\eta=\beta$. Write $p_0(x)\defeq\ell(\beta_0'\tilde x)$, $z(x;\gamma)\defeq(\tilde x',\psi(x;\gamma))'$, $J(\gamma)\defeq\E[p_0(1-p_0)z(x;\gamma)z(x;\gamma)']$, $I_0\defeq\E[p_0(1-p_0)\tilde x\tilde x']$,
\[
c(\gamma)\defeq I_0^{-1}\E[p_0(1-p_0)\tilde x\psi(x;\gamma)],\qquad
\omega(\gamma)^2\defeq\E\bigl[p_0(1-p_0)\bigl(\psi(x;\gamma)-c(\gamma)'\tilde x\bigr)^2\bigr],
\]
and $e(y,x;\gamma)\defeq(\psi(x;\gamma)-c(\gamma)'\tilde x)(y-p_0(x))/\omega(\gamma)$.

\begin{proposition}[One hidden unit]\label{prop:nn}
Let $\beta_0$ be interior to $B$ and suppose that $\E[\|x\|^4]<\infty$, $|\psi(x;\gamma)|\le C(1+\|x\|)$, the class $\{\psi(x;\gamma)(y-p_0(x)):\gamma\in\Gamma\}$ is $Q_0$-Donsker, the class $\{p_0(1-p_0)z(x;\gamma)z(x;\gamma)':\gamma\in\Gamma\}$ is $Q_0$-Glivenko--Cantelli, $J$ is continuous, and $\inf_{\gamma\in\Gamma}\lambda_{\min}(J(\gamma))>0$. Then, Assumptions~\ref{ass:null} and~\ref{ass:expansion} hold with $L=\{v'\tilde x(y-p_0):v\in\R^{d_x+1}\}$ and $\cE=\{\pm e(\cdot;\gamma):\gamma\in\Gamma\}$, so $\mathrm{LR}_n\dto\Lambda=\sup_{\gamma\in\Gamma}(\G e(\cdot;\gamma))^2$ under $Q_{\beta_0}$ and along every root-$n$ drift of $\beta$, and the restricted-fit bootstrap is consistent.
\end{proposition}

\noindent The same reasoning applies to tests for a threshold effect in regression with Gaussian errors \citep{Hansen1996}. If $x$ has a density positive on an open set met by the hyperplanes on which $\gamma'\tilde x=0$, the conditions cover a sigmoid $\varphi$, such as $\ell$, with slopes bounded away from zero; $\varphi(t)=\max(t,0)$, which for scalar $x$ gives a kink of unknown location \citep{Hansen2017}; and $\varphi(t)=1\{t>0\}$, which gives a discontinuity at an unknown cutoff and, interacted with a treatment indicator, a change-plane model for subgroup detection \citep{FanSongLu2017}. In each case the null hypothesis is that the term is absent; inference on its location when the term may be present is a different problem, with an entangled alternative (Section~\ref{sec:nonlin}). Without compactness of $\Gamma$, the LR statistic can diverge \citep{Fukumizu2003}; \citet{Rynkiewicz2008} derives its limit for regression networks with an overestimated number of hidden units.

\section{Failure and repair}\label{sec:fail}

The bootstrap fails exactly when the limit distribution varies on the support of the local null fit (Proposition~\ref{prop:dichotomy}). When \ref{G1} fails, through a boundary nuisance parameter or a singular null such as $\theta_1\theta_2=0$ in mediation analysis, the bootstrap generally fails \citep{Andrews2000,Drton2009,DrtonWilliams2011}, and the restricted fit may not be sufficient for the local null position; Proposition~\ref{prop:dichotomy} and Theorem~\ref{thm:diagnostic}\ref{thm:diagnostic:a} still apply, but the repairs below do not. This section treats the case in which \ref{G1} holds and \ref{G2} fails, as for the structural null in the IV example. The hypotheses are the usual ones, such as $\beta=\beta_0$, read as sets of distributions as in Section~\ref{sec:setup}; the conditional bootstrap keeps the usual LR statistic and changes only the critical value.

\subsection{Entangled alternatives and double counting}\label{sec:entangled}

The two IV nulls show what a repair must do. Both are regular, but for $H_0:\pi=0$ the alternative splits off the nuisance directions and the restricted-fit bootstrap is exact (Section~\ref{sec:suff}), whereas for $H_0:\gamma=0$ at $\pi=0$ the alternative is entangled with the first stage, the limit $T(z)$ depends on the first-stage score $t$, and the bootstrap is inconsistent (Section~\ref{sec:exact}). That limit is $\mathrm{CLR}(s,t)\defeq\sigma_1([s\ \,t])^2-\|t\|^2$, the conditional LR statistic of \citet{Moreira2003}. In a sample, $\mathrm{LR}_n=\mathrm{CLR}(S_n,T_n)+o_p(1)$ for his orthogonal statistics $S_n$ and $T_n$, the sample counterparts of $s$ and $t$, asymptotically independent, and only the mean of $T_n$ moves with the first stage. In this example, the conditional bootstrap below gives the conditional LR test of \citet{Moreira2003}, and orbit completion gives the \citet{AndersonRubin1949} statistic. Our contribution is to explain why the ordinary bootstrap fails although the null is regular, and to repair it in general. We use the representation of Section~\ref{sec:exact} with $C_0=L$.

\begin{assumption}[Score representation]\label{ass:score}
Assumption~\ref{ass:null} holds, and:
\begin{enumerate}[label=(S\arabic*)]
\item\label{S1} There are a finite-dimensional subspace $V\subset\Ltwo$ containing $L$ and a closed cone $C_1\subset\R^m$ containing $L$ such that $\mathrm{LR}_n=T(Z_n)+o_p(1)$ under $Q_0$, with $Z_n$ the standardized score of Section~\ref{sec:exact} and $T$ as in~\eqref{eq:coneLRdiagnostic} for $C_0=L$.
\item\label{S2} $n^{1/2}(\tilde\eta-\eta_0)=I_0^{-1}\Delta_n+o_p(1)$ under $Q_0$.
\end{enumerate}
\end{assumption}

\noindent \ref{S1} is the expansion of Section~\ref{sec:exact} with $C_0=L$, as for parameters on a boundary \citep{Andrews2001}. It holds under the conditions of Proposition~\ref{prop:model} with \ref{M3} replaced by the requirement that the limit points of $s_P$ as $\chi_P\to0$ lie in $\Sph(C_1)$ and a dense subset of $\Sph(C_1)$ consists of derivatives of paths in $\cP$: Lemma~\ref{lem:direxp} then gives $\sup_{\theta\in\Theta}L_n(\theta)-\overline{L}_n(\eta_0)=\tfrac12\Psi_{C_1}(Z_n)+o_p(1)$, and the quadratic term in \ref{A1} is $\tfrac12\Psi_L(Z_n)$. Propositions~\ref{prop:iv2} and~\ref{prop:hm} verify \ref{S1} directly; in the IV example at $\pi=0$, $z_L$ collects the covariance scores and $t$, and $z_\perp=s$. The local position of the restricted fit is then $\hat H_n=I_0^{1/2}n^{1/2}(\tilde\eta-\eta_0)=Z_{n,L}+o_p(1)$ (Section~\ref{sec:exact}), and we let $Z=(Z_L',Z_\perp')'\sim\mathcal N(0,I_m)$.

\begin{lemma}[Fitted shift]\label{lem:fittedshift}
Under Assumption~\ref{ass:score}, let $a_n\to a$ and $h\defeq I_0^{1/2}a$. Under $Q^n_{\eta_0+a_n/n^{1/2}}$, $(\mathrm{LR}_n,\hat H_n)\dto(T(h+Z),\,h+Z_L)$, and the convergence in $d_{\mathrm{BL}}$ is uniform over drifts with $\|a\|\le M$, for every $M$. Consequently, with $\hat H_n^*\defeq I_0^{1/2}n^{1/2}(\tilde\eta^*-\eta_0)$ computed from the bootstrap sample and $\varrho(h)\defeq\mathcal L(T(h+Z),\,h+Z_L)$,
\[
d_{\mathrm{BL}}\bigl(\mathcal L^*(\mathrm{LR}_n^*,\hat H_n^*),\,\varrho(\hat H_n)\bigr)\conp0.
\]
\end{lemma}

\noindent The ordinary bootstrap distribution of $\mathrm{LR}^*_n$ is thus asymptotically that of $T(\hat H_n+Z^*_L,Z^*_\perp)$ with $\hat H_n\approx h+Z_L$, whereas $\mathrm{LR}_n$ behaves like $T(h+Z_L,Z_\perp)$: the nuisance noise enters twice. The second copy arises in every parametric bootstrap from the restricted fit, because $\hat H_n-h\dto Z_L$: the local position of the null cannot be estimated consistently. In the IV example with $\pi$ local to zero, the bootstrap distribution is approximately that of $\mathrm{CLR}(S^*,T_n+T^*)$ with $(S^*,T^*)$ standard normal, whereas $\mathrm{LR}_n\approx\mathrm{CLR}(S_n,T_n)$ with $T_n$ centered at the local first stage \citep[for a residual bootstrap, see][Theorem~9]{MoreiraPorterSuarez2004}. The second copy is harmless if the distribution of $T(t+Z)$ does not depend on $t\in L$ (Theorem~\ref{thm:diagnostic}), as when the alternative splits. It matters when the alternative is entangled, as when a nuisance parameter such as the first stage governs how well the alternative is identified; weak identification enters only in this way, and the null remains regular. A repair then removes the second copy, by conditioning (Section~\ref{sec:fitcond}) or by changing the statistic (Section~\ref{sec:orbit}).

\subsection{The conditional bootstrap}\label{sec:fitcond}

For $t\in L$, let $F(x\,\vert\, t)\defeq\Pr(T(t,Z_\perp)\le x)$ and $c_\alpha(t)\defeq\inf\{x:F(x\,\vert\, t)\ge1-\alpha\}$. Because $Z_\perp$ is independent of $Z_L$ and its distribution does not depend on $h$, $F(\cdot\,\vert\, t)$ is the conditional distribution of $T(h+Z)$ given $h+Z_L=t$ for every $h\in L$: the nuisance score is sufficient for $h$. Let
\[
L_s\defeq\{h\in L:\ C_1+h=C_1\},\qquad L_e\defeq L\ominus L_s ,
\]
the \emph{harmless} and \emph{entangled} parts of the nuisance tangent space.

\begin{theorem}[Limit experiment]\label{thm:condlimit}
Let $Z\sim \mathcal N(h,I_m)$ with $h\in L$, and $C_1\supset L$ a closed cone.
\begin{enumerate}[label=(\alph*)]
\item\label{thm:condlimit:a} \emph{Similarity.} $\Pr_h(T(Z)>c_\alpha(Z_L))\le\alpha$ for every $h\in L$, with equality if $F(\cdot\,\vert\, t)$ is continuous at $c_\alpha(t)$ for almost every $t$.
\item\label{thm:condlimit:b} \emph{Bootstrap form.} For every $t\in L$, the conditional distribution of $T(t+Z)$ given $Z_L=0$ under $h=0$ is $F(\cdot\,\vert\, t)$.
\item\label{thm:condlimit:c} \emph{When conditioning is unnecessary.} The following are equivalent: $F(\cdot\,\vert\, t)$ does not depend on $t$; $T(Z)$ is independent of $Z_L$ under $h=0$; the distribution of $T(Z)$ does not depend on $h\in L$. Under the conditions of Theorem~\ref{thm:diagnostic}\ref{thm:diagnostic:b} each is equivalent to consistency of the ordinary bootstrap.
\item\label{thm:condlimit:d} \emph{Harmless directions.} $L_s$ is a subspace, $T(z+h)=T(z)$ for all $z$ and $h\in L_s$, and $F(\cdot\,\vert\, t)$ depends on $t$ only through $\Pi_{L_e}t$; \ref{G2} holds if and only if $L_s=L$.
\item\label{thm:condlimit:e} \emph{Necessity.} Every test $\varphi(Z)$ with $\E_h[\varphi(Z)]=\alpha$ for all $h\in L$ satisfies $\E[\varphi(Z)\,\vert\, Z_L]=\alpha$ almost surely.
\end{enumerate}
\end{theorem}

\noindent Part~\ref{thm:condlimit:e} is the Neyman-structure property of a complete sufficient statistic \citep[Theorem~4.3.2]{LehmannRomano2005}, which \citet{Moreira2009tests} uses to characterize the similar tests in the Gaussian IV model. Parts \ref{thm:condlimit:a}, \ref{thm:condlimit:b} and~\ref{thm:condlimit:e} hold for any statistic of $Z$ and part~\ref{thm:condlimit:c} for any continuous one. Part~\ref{thm:condlimit:d} confines conditioning to the entangled directions: in the IV example, $L_s$ contains the covariance scores and $L_e$ the first-stage ones, and $c_\alpha(t)$ is the conditional critical value of \citet{Moreira2003}.

\begin{corollary}[Uniqueness]\label{cor:unique}
Let $F(\cdot\,\vert\, t)$ be continuous and strictly increasing at $c_\alpha(t)$ for every $t\in L$. For $c:L\to\R$, the test that rejects when $T(Z)>c(Z_L)$ has rejection probability $\alpha$ for every $h\in L$ if and only if $c=c_\alpha$ almost everywhere.
\end{corollary}

\noindent A resampling scheme whose limiting rule has this form is either not similar or the conditional test. The corollary does not cover randomized or nonsimilar tests, such as those of \citet{ElliottMullerWatson2015}, nor other statistics: the completed statistic of Section~\ref{sec:orbit}, or the score statistic, which in the IV model is independent of the fit \citep{Kleibergen2002,Moreira2003} and whose residual bootstrap \citet{MoreiraPorterSuarez2009} validate.

\begin{theorem}[Power envelopes and weighted average power]\label{thm:condWAP}
In the experiment $Z\sim\mathcal N(h,I_m)$ with $H_0:h\in L$, let $\kappa=(\kappa_L',\kappa_\perp')'$ be an alternative mean.
\begin{enumerate}[label=(\alph*)]
\item\label{thm:condWAP:a} At every $\kappa$ with $\kappa_\perp\ne0$, the power envelopes over level-$\alpha$ tests, over similar tests and over tests with conditional null rejection probability $\alpha$ at every $Z_L=t$ all equal $1-\Phi(\Phi^{-1}(1-\alpha)-\|\kappa_\perp\|)$, attained by rejecting for large $\kappa_\perp'Z_\perp$.
\item\label{thm:condWAP:b} For a probability measure $\mathcal W$ on alternative means, let $A_{\mathcal W}(z)\defeq\int \exp(\kappa'z-\tfrac12\|\kappa\|^2)\,d\mathcal W(\kappa)$. Among similar level-$\alpha$ tests, the test that rejects when $A_{\mathcal W}(Z)$ exceeds its conditional $(1-\alpha)$ quantile given $Z_L$, randomizing at atoms, maximizes $\int\E_\kappa[\varphi(Z)]\,d\mathcal W(\kappa)$.
\end{enumerate}
\end{theorem}

\noindent The optimal test in~\ref{thm:condWAP:b} is the conditional LR test when $A_{\mathcal W}$ increases with $T$ given $Z_L$, but not in general; in the IV model, \citet{MoreiraMoreira2019} derive such tests in finite samples, and the conditional LR test is close to the two-sided power envelope for invariant similar tests \citep{AndrewsMoreiraStock2006}. Part~\ref{thm:condWAP:a} does not imply equal weighted-average-power bounds for similar and level-$\alpha$ tests, and~\ref{thm:condWAP:b} concerns the limit experiment, not finite samples. Its argument also covers the break-date experiment of \citet{ElliottMullerWatson2015}, where the best similar test is within $0.005$ of their estimated bound over all valid tests (Section~\ref{sec:practicalsim}).

\paragraph*{The conditional bootstrap in a sample.}
By Lemma~\ref{lem:fittedshift} and Theorem~\ref{thm:condlimit}\ref{thm:condlimit:b}, $c_\alpha(\hat H_n)$ is the $(1-\alpha)$ quantile of the limiting bootstrap distribution of $\mathrm{LR}^*_n$ given that the bootstrap restricted fit returns to the observed one. For continuously distributed estimates, this event has probability zero, so the procedure below weights the draws by a kernel in the difference between the bootstrap and the observed restricted fit, with a shrinking bandwidth. This weighting is that of \citet{BoothHallWood1992}, which \citet{MoreiraPorterSuarez2004} apply to the conditional LR test in the IV model. By Theorem~\ref{thm:condlimit}\ref{thm:condlimit:d}, it suffices to condition on components $\eta_e$ of $\eta$, of dimension $d_e\le d$, that carry the entangled part $L_e$, in the sense that $s_{0,j}\in L_s$ for every component $j$ not in $\eta_e$: the scores of the remaining components are harmless. The choice $\eta_e=\eta$ is always allowed; in the examples, $\eta_e$ is the first-stage coefficient vector in the IV model, the take-up effect $\beta$ in Section~\ref{sec:weakiv} and the coefficient $\beta$ of the nonlinear term in Section~\ref{sec:nonlin}, where exact simulation conditions on the whole restricted fit.

Let $\Sigma_e$ be the $\eta_e$ block of $I_0^{-1}$, $\hat\Sigma_e\conp\Sigma_e$, $\Upsilon\defeq\Sigma_e^{-1/2}[I_0^{-1/2}]_{e}$, where $[A]_e$ denotes the rows of $A$ indexed by $\eta_e$, and $D_n^*\defeq\hat\Sigma_e^{-1/2}n^{1/2}(\tilde\eta_e^*-\tilde\eta_e)$. For a bounded Lipschitz $K:\R^{d_e}\to[0,\infty)$ with compact support and $K(0)>0$ and a bandwidth $b>0$, define the random distribution $\hat\mu_{n,b}$, the kernel-weighted bootstrap distribution of $\mathrm{LR}_n^*$, by its integrals of bounded measurable functions $g$, and arbitrarily when the denominator below is zero,
\begin{equation}\label{eq:fitcondlaw}
\hat\mu_{n,b}(g)\defeq\frac{\E^*\bigl[K(D_n^*/b)\,g(\mathrm{LR}_n^*)\bigr]}{\E^*\bigl[K(D_n^*/b)\bigr]},
\end{equation}
with $(1-\alpha)$ quantile $\hat c_{n,b}$. With $B$ draws, the procedure is:
\begin{enumerate}[label=\arabic*.,leftmargin=*,itemsep=0pt,topsep=2pt]
\item Fit the null and the full model, and compute $\mathrm{LR}_n$, $\tilde\eta$ and $\hat\Sigma_e$.
\item For $j=1,\dots,B$, draw a sample of size $n$ from $Q_{\tilde\eta}$ and compute $\mathrm{LR}^*_j$ and the restricted fit $\tilde\eta^*_j$ as for the data.
\item Set $D^*_j\defeq\hat\Sigma_e^{-1/2}n^{1/2}(\tilde\eta^*_{e,j}-\tilde\eta_e)$ and $w_j\defeq K(D^*_j/b)$.
\item Reject when $\sum_j w_j1\{\mathrm{LR}^*_j\ge\mathrm{LR}_n\}/\sum_jw_j\le\alpha$, the finite-$B$ version of~\eqref{eq:fitcondlaw}, and report the effective number of draws $(\sum_jw_j)^2/\sum_jw_j^2$; if every $w_j$ is zero, increase $b$ or $B$. Adding $K(0)$ to the numerator and the denominator, which counts the observed sample as a draw, is an optional finite-$B$ adjustment (Section~\ref{app:implement}).
\end{enumerate}
Steps 1--2 are the ordinary bootstrap, so kernel reweighting needs no additional model fits; with all weights equal to one, step~4 gives the ordinary bootstrap test. Section~\ref{app:implement} discusses the choice of $b$ and two alternatives that need no reweighting: exact conditional simulation in Gaussian models, and projection draws, a shortcut not covered by Theorem~\ref{thm:fitcond}, which perturb an asymptotically normal reduced form only orthogonally to the null tangent space at the fit.

\begin{theorem}[Conditional bootstrap]\label{thm:fitcond}
Suppose Assumption~\ref{ass:score} holds, $\eta_e$ is as above, and $F(\cdot\,\vert\, t)$ is continuous and strictly increasing at $c_\alpha(t)$ for every $t\in L$.
\begin{enumerate}[label=(\alph*)]
\item\label{thm:fitcond:a} For each $b>0$, $d_{\mathrm{BL}}(\hat\mu_{n,b},\mu_b(\hat H_n))\conp0$, where, with $Z\sim\mathcal N(0,I_m)$, $\mu_b(t)(g)\defeq\E[K(\Upsilon Z_L/b)\,g(T(t+Z_L,Z_\perp))]/\E[K(\Upsilon Z_L/b)]$, and $d_{\mathrm{BL}}(\mu_b(t),F(\cdot\,\vert\, t))\to0$ as $b\downarrow0$, uniformly on compact sets of $t$.
\item\label{thm:fitcond:b} There is a sequence $b_n\downarrow0$ such that $\hat c_n\defeq\hat c_{n,b_n}$ satisfies $\hat c_n-c_\alpha(\hat H_n)\conp0$.
\item\label{thm:fitcond:c} For every compact $A\subset\R^d$, $\sup_{a\in A}|Q^n_{\eta_0+a/n^{1/2}}(\mathrm{LR}_n>\hat c_n)-\alpha|\to0$.
\item\label{thm:fitcond:d} If $T(Z)\indep Z_L$, then $\hat c_n-c_n^*(\alpha)\conp0$, with $c_n^*(\alpha)$ the ordinary bootstrap critical value of Theorem~\ref{thm:boot}.
\end{enumerate}
\end{theorem}

\noindent Part~\ref{thm:fitcond:c} holds whether or not \ref{G2} holds, so the procedure does not require knowing which case obtains, and by part~\ref{thm:fitcond:d} the two critical values are asymptotically equivalent when the ordinary bootstrap is valid. Part~\ref{thm:fitcond:b} shows that a suitable $b_n$ exists, possibly depending on $Q_0$; it gives no rate and no data-driven rule.

\paragraph*{Prepivoting.}
In the experiment of Theorem~\ref{thm:condlimit}, the ordinary bootstrap $p$-value $\hat p\defeq\Pr(T(Z_L+Z^*)\ge T(Z)\,\vert\, Z)$, $Z^*\sim\mathcal N(0,I_m)$ independent of $Z$, decreases in $T(Z)$ given $Z_L$, so a rule rejecting when $\hat p$ falls below a function of $Z_L$, as prepivoting \citep{Beran1987} and the double bootstrap do, is similar only if it is the conditional test (Corollary~\ref{cor:unique}). When $T(z)=a'z_L+f(z_\perp)$, as with the additive bias in \citet{CavaliereGoncalvesNielsenZanelli2024}, $\hat p$ depends only on $Z_\perp$ and prepivoting gives the conditional test.

\subsection{Orbit completion}\label{sec:orbit}

The second repair changes the statistic. Let
\[
C_1^{\orb}\defeq\cl(C_1+L)=L\oplus\cl(\Pi_{L^\perp}C_1),\qquad T^{\orb}(z)\defeq\Psi_{C_1^{\orb}}(z)-\Psi_L(z).
\]

\begin{theorem}[Orbit envelope]\label{thm:orbit}
Let $L\subset\R^m$ be a linear subspace and $C_1\supset L$ a closed set. For all $z$,
\[
T^{\orb}(z)=\sup_{h\in L}T(z+h)=\sup_{t\in L}T(t,z_\perp)=\|z_\perp\|^2-d\bigl(z_\perp,\cl(\Pi_{L^\perp}C_1)\bigr)^2 .
\]
Hence, $T^{\orb}$ is invariant under translations in $L$ and $T^{\orb}\ge T$; every $L$-invariant function that dominates $T$ dominates $T^{\orb}$; and $C_1^{\orb}$ is the smallest closed set containing $C_1$ that is stable under translations in $L$.
\end{theorem}

\noindent The second expression relates the two repairs: the conditional test uses the distribution of $T(t,Z_\perp)$ at the observed $t$, and the completed statistic is the envelope of $T(t,z_\perp)$ over $t$, which does not use the nuisance score. In the IV example, $\Pi_{L^\perp}C_1$ is the $s$-space and $T^{\orb}(z)=\|s\|^2$, the limit of the Anderson--Rubin statistic.

\paragraph*{The completed statistic in a sample.}
In a sample, the completed statistic has two forms: an LR form, which needs a completed model, and a score form, which does not. A \emph{completed model} is a parameter set $\Theta^{\orb}\supset\Theta_0$ whose likelihood satisfies \ref{S1} with $C_1^{\orb}$ in place of $C_1$, so that
\[
\mathrm{LR}_n^{\orb}\defeq2\Bigl[\sup_{\theta\in\Theta^{\orb}}L_n(\theta)-L_n(\tilde\theta)\Bigr]=T^{\orb}(Z_n)+o_p(1)\quad\text{under }Q_0 .
\]
In the IV example, $\Theta^{\orb}$ leaves $\gamma$ unrestricted, and $\mathrm{LR}_n^{\orb}$ is the Anderson--Rubin statistic; in general, replacing $C_1^{\orb}$ by $\R^m$, the whole of $V$, gives the LR statistic against the saturated local model, such as $\mathrm{AR}_n$ in Section~\ref{sec:weakiv}. The score form needs no completed model, because $T^{\orb}$ depends on the data only through $z_\perp$, the efficient score, which can be evaluated at the restricted fit. Let $d_\perp\defeq m-d$, with $m$ the dimension of $V$ in \ref{S1}, and let $s_\perp(\cdot;\eta)\in\R^{d_\perp}$ be functions with $\E_{Q_\eta}[s_\perp(\cdot;\eta)]=0$ for $\eta$ near $\eta_0$, whose value $s_\perp\defeq s_\perp(\cdot;\eta_0)$ at $\eta_0$ is the vector of Section~\ref{sec:exact}, with components spanning $V\ominus L$. For a closed cone $D\subset\R^{d_\perp}$ and the restricted-fit parameter $\tilde\eta$ of Section~\ref{sec:assumptions}, let
\begin{gather*}
\mathrm{C}_n^{\orb}\defeq\sup_{\delta\in D}(2\delta'U_n-\delta'\hat J\delta),\\
U_n\defeq n^{-1/2}\sum_{i\le n} s_\perp(X_i;\tilde\eta),\qquad \hat J\defeq n^{-1}\sum_{i\le n} s_\perp(X_i;\tilde\eta)\,s_\perp(X_i;\tilde\eta)'.
\end{gather*}

\begin{proposition}[Completed statistic in a sample]\label{prop:orbitsample}
Suppose Assumption~\ref{ass:score} holds and, under $Q_0$, $U_n=\G_n s_\perp+o_p(1)$ and $\hat J\conp J\defeq\E_{Q_0}[s_\perp s_\perp']$. Then, $J$ is nonsingular, and along every root-$n$ drift of the null
\[
\mathrm{C}_n^{\orb}\dto\sup_{\delta\in D}(2\delta'U-\delta'J\delta),\qquad U\sim\mathcal N(0,J),
\]
a distribution that does not depend on the drift. If $\{\delta' s_\perp:\delta\in D\}=\cl(\Pi_{L^\perp}C_1)$, that is, if $D$ describes the completed cone in terms of $s_\perp$, then $\mathrm{C}_n^{\orb}=T^{\orb}(Z_n)+o_p(1)$, and, when a completed model exists, $\mathrm{C}_n^{\orb}-\mathrm{LR}_n^{\orb}\conp0$ along every root-$n$ drift of the null. If $D=\R^{d_\perp}$, then $\mathrm{C}_n^{\orb}=U_n'\hat J^{-1}U_n$ whenever $\hat J$ is nonsingular, the $C(\alpha)$ statistic of \citet{Neyman1959}, with a $\chi^2_{d_\perp}$ limit.
\end{proposition}

\noindent The cone $D$ can be any closed cone fixed in advance, so the score test is asymptotically similar wherever the conditions hold and the limit distribution is continuous at the critical value; $D=\R^{d_\perp}$ corresponds to replacing $C_1^{\orb}$ by $\R^m$, and the saturated completed test also tests the rest of the specification. The expansion of $U_n$ rests on Neyman orthogonality: differentiating $\E_{Q_\eta}[s_\perp(\cdot;\eta)]=0$ gives $\E_{Q_0}[\partial_\eta s_\perp]=-\E_{Q_0}[s_\perp s_0']=0$, so replacing $\eta_0$ by $\tilde\eta$ changes $U_n$ by $o_p(1)$ under the usual smoothness conditions \citep{Neyman1959}. Critical values are obtained from the distribution of $\sup_{\delta\in D}(2\delta'U-\delta'\hat J\delta)$ with $U\sim\mathcal N(0,\hat J)$ or from the ordinary bootstrap. In the IV example, $D=\R^k$ and $\mathrm{C}_n^{\orb}$ is the score form of the Anderson--Rubin statistic.

\section{Applications}\label{sec:apps}

\subsection{An endogenous binary treatment}\label{sec:weakiv}

In the threshold-crossing model of \citet[Example~2.3]{HanMcCloskey2019},
\[
Y_i=1\{\varepsilon_i\le\pi_1+(\pi_2-\pi_1)D_i\},\qquad D_i=1\{v_i\le\zeta+\beta Z_i\},\qquad Z_i\in\{0,1\},
\]
the errors $\varepsilon_i,v_i$ are uniform on $(0,1)$ with copula $C(\cdot,\cdot;\pi_3)$, $|\pi_3|\le\bar\rho<1$, which is Gaussian or the Ali--Mikhail--Haq copula used there, and independent of the instrument $Z_i$, with $\Pr(Z_i=1)=\phi_1\in(0,1)$. Thus, $\pi_2-\pi_1$ is the average treatment effect and $\beta$ the effect of the instrument on take-up. When $\beta=0$, the conditional distributions of $(Y_i,D_i)$ given $Z_i=0$ and $Z_i=1$ coincide and $\pi_2$ is not identified. We test $H_0:\pi_2=\pi_{2,0}$; at $\beta=0$, the distribution is compatible with every $\pi_{2,0}$ in an interval, as in Section~\ref{sec:setup}.

\begin{lemma}[Regular null]\label{lem:hm-null}
At every interior point of the parameter space, including points with $\beta=0$, the map from $(\pi_1,\pi_3,\zeta,\beta)'$ to the two conditional cell-probability vectors of the null model has a Jacobian of full column rank.
\end{lemma}

\noindent At $\beta=0$, the Jacobian with respect to $(\pi_1,\pi_2,\pi_3,\zeta)'$ has rank three \citep[Example~2.3]{HanMcCloskey2019}; fixing $\pi_2$ removes the deficiency, so \ref{G1} holds and the singularity lies in the alternative. The parameters observationally equivalent at $\beta=0$ to the true one have $\zeta=\zeta^0$, its true value, and $\pi\defeq(\pi_1,\pi_2,\pi_3)'$ equals $\pi(\pi_2)\defeq(\pi_1(\pi_2),\pi_2,\pi_3(\pi_2))'$, which traces a curve as $\pi_2$ ranges over a compact interval $\mathcal I_2$. The probabilities of the cells $(Y_i,D_i)=(1,1),(1,0),(0,1)$ given $Z_i=z$ are $q(\pi,\zeta+\beta z)$ for a function $q(\pi,\cdot)$ with values in $\R^3$ (Section~\ref{app:apps}). Let $\hat q_z$ be the cell frequencies among the observations with $Z_i=z$, and let $X_n\defeq n^{1/2}\Omega^{-1/2}(\hat q_1-\hat q_0)$, where $\Omega$ is the asymptotic variance of $n^{1/2}(\hat q_1-\hat q_0)$ at $\beta=0$. Let $g(\pi_2)\defeq\Omega^{-1/2}\partial q(\pi(\pi_2),\zeta^0)/\partial\zeta$, the direction in which the mean of $X_n$ moves when $\beta$ departs from zero at $\pi(\pi_2)$, $u(\pi_2)\defeq g(\pi_2)/\|g(\pi_2)\|$, $g_0\defeq g(\pi_{2,0})$ and $u_0\defeq u(\pi_{2,0})$.

\begin{proposition}[Endogenous binary treatment]\label{prop:hm}
Let the true parameter at $\beta=0$ be interior to the parameter space of Section~\ref{app:apps}, with $\pi_{2,0}$ interior to $\mathcal I_2$. Under $H_0$ and $\beta=h/n^{1/2}$, $X_n\dto X\sim\mathcal N(hg_0,I_3)$, and
\[
\mathrm{LR}_n\dto\sup_{\pi_2\in\mathcal I_2}(u(\pi_2)'X)^2-(u_0'X)^2,\qquad \mathrm{AR}_n\dto\|X\|^2-(u_0'X)^2\sim\chi^2_2,
\]
where $\mathrm{AR}_n$ is the LR statistic against the saturated multinomial model. The nuisance score $t\defeq u_0'X$ is independent of $X-tu_0$. If $\dot u(\pi_{2,0})\neq0$ and no $u(\pi_2)$ is orthogonal to $u_0$, then \ref{G2} fails and the restricted-fit bootstrap is inconsistent at $\beta=0$.
\end{proposition}

\noindent The alternative is the fan $\bigcup_{\pi_2}\operatorname{span}\{u(\pi_2)\}$, as in the right panel of Figure~\ref{fig:cones}. In the designs we consider, the lines $\operatorname{span}\{u(\pi_2)\}$ project onto only a sector of the plane orthogonal to $u_0$ (Section~\ref{sec:practicalsim}), so the completed cone is not the whole space, and $\mathrm{AR}_n$ is the saturated version of the completed statistic.

We use the design of \citet[Section~7]{HanMcCloskey2019} with a Gaussian copula, which gives the bivariate probit model and allows stronger dependence than the Ali--Mikhail--Haq copula. Table~\ref{tab:hm} reports rejection frequencies at $n=1000$; the conditional test uses projection draws of the cell frequencies, which keep the restricted fit unchanged to first order, a shortcut described in Section~\ref{app:implement} that gives results close to kernel reweighting with $\eta_e=\beta$ as in Theorem~\ref{thm:fitcond}. The restricted-fit bootstrap is inconsistent, but here only mildly: over $32$ designs, including this one, its limiting null rejection probability lies between $0.023$ and $0.054$ when the mean of $t$ is at most $3$ (Section~\ref{sec:practicalsim}). The two resampling LR tests have nearly equal power, up to $10$ points above the $\chi^2_2$ test based on $\mathrm{AR}_n$, because $u(\mathcal I_2)$ lies close to a plane, so here the gain of the conditional test over the restricted-fit bootstrap is in size, not power.

In the original design of \citet{HanMcCloskey2019}, whose copula restricts Kendall's $\tau$ to about $[-0.18,1/3]$, the conditional LR test has nearly nominal size and more power than their robust Wald test, partly because the latter is conservative under weak identification (Section~\ref{sec:practicalsim}).

\begin{table}[t]\begin{center}
\caption{Rejection frequencies at level $0.05$ in the threshold-crossing model}\label{tab:hm}
\begin{tabular}{cccccc}\toprule
& & \multicolumn{3}{c}{$\mathrm{LR}$} & $\mathrm{AR}$\\\cmidrule(lr){3-5}\cmidrule(lr){6-6}
$\beta$ & True $\pi_2$ & $\chi^2_1$ & Restricted fit & Conditional & $\chi^2_2$\\\midrule
0 & 0.4 & 0.015 & 0.042 & 0.046 & 0.047\\
0.025 & 0.4 & 0.028 & 0.047 & 0.049 & 0.055\\
0.05 & 0.4 & 0.036 & 0.048 & 0.049 & 0.051\\
0.1 & 0.4 & 0.050 & 0.053 & 0.051 & 0.046\\
0.2 & 0.4 & 0.057 & 0.054 & 0.054 & 0.049\\
0.4 & 0.4 & 0.045 & 0.049 & 0.049 & 0.049\\
\midrule
0.2 & 0.2 & 0.613 & 0.608 & 0.604 & 0.517\\
0.2 & 0.3 & 0.189 & 0.192 & 0.188 & 0.148\\
0.2 & 0.5 & 0.161 & 0.152 & 0.159 & 0.124\\
0.2 & 0.6 & 0.440 & 0.421 & 0.420 & 0.349\\
0.2 & 0.7 & 0.725 & 0.716 & 0.716 & 0.625\\
0.4 & 0.2 & 0.998 & 0.997 & 0.998 & 0.994\\
0.4 & 0.3 & 0.642 & 0.634 & 0.628 & 0.531\\
0.4 & 0.5 & 0.594 & 0.589 & 0.594 & 0.495\\
0.4 & 0.6 & 0.988 & 0.989 & 0.988 & 0.977\\
0.4 & 0.7 & 1.000 & 1.000 & 1.000 & 1.000\\
\bottomrule\end{tabular}\end{center}
\floatnotes{Design of \citet{HanMcCloskey2019} with a Gaussian copula: $n=1000$, $H_0:\pi_2=0.4$, $\zeta=0.2$, $\pi_1=0.6$, $\pi_3=0.4$, $\phi_1=1/2$, $|\pi_3|\le0.99$; $3{,}000$ replications; $B=199$ draws for both resampling tests, with projection draws for the conditional test. Top: size; bottom: power.}
\end{table}

\subsection{Confidence sets for a nonlinear parameter}\label{sec:nonlin}

Consider
\[
y_i=w_i'\zeta+g(x_i;\pi)'\beta+\varepsilon_i,\quad i\le n,\qquad (\varepsilon_1,\dots,\varepsilon_n)'\sim \mathcal N_n(0,\sigma^2I_n),
\]
with fixed regressors, a known function $g$, $\beta\in\R^{d_\beta}$ and $\pi\in\mathcal I\subset\R^{d_\pi}$. A confidence set for $\pi$ inverts tests of $H_0:\pi=\pi_0$; when $\beta=0$, $\pi$ is not identified, and when $\beta$ is of order $n^{-1/2}$, it is weakly identified \citep{AndrewsCheng2012}. The model covers the decay parameters of the yield-curve models of \citet{NelsonSiegel1987} and \citet{Svensson1994}; an index threshold or hidden unit, $g=\varphi(\pi'x_i)$; and locations such as a break date or threshold, $g=1\{q_i>\pi\}$ \citep{Bai1997,Hansen2000}, a kink, $g=(q_i-\pi)_+$ \citep{Hansen2017}, and the cutpoint of a subgroup with a different treatment effect, $g=d_i1\{q_i>\pi\}$ with $d_i$ a treatment indicator \citep{JiangFreidlinSimon2007}. For the location models, Section~\ref{sec:examples} tests whether the term is present; here we test its location.

The null model is the Gaussian linear regression on $X_0\defeq[W\ \,g(\cdot;\pi_0)]$, where $W$ and $g(\cdot;\pi)$ have rows $w_i'$ and $g(x_i;\pi)'$. When $X_0$ has full column rank, its coefficients and $\sigma^2$ are identified, with nonsingular information, also where $\beta=0$, so the null model has no singular points. At $\beta=0$, however, the alternative is in general entangled with the nuisance direction $g(\cdot;\pi_0)$: for $\pi\neq\pi_0$, a departure $g(\cdot;\pi)'b$ plus $g(\cdot;\pi_0)'b_0$ is not a local direction of the full model. The test of no effect, $H_0:\beta=0$, is different: as for the hidden unit of Section~\ref{sec:nn}, the possible departures $g(\cdot;\pi)'b$ do not depend on the null parameters, and the alternative splits. Tests invariant to the coefficients of the null regression depend on the data only through its residuals, as the completed statistic does. \citet{ElliottMuller2007} and \citet{KurozumiYamamoto2015} invert such tests to obtain confidence sets for a break date, and \citet{Dufour1990} to obtain exact confidence sets for an autocorrelation coefficient. Let $P_0$ be the projection onto the columns of $X_0$, $M_0\defeq I-P_0$ and $\hat e_0\defeq M_0y$.

\begin{proposition}[Exact conditional test for the nonlinear parameter]\label{prop:nonlin}
Let $n>\operatorname{rank}(X_0)$. Under $H_0:\pi=\pi_0$, for every value of $(\zeta',\beta',\sigma)'$ with $\sigma>0$ the conditional distribution of $y$ given $P_0y$ and $\|\hat e_0\|$ is that of $y^*\defeq P_0y+\|\hat e_0\|\,M_0\xi/\|M_0\xi\|$ with $\xi\sim \mathcal N(0,I_n)$. Consequently, for any statistic $S(y)$, the test that rejects when $(1+\sum_{b\le B}1\{S(y^*_b)\ge S(y)\})/(B+1)\le\alpha$, with independent draws $y^*_b$, has size at most $\alpha$ for every $n$ and every value of the nuisance parameter, and exactly $\alpha$ when $\alpha(B+1)$ is an integer and ties have probability zero \citep{Dufour2006}.
\end{proposition}

\noindent The draws keep the observed restricted fit, coefficients and residual variance alike, whereas the ordinary bootstrap redraws both. We take $S=\mathrm{LR}_n(\pi_0)=n\log(SSR_0/\min_\pi SSR(\pi))$, where $SSR(\pi)$ is the residual sum of squares of the regression on $[W\ \,g(\cdot;\pi)]$ and $SSR_0\defeq SSR(\pi_0)$.

\paragraph*{A second curvature term in the yield curve.}
We use the monthly unsmoothed Fama--Bliss zero-coupon yields of \citet{DieboldLi2006}, 1985:01--2000:12, at $n=17$ maturities $\tau$ from $3$ to $120$ months. They fit the \citet{NelsonSiegel1987} curve, with loadings $f_1(\tau;\lambda)\defeq(1-e^{-\lambda\tau})/(\lambda\tau)$ and $f_2(\tau;\lambda)\defeq f_1(\tau;\lambda)-e^{-\lambda\tau}$, and fix the decay at $\lambda_1=0.0609$, which places the peak of $f_2$ at $29$ months. \citet{Svensson1994} adds a second curvature term,
\[
y_t(\tau)=\beta_{1t}+\beta_{2t}f_1(\tau;\lambda_1)+\beta_{3t}f_2(\tau;\lambda_1)+\beta_{4t}f_2(\tau;\lambda_2)+\varepsilon_t(\tau),
\]
whose decay $\lambda_2$ is not identified when $\beta_{4t}=0$. The term may have either sign, so it may not produce a second hump in the curve; $\lambda_2$ locates the peak of its loading. With independent $\mathcal N(0,\sigma_t^2)$ errors across maturities, each month is the model above with $g=f_2(\cdot;\lambda_2)$ and $\pi=\lambda_2$, which we restrict to place the peak of $f_2(\cdot;\lambda_2)$ between $3$ and $120$ months and at least a factor $1.5$ from that of $f_2(\cdot;\lambda_1)$.

We evaluate the conditional test, with $B=999$ draws, at $95$ log-spaced values of $\lambda_2$ in this set and report the accepted values; the statistic minimizes $SSR(\pi)$ over $\pi_0$ and $314$ such values, which leaves each test exact. The completed statistic of Section~\ref{sec:orbit} is here the LR statistic for adding $f_2(\cdot;\lambda_2)$ to the null regression. It depends on the data only through the direction $\hat e_0/\|\hat e_0\|$ of the null residuals, which under $H_0$ is uniform on the unit sphere of the range of $M_0$ whatever the nuisance parameters (proof of Proposition~\ref{prop:nonlin}), so those draws also calibrate it exactly.

\begin{table}[t]\begin{center}
\caption{$95\%$ confidence sets for the second Svensson decay parameter}\label{tab:svensson}
\setlength{\tabcolsep}{3.5pt}
\begin{tabular}{lccccc}\toprule
& Conditional & Completed & Restricted fit & $\chi^2_1$ & Wald\\\midrule
\multicolumn{6}{l}{Share of months accepting every grid value}\\
\quad All months ($192$) & 0.36 & 0.31 & 0.38 & 0.16 & 0.01\\
\quad $\beta_{4t}=0$ not rejected ($94$) & 0.74 & 0.64 & 0.77 & 0.32 & 0.02\\\midrule
\multicolumn{6}{l}{Coverage in calibrated simulations}\\
\quad Gaussian errors & 0.950 & 0.950 & 0.947 & 0.872 & 0.760\\
\quad Resampled residuals & 0.950 & 0.949 & 0.947 & 0.871 & 0.753\\
\quad Correlated errors & 0.974 & 0.967 & 0.973 & 0.927 & 0.804\\
\bottomrule\end{tabular}\end{center}
\floatnotes{Monthly yields of \citet{DieboldLi2006}, 1985:01--2000:12, at $17$ maturities. Each test is inverted over $95$ grid values of $\lambda_2$, with $999$ draws for the resampling tests; the second row covers the $94$ months in which the exact test of $\beta_{4t}=0$ does not reject at level $0.05$. Coverages use $3{,}000$ replications per month from the fitted curve, each testing the true $\lambda_2$ with $199$ draws (Section~\ref{sec:practicalsim}).}
\end{table}

The exact test of $\beta_{4t}=0$, whose alternative splits, rejects at level $0.05$ in $51\%$ of the months. In the other $94$ months, the conditional test accepts every grid value in $74\%$ of cases and the completed test in $64\%$, against $32\%$ for the $\chi^2_1$ critical value applied to the same statistic and $2\%$ for the Wald interval: the standard sets are short even when the exact sets are the whole admissible range (Table~\ref{tab:svensson}).

In simulations calibrated to each month, the conditional and completed sets cover with probability $0.950$ and the $\chi^2_1$ and Wald sets undercover (Table~\ref{tab:svensson}); the restricted-fit bootstrap, which is not guaranteed to be consistent here, covers with probability $0.947$. Exactness assumes independent homoskedastic Gaussian errors across maturities within a month. Resampling a month's residuals leaves the coverages nearly unchanged; with Gaussian errors whose cross-maturity covariance is estimated from the pooled residuals, the conditional sets are conservative, covering with probability $0.974$ (Section~\ref{sec:practicalsim}); this is a calibration to these data, not robustness to correlated errors in general.

\section{Conclusion}\label{sec:discussion}

On the quotient, the restricted-fit bootstrap of the LR statistic works when the alternative splits off the nuisance directions, and conditioning on the restricted fit gives asymptotically similar tests whether or not it does. The theory is local at a fixed regular point of the null; boundary nuisance parameters, singular nulls and dependent data are outside it. The exact test of Proposition~\ref{prop:nonlin} rests instead on the complete sufficient statistic of a Gaussian linear null model and holds in finite samples.

\appendix
\renewcommand{\thesection}{A}
\setcounter{equation}{0}
\section{Proofs}\label{app:proofs}

Since $\Psi_C(z)=\|z\|^2-d(z,C)^2$, every $T$ of the form~\eqref{eq:coneLRdiagnostic} is continuous.

\subsection{Splitting}

\begin{lemma}[Orthogonal splitting]\label{lem:split}
Let $V$ be a real inner-product space, $L\subset V$ a finite-dimensional subspace, $K\subset L^\perp$ a cone and $\psi:V\to\R$ linear. With $\sup\emptyset=0$,
\[
\sup_{h\in\Sph(L\oplus K)}\plus{\psi(h)}^2=\sup_{u\in\Sph(L)}\psi(u)^2+\sup_{k\in\Sph(K)}\plus{\psi(k)}^2,
\]
both sides possibly infinite. For any cone $K_0\subset V$, $L+K_0=L\oplus\Pi^\perp K_0$.
\end{lemma}

\begin{proof}
If $L=\{0\}$, both sides equal $\sup_{\Sph(K)}\plus{\psi}^2$, so let $\dim L\ge1$. Each $h\in L\oplus K$ is $h=u+k$ with $u=\Pi_Lh$, $k\in K$ and $\|h\|^2=\|u\|^2+\|k\|^2$. Let $a\defeq\sup_{\Sph(L)}\psi$, which is $\ge0$ and equals $\sup_{\Sph(L)}|\psi|$ because $L$ is symmetric, and $r\defeq\sup_{\Sph(K)}\psi$, $r_+\defeq\max(r,0)$. Then, $\psi(u)\le a\|u\|$ and $\psi(k)\le r_+\|k\|$, so for $\|h\|=1$, $\psi(h)\le a\|u\|+r_+\|k\|\le(a^2+r_+^2)^{1/2}$. Conversely, for $\epsilon>0$ take $u_\epsilon\in\Sph(L)$ with $\psi(u_\epsilon)\ge a-\epsilon$; if $r_+=0$ use $h=u_\epsilon$, and otherwise take $k_\epsilon\in\Sph(K)$ with $\psi(k_\epsilon)\ge r-\epsilon$ and $h=(au_\epsilon+rk_\epsilon)/(a^2+r^2)^{1/2}\in\Sph(L\oplus K)$, which gives $\psi(h)\ge(a^2+r^2)^{1/2}-2\epsilon$. If $a$ or $r$ is infinite, the same choices show that both sides are infinite. Finally, $u+k=(u+\Pi_Lk)+\Pi^\perp k$, and $\Pi^\perp K_0$ is a cone in $L^\perp$.
\end{proof}

\begin{proof}[Proof of Proposition~\ref{prop:split}]
Apply Lemma~\ref{lem:split} at fixed $n$ with $\psi=\G_n$ and $K=\Pi^\perp K_0$. For $u=c's_0$, $\|u\|^2=c'I_0c$ and $\G_nu=c'\Delta_n$, so $\sup_{\Sph(L)}(\G_nu)^2=\Delta_n'I_0^{-1}\Delta_n$, and
$\sup_{\Sph(L+K_0)}\plus{\G_nh}^2=\Delta_n'I_0^{-1}\Delta_n+\sup_{\cE_0}\plus{\G_ne}^2$ with $\cE_0\defeq\Sph(\Pi^\perp K_0)$. The class $\cE=\cl\,\cE_0$ is Donsker, hence totally bounded, and closed, hence compact; so $\mathcal A(\cE)\defeq\{u\cos\omega+e\sin\omega:u\in\Sph(L),e\in\cE,\omega\in[0,\pi/2]\}=\Sph(L\oplus\operatorname{cone}\cE)$, with $\operatorname{cone}\cE\defeq\{ce:c\ge0,e\in\cE\}$, is compact and $\Sph(L+K_0)\subset\cD\subset\mathcal A(\cE)$; if $L=\{0\}$, set $\mathcal A(\cE)\defeq\cE$. Lemma~\ref{lem:split} applied to $L\oplus\operatorname{cone}\cE$ therefore gives
\[
\Delta_n'I_0^{-1}\Delta_n+\sup_{\cE_0}\plus{\G_ne}^2\le\sup_{\cD}\plus{\G_nh}^2\le\Delta_n'I_0^{-1}\Delta_n+\sup_{\cE}\plus{\G_ne}^2 .
\]
Since $\cE$ is Donsker, $\G_n$ is asymptotically uniformly $L^2(Q_0)$-equicontinuous on $\cE$ \citep[Section~2.1.2]{vanderVaartWellner1996}, and $\cE_0$ is dense in $\cE$, so $\sup_\cE\G_ne-\sup_{\cE_0}\G_ne=o_p(1)$; as $\sup_\cE\G_ne=O_p(1)$ and $x\mapsto x_+^2$ is locally Lipschitz, the two bounds differ by $o_p(1)$. Combining with~\eqref{eq:directional} gives the expansion, and subtracting \ref{A1} gives~\eqref{eq:efficientrep}. Each $e\in\cE$ is orthogonal to $L$ as a limit of elements of $\Pi^\perp K_0$.
\end{proof}

\subsection{Bootstrap validity}

\begin{proof}[Proof of Theorem~\ref{thm:drift}]
Under $Q_0^n$, $g(z)\defeq(\sup_{e\in\cE}z(e))_+^2$ is continuous on $\ell^\infty(\cE)$ and $\G_n\dto\G$ in $\ell^\infty(\cE)$, so $g(\G_n)\dto\Lambda$, and $\mathrm{LR}_n=g(\G_n)+o_p(1)$ by~\eqref{eq:efficientrep}. For general $a$, by DQM and \citet[Theorem~7.2]{vanderVaart1998}, $\log(dQ_{\eta_n}^n/dQ_0^n)=a'\Delta_n-\tfrac12a'I_0a+o_p(1)$ under $Q_0^n$, so the two sequences are mutually contiguous, and $\mathrm{LR}_n-g(\G_n)\to0$ in $Q^n_{\eta_n}$-probability. For $f_1,\dots,f_J\in\cE$ the vector $(\G_nf_1,\dots,\G_nf_J,a'\Delta_n)$ is asymptotically normal under $Q_0^n$ with $\operatorname{Cov}(\G f_j,a'\G s_0)=\ip{f_j}{a's_0}=0$, so by Le Cam's third lemma the marginals of $\G_n$ have the same limits under $Q^n_{\eta_n}$. Contiguity transfers asymptotic equicontinuity on the totally bounded $(\cE,\|\cdot\|)$, so $\G_n\dto\G$ under $Q^n_{\eta_n}$ \citep[Theorems~1.5.4 and~1.5.7]{vanderVaartWellner1996}, giving~\ref{thm:drift:a}; \ref{thm:drift:b} follows by the CMT.
\end{proof}

\begin{proof}[Proof of Theorem~\ref{thm:boot}]
Let $\mu_n(a)$ be the distribution of $\mathrm{LR}_n$ under $Q^n_{\eta_0+a/n^{1/2}}$. For every $M$, $\sup_{\|a\|\le M}d_{\mathrm{BL}}(\mu_n(a),\mathcal L(\Lambda))\to0$: else a subsequence $a_{n'}\to a$ would contradict Theorem~\ref{thm:drift}. By Lemma~\ref{lem:quotient} and \ref{A0}, $\mathcal L^*(\mathrm{LR}_n^*)=\mu_n(\hat a_n)$ with $\hat a_n\defeq n^{1/2}(\tilde\eta-\eta_0)$ w.p.a.1, and $\hat a_n$ is tight by \ref{A1}; hence the first claim. By the subsequence principle and \citet[Lemma~21.2]{vanderVaart1998}, the continuity condition gives $c_n^*(\alpha)\conp c_\alpha$ and, by Theorem~\ref{thm:drift}, $Q_0^n(\mathrm{LR}_n>c_n^*(\alpha))\to\alpha$.
\end{proof}

\begin{proof}[Proof of Proposition~\ref{prop:dichotomy}]
Let $\mu_n(h)$ be the distribution of $\mathrm{LR}_n$ under $P^n_{\theta_n(h)}$. \emph{Continuity:} if $h_j\to h$, choose $n_1<n_2<\cdots$ with $d_{\mathrm{BL}}(\mu_{n_j}(h_j),\mu(h_j))<1/j$; the sequence equal to $h_j$ for $n_j\le n<n_{j+1}$ converges to $h$, so $\mu_{n_j}(h_j)\to\mu(h)$ and $\mu(h_j)\to\mu(h)$. \emph{Uniformity:} for every $M$, $\sup\{d_{\mathrm{BL}}(\mu_n(h),\mu(h)):h\in\mathcal K,\|h\|\le M\}\to0$, since otherwise a subsequence $h_{n'}\to h$ would give two limits. (a)~follows from Lemma~\ref{lem:quotient} and tightness of $\hat H_n$. (b)~By continuity, $d_{\mathrm{BL}}(\mu(\hat H_n),\mu(0))\dto d_{\mathrm{BL}}(\mu(H),\mu(0))$, which with~\ref{prop:dichotomy:a} is degenerate at zero if and only if $\mu(H)=\mu(0)$ almost surely.
\end{proof}

\begin{proof}[Proof of Theorem~\ref{thm:diagnostic}]
(a) By Proposition~\ref{prop:dichotomy}\ref{prop:dichotomy:b}, consistency holds if and only if $\mu(H)=\mu(0)$ almost surely, which by continuity of $\mu$ and the definition of the support is equivalent to $\mu(h)=\mu(0)$ for all $h\in\operatorname{supp}H$. If $H$ is a projection $\operatorname{pr}(Z)$ of $Z$ onto $C_0$, then for $y\in C_0$, $\|\operatorname{pr}(z)-y\|\le\|\operatorname{pr}(z)-z\|+\|z-y\|\le2\|z-y\|$, so every ball around $y$ has positive probability under $H$; hence $\operatorname{supp}H=C_0$.

(b) For $h\in L$, $h'Z=h'Z_L$, and for bounded Borel $b$, $\E[b(T(Z+h))]=\E[b(T(Z))\exp(h'Z_L-\tfrac12\|h\|^2)]$. If $T(Z)$ is independent of $Z_L$, this equals $\E[b(T(Z))]$. Conversely, if the distributions agree for all $h\in L$, then $g(Z_L)\defeq\E[b(T(Z))\,\vert\, Z_L]-\E[b(T(Z))]$ satisfies $\E[g(Z_L)\exp(h'Z_L-\tfrac12\|h\|^2)]=0$ for all $h$, so, by completeness of the normal location family, $g(Z_L)=0$ almost surely and $T(Z)\indep Z_L$.

(c) Here, $T(z)=\|\Pi_{L^\perp}z\|^2-d(z,C_1)^2$. If $C_1+L=C_1$, then $T(z+u)=T(z)$ for $u\in L$. Conversely, if $T(z+u)=T(z)$ for all $z,u$, then for $z\in C_1$, $d(z+u,C_1)=d(z,C_1)=0$, so $C_1+L=C_1$. If $C_1+L=C_1$ and $z\in C_1$, then $z-\Pi_Lz\in C_1\cap L^\perp$, so $C_1=L+(C_1\cap L^\perp)$; the converse is immediate. Then, $T(z)$ depends only on $\Pi_{L^\perp}z$, independent of $Z_L$. A convex $C_1\supset L$ has $C_1+L\subset C_1+C_1=C_1$.
\end{proof}

\subsection{Conditioning}

\begin{proof}[Proof of Lemma~\ref{lem:fittedshift}]
Under $Q_0^n$, $Z_n\dto\mathcal N(0,I_m)$, and by DQM (Definition~\ref{def:nullpar}) the log-likelihood ratio of $Q^n_{\eta_0+a_n/n^{1/2}}$ to $Q^n_0$ is $h'Z_{n,L}-\tfrac12\|h\|^2+o_p(1)$. By Le Cam's third lemma $Z_n\dto\mathcal N((h',0')',I_m)$ under the drifting distributions, and by contiguity \ref{S1}--\ref{S2} continue to hold, so $(\mathrm{LR}_n,\hat H_n)=(T(Z_n),Z_{n,L})+o_p(1)$ and the first claim follows by the CMT; uniformity over $\|a_n\|\le M$ follows from continuity of $h\mapsto\varrho(h)$ for $d_{\mathrm{BL}}$ as in the proof of Proposition~\ref{prop:dichotomy}. As in Lemma~\ref{lem:quotient}, the bootstrap distribution of $(\mathrm{LR}_n^*,\tilde\eta^*)$ is that of $(\mathrm{LR}_n,\tilde\eta)$ under $Q^n_{\tilde\eta}$, and $\hat H_n$ is tight; hence the second claim.
\end{proof}

\begin{proof}[Proof of Theorem~\ref{thm:condlimit}]
(a) Write $Z=((h+W_L)',W_\perp')'$ with $W\sim\mathcal N(0,I_m)$. Since $W_\perp$ is independent of $Z_L$ and its distribution is free of $h$, $\Pr_h(T(Z)>c_\alpha(Z_L)\,\vert\, Z_L=t)=1-F(c_\alpha(t)\,\vert\, t)\le\alpha$, with equality under continuity; integrate over $Z_L$. (b) $T(t+Z)=T(t+Z_L,Z_\perp)$ and $Z_\perp\indep Z_L$. (c) Under $h=0$, $F(\cdot\,\vert\, t)$ is a version of the conditional distribution of $T(Z)$ given $Z_L=t$, so independence holds if and only if $F(\cdot\,\vert\, t)$ is the same for almost every $t$, and, $t\mapsto F(\cdot\,\vert\, t)$ being continuous for $d_{\mathrm{BL}}$, for every $t$. Equivalence with constancy in $h$ of the distribution of $T(Z)$ is proved as Theorem~\ref{thm:diagnostic}\ref{thm:diagnostic:b}, which also gives the last statement. (d) If $C_1+h=C_1$ then $C_1-h=C_1$, and $C_1+\lambda h=\lambda(C_1+h)=C_1$ for $\lambda>0$ because $C_1$ is a cone, so $L_s$ is a subspace. For $h\in L_s$, $d(z+h,C_1)=d(z,C_1)$ and $\Pi_{L^\perp}(z+h)=\Pi_{L^\perp}z$, so $T(z+h)=T(z)$ and $T(t,z_\perp)=T(\Pi_{L_e}t,z_\perp)$. \ref{G2} is $C_1+L=C_1$. (e) $Z_L$ is complete sufficient for $h$, so $\E_h[\E[\varphi\,\vert\, Z_L]-\alpha]=0$ for all $h$ gives $\E[\varphi\,\vert\, Z_L]=\alpha$ almost surely.
\end{proof}

\begin{proof}[Proof of Corollary~\ref{cor:unique}]
Sufficiency is Theorem~\ref{thm:condlimit}\ref{thm:condlimit:a}. If $1\{T(Z)>c(Z_L)\}$ is similar, Theorem~\ref{thm:condlimit}\ref{thm:condlimit:e} gives $1-F(c(t)\,\vert\, t)=\alpha$ for almost every $t$, hence $c(t)=c_\alpha(t)$.
\end{proof}

\begin{proof}[Proof of Theorem~\ref{thm:condWAP}]
(a) The simple null $\mathcal N((\kappa_L',0')',I_m)$ belongs to the composite null, and by the Neyman--Pearson lemma no level-$\alpha$ test has power above that of the test of this simple null against $\kappa$, which rejects when $\kappa_\perp'Z_\perp>\|\kappa_\perp\|\Phi^{-1}(1-\alpha)$ and has power $1-\Phi(\Phi^{-1}(1-\alpha)-\|\kappa_\perp\|)$. This test has conditional level $\alpha$ given every $Z_L=t$, so the three envelopes coincide. (b) By completeness every similar test has $\E[\varphi\,\vert\, Z_L=t]=\alpha$ for almost every $t$. Under the weighted alternative the density of $Z$ relative to $\mathcal N(0,I_m)$ is $A_{\mathcal W}$, so given $Z_L=t$ the alternative density of $Z_\perp$ relative to its null density is $A_{\mathcal W}(t,z_\perp)/\E[A_{\mathcal W}(t,U)]$ with $U\sim\mathcal N(0,I_{m-d})$. The conditional Neyman--Pearson lemma maximizes conditional power for almost every $t$, and integrating over the distribution of $Z_L$ under the alternative gives the claim.
\end{proof}

\begin{proof}[Proof of Theorem~\ref{thm:fitcond}]
(a) Since $n^{1/2}(\tilde\eta^*_e-\tilde\eta_e)=[I_0^{-1/2}]_e(\hat H_n^*-\hat H_n)$, $D_n^*=\hat\Upsilon(\hat H_n^*-\hat H_n)$ with $\hat\Upsilon\defeq\hat\Sigma_e^{-1/2}\Sigma_e^{1/2}\Upsilon\conp \Upsilon$. By Lemma~\ref{lem:fittedshift}, the bootstrap distribution of $(\mathrm{LR}_n^*,\hat H_n^*-\hat H_n)$ is within $o_p(1)$ in $d_{\mathrm{BL}}$ of $\mathcal L(T(t+Z),Z_L)$ at $t=\hat H_n$; in particular, $\hat H_n^*-\hat H_n$ is bootstrap-tight, so replacing $\hat\Upsilon$ by $\Upsilon$ in $K(D_n^*/b)$ changes the numerator and denominator of~\eqref{eq:fitcondlaw} by $o_p(1)$. As $\Upsilon\Upsilon'=I_{d_e}$, for $g$ bounded by one with Lipschitz constant one the function $(x,y)\mapsto K(\Upsilon y/b)g(x)$ is bounded by $\|K\|_\infty\defeq\sup|K|$ with Lipschitz constant at most $\|K\|_\infty+\|K\|_{\mathrm{Lip}}/b$, $\|K\|_{\mathrm{Lip}}$ being that of $K$, so the numerator and denominator of~\eqref{eq:fitcondlaw} converge, uniformly in such $g$, to those of $\mu_b(\hat H_n)$, the denominator to $\E[K(\Upsilon Z_L/b)]>0$; this is the first claim. For the second, $\Upsilon'\Upsilon$ is the orthogonal projection onto the row space of $\Upsilon$, whose orthogonal complement $\ker\Upsilon$ is spanned by the vectors $I_0^{1/2}e_j$, $j\notin\eta_e$, that represent the scores $s_{0,j}$, which lie in $L_s$; here $e_j$ is the $j$th unit vector of $\R^d$. Hence, by Theorem~\ref{thm:condlimit}\ref{thm:condlimit:d}, $T(t+Z_L,Z_\perp)=T(t+\Upsilon'\Upsilon Z_L,Z_\perp)$. On the support of $K(\Upsilon y/b)$, $\|\Upsilon'\Upsilon y\|=\|\Upsilon y\|\le\rho b$ for some $\rho<\infty$, and since $|d(x+w,C)^2-d(x,C)^2|\le\|w\|(2\|x\|+\|w\|)$,
\[
\bigl|g(T(t+\Upsilon'\Upsilon Z_L,Z_\perp))-g(T(t,Z_\perp))\bigr|\le\rho b\,(2\|t\|+2\|Z_\perp\|+\rho b).
\]
As $Z_\perp$ is independent of $Z_L$, $|\mu_b(t)(g)-\int g\,dF(\cdot\,\vert\, t)|\le\rho b(2\|t\|+2\E[\|Z_\perp\|]+\rho b)$.

(b) By~\ref{thm:fitcond:a} choose $n_1<n_2<\cdots$ with $\Pr(d_{\mathrm{BL}}(\hat\mu_{n,1/j},\mu_{1/j}(\hat H_n))>1/j)<1/j$ for $n\ge n_j$ and set $b_n\defeq1/j$ for $n_j\le n<n_{j+1}$. Tightness of $\hat H_n$ and~\ref{thm:fitcond:a} give $d_{\mathrm{BL}}(\hat\mu_{n,b_n},F(\cdot\,\vert\,\hat H_n))\conp0$. The map $t\mapsto F(\cdot\,\vert\, t)$ is continuous for $d_{\mathrm{BL}}$, and continuity and strict increase of $F(\cdot\,\vert\, t)$ at $c_\alpha(t)$ make the $(1-\alpha)$ quantile continuous for $d_{\mathrm{BL}}$ at $F(\cdot\,\vert\, t)$, uniformly for $t$ in compact sets; hence $\hat c_n-c_\alpha(\hat H_n)\conp0$.

(c) For $a_n\to a$ in $A$, contiguity carries~\ref{thm:fitcond:b} to $Q^n_{\eta_0+a_n/n^{1/2}}$, and Lemma~\ref{lem:fittedshift} gives $\mathrm{LR}_n-\hat c_n\dto T(h+Z)-c_\alpha(h+Z_L)$ with $h=I_0^{1/2}a$. Given $Z_L$, the limit has no atom at zero because $F(\cdot\,\vert\, t)$ is continuous at $c_\alpha(t)$, so the rejection probability tends to $\alpha$ by Theorem~\ref{thm:condlimit}\ref{thm:condlimit:a}; uniformity over $A$ follows by passing to convergent subsequences.

(d) By Theorem~\ref{thm:condlimit}\ref{thm:condlimit:c}, $c_\alpha(t)$ is then a constant, the limit of $c_n^*(\alpha)$ by Lemma~\ref{lem:fittedshift}.
\end{proof}

\subsection{Orbit completion}

\begin{proof}[Proof of Theorem~\ref{thm:orbit}]
For $h\in L$, $T(z+h)=\|z_\perp\|^2-d(z+h,C_1)^2$, so $\sup_{h\in L}T(z+h)=\|z_\perp\|^2-d(z,C_1+L)^2$. As $C_1+L=L+\Pi_{L^\perp}C_1$, this distance is $d(z_\perp,\cl(\Pi_{L^\perp}C_1))$, and completing the square gives $\Psi_{C_1^{\orb}}(z)-\Psi_L(z)$; the second expression is the first with $t=z_L+h$. Invariance and $T^{\orb}\ge T$ are immediate. If $R$ is $L$-invariant and $R\ge T$, then $R(z)=R(z+h)\ge T(z+h)$ for all $h\in L$, so $R\ge T^{\orb}$. A closed set that contains $C_1$ and is stable under translations in $L$ contains $\cl(C_1+L)$.
\end{proof}

\begin{proof}[Proof of Proposition~\ref{prop:orbitsample}]
The $d_\perp$ components of $s_\perp$ span the $d_\perp$-dimensional space $V\ominus L$, so $J$ is invertible. The map $\Psi_D(u,A)\defeq\sup_{\delta\in D}(2\delta'u-\delta'A\delta)=u'A^{-1}u-\min_{\delta\in D}\|A^{-1/2}u-A^{1/2}\delta\|^2$ is continuous on $\R^{d_\perp}\times\{A\succ0\}$, because $0\in D$ restricts any minimizer to $\|A^{1/2}\delta\|\le2\|A^{-1/2}u\|$, a set that is locally bounded in $(u,A)$. Under $Q_0$, $\mathrm{C}_n^{\orb}=\Psi_D(U_n,\hat J)=\Psi_D(\G_n s_\perp,J)+o_p(1)$, and by contiguity this holds along every root-$n$ drift of the null. By the CLT and Le Cam's third lemma, $\G_n s_\perp\dto\mathcal N(\E_{Q_0}[s_\perp\,a's_0],J)=\mathcal N(0,J)$ along the drift $a$, because $(V\ominus L)\perp L$. This gives the limit. In the representation of Section~\ref{sec:exact}, $Z_{n,\perp}=J^{-1/2}\G_n s_\perp$ and $\delta's_\perp$ is represented by $J^{1/2}\delta$, so $\cl(\Pi_{L^\perp}C_1)=J^{1/2}D$, and Theorem~\ref{thm:orbit} gives $T^{\orb}(Z_n)=\sup_{c\in J^{1/2}D}(2c'Z_{n,\perp}-\|c\|^2)=\Psi_D(\G_n s_\perp,J)$. If a completed model exists, $\mathrm{LR}_n^{\orb}=T^{\orb}(Z_n)+o_p(1)$ under $Q_0$ by definition, so $\mathrm{C}_n^{\orb}-\mathrm{LR}_n^{\orb}=o_p(1)$ under $Q_0$, and by contiguity along every root-$n$ drift. For $D=\R^{d_\perp}$, the supremum defining $\Psi_D(u,A)$ is attained at $\delta=A^{-1}u$, so $\Psi_D(u,A)=u'A^{-1}u$; this gives the $C(\alpha)$ form, whose limit $U'J^{-1}U$ is $\chi^2_{d_\perp}$.
\end{proof}

\paragraph*{Use of AI tools.}Generative AI tools, Claude and ChatGPT, assisted with drafting, editing, writing code and checking the proofs. The author reviewed the paper and takes full responsibility for its content.

\setcitestyle{round}
\bibliographystyle{plainnat}
\bibliography{references}

\begin{thebibliography}{75}
\providecommand{\natexlab}[1]{#1}
\providecommand{\url}[1]{\texttt{#1}}
\expandafter\ifx\csname urlstyle\endcsname\relax
  \providecommand{\doi}[1]{doi: #1}\else
  \providecommand{\doi}{doi: \begingroup \urlstyle{rm}\Url}\fi

\bibitem[Anderson(1951)]{Anderson1951}
T.~W. Anderson.
\newblock Estimating linear restrictions on regression coefficients for
  multivariate normal distributions.
\newblock \emph{Annals of Mathematical Statistics}, 22:\penalty0 327--351,
  1951.

\bibitem[Anderson and Rubin(1949)]{AndersonRubin1949}
T.~W. Anderson and H.~Rubin.
\newblock Estimation of the parameters of a single equation in a complete
  system of stochastic equations.
\newblock \emph{Annals of Mathematical Statistics}, 20:\penalty0 46--63, 1949.

\bibitem[Andrews(1993)]{Andrews1993}
D.~W.~K. Andrews.
\newblock Tests for parameter instability and structural change with unknown
  change point.
\newblock \emph{Econometrica}, 61:\penalty0 821--856, 1993.

\bibitem[Andrews(2000)]{Andrews2000}
D.~W.~K. Andrews.
\newblock Inconsistency of the bootstrap when a parameter is on the boundary of
  the parameter space.
\newblock \emph{Econometrica}, 68:\penalty0 399--405, 2000.

\bibitem[Andrews(2001)]{Andrews2001}
D.~W.~K. Andrews.
\newblock Testing when a parameter is on the boundary of the maintained
  hypothesis.
\newblock \emph{Econometrica}, 69:\penalty0 683--734, 2001.

\bibitem[Andrews and Cheng(2012)]{AndrewsCheng2012}
D.~W.~K. Andrews and X.~Cheng.
\newblock Estimation and inference with weak, semi-strong, and strong
  identification.
\newblock \emph{Econometrica}, 80:\penalty0 2153--2211, 2012.

\bibitem[Andrews and Ploberger(1994)]{AndrewsPloberger1994}
D.~W.~K. Andrews and W.~Ploberger.
\newblock Optimal tests when a nuisance parameter is present only under the
  alternative.
\newblock \emph{Econometrica}, 62:\penalty0 1383--1414, 1994.

\bibitem[Andrews et~al.(2006)Andrews, Moreira, and
  Stock]{AndrewsMoreiraStock2006}
D.~W.~K. Andrews, M.~J. Moreira, and J.~H. Stock.
\newblock Optimal two-sided invariant similar tests for instrumental variables
  regression.
\newblock \emph{Econometrica}, 74:\penalty0 715--752, 2006.

\bibitem[Andrews and Mikusheva(2016)]{AndrewsMikusheva2016}
I.~Andrews and A.~Mikusheva.
\newblock Conditional inference with a functional nuisance parameter.
\newblock \emph{Econometrica}, 84:\penalty0 1571--1612, 2016.

\bibitem[Aza\"is et~al.(2009)Aza\"is, Gassiat, and
  Mercadier]{AzaisGassiatMercadier2009}
J.-M. Aza\"is, \'E. Gassiat, and C.~Mercadier.
\newblock The likelihood ratio test for general mixture models with or without
  structural parameter.
\newblock \emph{ESAIM: Probability and Statistics}, 13:\penalty0 301--327,
  2009.

\bibitem[Bai(1997)]{Bai1997}
J.~Bai.
\newblock Estimation of a change point in multiple regression models.
\newblock \emph{Review of Economics and Statistics}, 79:\penalty0 551--563,
  1997.

\bibitem[Barankin(1960)]{Barankin1960}
E.~W. Barankin.
\newblock Sufficient parameters: solution of the minimal dimensionality
  problem.
\newblock \emph{Annals of the Institute of Statistical Mathematics},
  12:\penalty0 91--118, 1960.

\bibitem[Beran(1987)]{Beran1987}
R.~Beran.
\newblock Prepivoting to reduce level error of confidence sets.
\newblock \emph{Biometrika}, 74:\penalty0 457--468, 1987.

\bibitem[Beran(1997)]{Beran1997}
R.~Beran.
\newblock Diagnosing bootstrap success.
\newblock \emph{Annals of the Institute of Statistical Mathematics},
  49:\penalty0 1--24, 1997.

\bibitem[Booth et~al.(1992)Booth, Hall, and Wood]{BoothHallWood1992}
J.~G. Booth, P.~Hall, and A.~T.~A. Wood.
\newblock Bootstrap estimation of conditional distributions.
\newblock \emph{Annals of Statistics}, 20:\penalty0 1594--1610, 1992.

\bibitem[Cattaneo et~al.(2012)Cattaneo, Crump, and
  Jansson]{CattaneoCrumpJansson2012}
M.~D. Cattaneo, R.~K. Crump, and M.~Jansson.
\newblock Optimal inference for instrumental variables regression with
  non-{Gaussian} errors.
\newblock \emph{Journal of Econometrics}, 167:\penalty0 1--15, 2012.

\bibitem[Cavaliere and Georgiev(2020)]{CavaliereGeorgiev2020}
G.~Cavaliere and I.~Georgiev.
\newblock Inference under random limit bootstrap measures.
\newblock \emph{Econometrica}, 88:\penalty0 2547--2574, 2020.

\bibitem[Cavaliere et~al.(2012)Cavaliere, Rahbek, and
  Taylor]{CavaliereRahbekTaylor2012}
G.~Cavaliere, A.~Rahbek, and A.~M.~R. Taylor.
\newblock Bootstrap determination of the co-integration rank in vector
  autoregressive models.
\newblock \emph{Econometrica}, 80:\penalty0 1721--1740, 2012.

\bibitem[Cavaliere et~al.(2024)Cavaliere, Gon\c{c}alves, Nielsen, and
  Zanelli]{CavaliereGoncalvesNielsenZanelli2024}
G.~Cavaliere, S.~Gon\c{c}alves, M.~{\O}. Nielsen, and E.~Zanelli.
\newblock Bootstrap inference in the presence of bias.
\newblock \emph{Journal of the American Statistical Association}, 119:\penalty0
  2908--2918, 2024.

\bibitem[Chen and Fang(2019)]{ChenFang2019}
Q.~Chen and Z.~Fang.
\newblock Improved inference on the rank of a matrix.
\newblock \emph{Quantitative Economics}, 10:\penalty0 1787--1824, 2019.

\bibitem[Chen et~al.(2018)Chen, Christensen, and
  Tamer]{ChenChristensenTamer2018}
X.~Chen, T.~M. Christensen, and E.~Tamer.
\newblock {Monte} {Carlo} confidence sets for identified sets.
\newblock \emph{Econometrica}, 86:\penalty0 1965--2018, 2018.

\bibitem[Cho and White(2018)]{ChoWhite2018}
J.~S. Cho and H.~White.
\newblock Directionally differentiable econometric models.
\newblock \emph{Econometric Theory}, 34:\penalty0 1101--1131, 2018.

\bibitem[Dacunha-Castelle and Gassiat(1999)]{DacunhaCastelleGassiat1999}
D.~Dacunha-Castelle and \'E. Gassiat.
\newblock Testing the order of a model using locally conic parametrization:
  population mixtures and stationary {ARMA} processes.
\newblock \emph{Annals of Statistics}, 27:\penalty0 1178--1209, 1999.

\bibitem[Davidson and MacKinnon(2008)]{DavidsonMacKinnon2008}
R.~Davidson and J.~G. MacKinnon.
\newblock Bootstrap inference in a linear equation estimated by instrumental
  variables.
\newblock \emph{Econometrics Journal}, 11:\penalty0 443--477, 2008.

\bibitem[Davies(1977)]{Davies1977}
R.~B. Davies.
\newblock Hypothesis testing when a nuisance parameter is present only under
  the alternative.
\newblock \emph{Biometrika}, 64:\penalty0 247--254, 1977.

\bibitem[Davies(1987)]{Davies1987}
R.~B. Davies.
\newblock Hypothesis testing when a nuisance parameter is present only under
  the alternative.
\newblock \emph{Biometrika}, 74:\penalty0 33--43, 1987.

\bibitem[DiCiccio and Young(2008)]{DiCiccioYoung2008}
T.~J. DiCiccio and G.~A. Young.
\newblock Conditional properties of unconditional parametric bootstrap
  procedures for inference in exponential families.
\newblock \emph{Biometrika}, 95:\penalty0 747--758, 2008.

\bibitem[Diebold and Li(2006)]{DieboldLi2006}
F.~X. Diebold and C.~Li.
\newblock Forecasting the term structure of government bond yields.
\newblock \emph{Journal of Econometrics}, 130:\penalty0 337--364, 2006.

\bibitem[Drton(2009)]{Drton2009}
M.~Drton.
\newblock Likelihood ratio tests and singularities.
\newblock \emph{Annals of Statistics}, 37:\penalty0 979--1012, 2009.

\bibitem[Drton and Williams(2011)]{DrtonWilliams2011}
M.~Drton and B.~Williams.
\newblock Quantifying the failure of bootstrap likelihood ratio tests.
\newblock \emph{Biometrika}, 98:\penalty0 919--934, 2011.

\bibitem[Dufour(1990)]{Dufour1990}
J.-M. Dufour.
\newblock Exact tests and confidence sets in linear regressions with
  autocorrelated errors.
\newblock \emph{Econometrica}, 58:\penalty0 475--494, 1990.

\bibitem[Dufour(2006)]{Dufour2006}
J.-M. Dufour.
\newblock Monte {Carlo} tests with nuisance parameters: a general approach to
  finite-sample inference and nonstandard asymptotics.
\newblock \emph{Journal of Econometrics}, 133:\penalty0 443--477, 2006.

\bibitem[Elliott and M{\"u}ller(2007)]{ElliottMuller2007}
G.~Elliott and U.~K. M{\"u}ller.
\newblock Confidence sets for the date of a single break in linear time series
  regressions.
\newblock \emph{Journal of Econometrics}, 141:\penalty0 1196--1218, 2007.

\bibitem[Elliott et~al.(2015)Elliott, M{\"u}ller, and
  Watson]{ElliottMullerWatson2015}
G.~Elliott, U.~K. M{\"u}ller, and M.~W. Watson.
\newblock Nearly optimal tests when a nuisance parameter is present under the
  null hypothesis.
\newblock \emph{Econometrica}, 83:\penalty0 771--811, 2015.

\bibitem[Fan et~al.(2017)Fan, Song, and Lu]{FanSongLu2017}
A.~Fan, R.~Song, and W.~Lu.
\newblock Change-plane analysis for subgroup detection and sample size
  calculation.
\newblock \emph{Journal of the American Statistical Association}, 112:\penalty0
  769--778, 2017.

\bibitem[Fang and Santos(2019)]{FangSantos2019}
Z.~Fang and A.~Santos.
\newblock Inference on directionally differentiable functions.
\newblock \emph{Review of Economic Studies}, 86:\penalty0 377--412, 2019.

\bibitem[Feng and McCulloch(1996)]{FengMcCulloch1996}
Z.~D. Feng and C.~E. McCulloch.
\newblock Using bootstrap likelihood ratios in finite mixture models.
\newblock \emph{Journal of the Royal Statistical Society: Series B},
  58:\penalty0 609--617, 1996.

\bibitem[Fukumizu(2003)]{Fukumizu2003}
K.~Fukumizu.
\newblock Likelihood ratio of unidentifiable models and multilayer neural
  networks.
\newblock \emph{Annals of Statistics}, 31:\penalty0 833--851, 2003.

\bibitem[Gassiat(2002)]{Gassiat2002}
\'E. Gassiat.
\newblock Likelihood ratio inequalities with applications to various mixtures.
\newblock \emph{Annales de l'Institut Henri Poincar\'e, Probabilit\'es et
  Statistiques}, 38:\penalty0 897--906, 2002.

\bibitem[Gu et~al.(2018)Gu, Koenker, and Volgushev]{GuKoenkerVolgushev2018}
J.~Gu, R.~Koenker, and S.~Volgushev.
\newblock Testing for homogeneity in mixture models.
\newblock \emph{Econometric Theory}, 34:\penalty0 850--895, 2018.

\bibitem[Han and McCloskey(2019)]{HanMcCloskey2019}
S.~Han and A.~McCloskey.
\newblock Estimation and inference with a (nearly) singular {Jacobian}.
\newblock \emph{Quantitative Economics}, 10:\penalty0 1019--1068, 2019.

\bibitem[Hansen(1996)]{Hansen1996}
B.~E. Hansen.
\newblock Inference when a nuisance parameter is not identified under the null
  hypothesis.
\newblock \emph{Econometrica}, 64:\penalty0 413--430, 1996.

\bibitem[Hansen(2000)]{Hansen2000}
B.~E. Hansen.
\newblock Sample splitting and threshold estimation.
\newblock \emph{Econometrica}, 68:\penalty0 575--603, 2000.

\bibitem[Hansen(2017)]{Hansen2017}
B.~E. Hansen.
\newblock Regression kink with an unknown threshold.
\newblock \emph{Journal of Business \& Economic Statistics}, 35:\penalty0
  228--240, 2017.

\bibitem[Hansen(2022)]{Hansen2022}
B.~E. Hansen.
\newblock \emph{Probability and Statistics for Economists}.
\newblock Princeton University Press, Princeton, 2022.

\bibitem[Higgins and Jochmans(2024)]{HigginsJochmans2024}
A.~Higgins and K.~Jochmans.
\newblock Bootstrap inference for fixed-effect models.
\newblock \emph{Econometrica}, 92:\penalty0 411--427, 2024.

\bibitem[Hirano and Porter(2012)]{HiranoPorter2012}
K.~Hirano and J.~R. Porter.
\newblock Impossibility results for nondifferentiable functionals.
\newblock \emph{Econometrica}, 80:\penalty0 1769--1790, 2012.

\bibitem[Hjort and Pollard(1993)]{HjortPollard1993}
N.~L. Hjort and D.~Pollard.
\newblock Asymptotics for minimisers of convex processes.
\newblock Technical report, Yale University, 1993.
\newblock arXiv:1107.3806.

\bibitem[Jansson and Moreira(2006)]{JanssonMoreira2006}
M.~Jansson and M.~J. Moreira.
\newblock Optimal inference in regression models with nearly integrated
  regressors.
\newblock \emph{Econometrica}, 74:\penalty0 681--714, 2006.

\bibitem[Jiang et~al.(2007)Jiang, Freidlin, and Simon]{JiangFreidlinSimon2007}
W.~Jiang, B.~Freidlin, and R.~Simon.
\newblock Biomarker-adaptive threshold design: a procedure for evaluating
  treatment with possible biomarker-defined subset effect.
\newblock \emph{Journal of the National Cancer Institute}, 99:\penalty0
  1036--1043, 2007.

\bibitem[Kaji(2021)]{Kaji2021}
T.~Kaji.
\newblock Theory of weak identification in semiparametric models.
\newblock \emph{Econometrica}, 89:\penalty0 733--763, 2021.

\bibitem[Kasahara and Shimotsu(2019)]{KasaharaShimotsu2019}
H.~Kasahara and K.~Shimotsu.
\newblock Testing the order of multivariate normal mixture models.
\newblock arXiv:1902.02920, 2019.

\bibitem[Kleibergen(2002)]{Kleibergen2002}
F.~Kleibergen.
\newblock Pivotal statistics for testing structural parameters in instrumental
  variables regression.
\newblock \emph{Econometrica}, 70:\penalty0 1781--1803, 2002.

\bibitem[Kleibergen(2005)]{Kleibergen2005}
F.~Kleibergen.
\newblock Testing parameters in {GMM} without assuming that they are
  identified.
\newblock \emph{Econometrica}, 73:\penalty0 1103--1123, 2005.

\bibitem[Kurozumi and Yamamoto(2015)]{KurozumiYamamoto2015}
E.~Kurozumi and Y.~Yamamoto.
\newblock Confidence sets for the break date based on optimal tests.
\newblock \emph{Econometrics Journal}, 18:\penalty0 412--435, 2015.

\bibitem[Lehmann and Romano(2005)]{LehmannRomano2005}
E.~L. Lehmann and J.~P. Romano.
\newblock \emph{Testing Statistical Hypotheses}.
\newblock Springer, New York, third edition, 2005.

\bibitem[Lindqvist and Taraldsen(2005)]{LindqvistTaraldsen2005}
B.~H. Lindqvist and G.~Taraldsen.
\newblock Monte {Carlo} conditioning on a sufficient statistic.
\newblock \emph{Biometrika}, 92:\penalty0 451--464, 2005.

\bibitem[Liu and Shao(2003)]{LiuShao2003}
X.~Liu and Y.~Shao.
\newblock Asymptotics for likelihood ratio tests under loss of identifiability.
\newblock \emph{Annals of Statistics}, 31:\penalty0 807--832, 2003.

\bibitem[McLachlan(1987)]{McLachlan1987}
G.~J. McLachlan.
\newblock On bootstrapping the likelihood ratio test statistic for the number
  of components in a normal mixture.
\newblock \emph{Journal of the Royal Statistical Society: Series C (Applied
  Statistics)}, 36:\penalty0 318--324, 1987.

\bibitem[Moreira and Moreira(2019)]{MoreiraMoreira2019}
H.~Moreira and M.~J. Moreira.
\newblock Optimal two-sided tests for instrumental variables regression with
  heteroskedastic and autocorrelated errors.
\newblock \emph{Journal of Econometrics}, 213:\penalty0 398--433, 2019.

\bibitem[Moreira(2003)]{Moreira2003}
M.~J. Moreira.
\newblock A conditional likelihood ratio test for structural models.
\newblock \emph{Econometrica}, 71:\penalty0 1027--1048, 2003.

\bibitem[Moreira(2009)]{Moreira2009tests}
M.~J. Moreira.
\newblock Tests with correct size when instruments can be arbitrarily weak.
\newblock \emph{Journal of Econometrics}, 152:\penalty0 131--140, 2009.

\bibitem[Moreira et~al.(2004)Moreira, Porter, and
  Suarez]{MoreiraPorterSuarez2004}
M.~J. Moreira, J.~R. Porter, and G.~A. Suarez.
\newblock Bootstrap and higher-order expansion validity when instruments may be
  weak.
\newblock Technical Working Paper 302, National Bureau of Economic Research,
  2004.

\bibitem[Moreira et~al.(2009)Moreira, Porter, and
  Suarez]{MoreiraPorterSuarez2009}
M.~J. Moreira, J.~R. Porter, and G.~A. Suarez.
\newblock Bootstrap validity for the score test when instruments may be weak.
\newblock \emph{Journal of Econometrics}, 149:\penalty0 52--64, 2009.

\bibitem[Nelson and Siegel(1987)]{NelsonSiegel1987}
C.~R. Nelson and A.~F. Siegel.
\newblock Parsimonious modeling of yield curves.
\newblock \emph{Journal of Business}, 60:\penalty0 473--489, 1987.

\bibitem[Neyman(1959)]{Neyman1959}
J.~Neyman.
\newblock Optimal asymptotic tests of composite statistical hypotheses.
\newblock In U.~Grenander, editor, \emph{Probability and Statistics: The
  {Harald Cram\'er} Volume}, pages 213--234. Almqvist and Wiksell, Stockholm,
  1959.

\bibitem[Picci(1977)]{Picci1977}
G.~Picci.
\newblock Some connections between the theory of sufficient statistics and the
  identifiability problem.
\newblock \emph{SIAM Journal on Applied Mathematics}, 33:\penalty0 383--398,
  1977.

\bibitem[Portier and Delyon(2014)]{PortierDelyon2014}
F.~Portier and B.~Delyon.
\newblock Bootstrap testing of the rank of a matrix via least-squared
  constrained estimation.
\newblock \emph{Journal of the American Statistical Association}, 109:\penalty0
  160--172, 2014.

\bibitem[Redner(1981)]{Redner1981}
R.~Redner.
\newblock Note on the consistency of the maximum likelihood estimate for
  nonidentifiable distributions.
\newblock \emph{Annals of Statistics}, 9:\penalty0 225--228, 1981.

\bibitem[Rynkiewicz(2008)]{Rynkiewicz2008}
J.~Rynkiewicz.
\newblock Asymptotic law of likelihood ratio for multilayer perceptron models.
\newblock In \emph{Advances in Neural Networks -- {ISNN} 2008}, volume 5263 of
  \emph{Lecture Notes in Computer Science}. Springer, Berlin, 2008.

\bibitem[Svensson(1994)]{Svensson1994}
L.~E.~O. Svensson.
\newblock Estimating and interpreting forward interest rates: {S}weden
  1992--1994.
\newblock Working Paper 4871, National Bureau of Economic Research, 1994.

\bibitem[Tsirel'son(1975)]{Tsirelson1975}
B.~S. Tsirel'son.
\newblock The density of the distribution of the maximum of a {Gaussian}
  process.
\newblock \emph{Theory of Probability and its Applications}, 20:\penalty0
  847--856, 1975.

\bibitem[Tuvaandorj(2025)]{Tuvaandorj2025}
P.~Tuvaandorj.
\newblock Robust permutation tests in linear instrumental variables regression.
\newblock \emph{Journal of the American Statistical Association}, 120:\penalty0
  1294--1304, 2025.

\bibitem[{van der Vaart}(1998)]{vanderVaart1998}
A.~W. {van der Vaart}.
\newblock \emph{Asymptotic Statistics}.
\newblock Cambridge University Press, Cambridge, 1998.

\bibitem[{van der Vaart} and Wellner(1996)]{vanderVaartWellner1996}
A.~W. {van der Vaart} and J.~A. Wellner.
\newblock \emph{Weak Convergence and Empirical Processes}.
\newblock Springer, New York, 1996.

\end{thebibliography}

\clearpage
\setcounter{section}{0}\setcounter{table}{0}\setcounter{figure}{0}\setcounter{equation}{0}
\renewcommand{\thesection}{S.\arabic{section}}
\renewcommand{\thetable}{S.\arabic{table}}
\renewcommand{\thefigure}{S.\arabic{figure}}
\renewcommand{\theHsection}{S.\arabic{section}}
\renewcommand{\theHtable}{S.\arabic{table}}
\renewcommand{\theHfigure}{S.\arabic{figure}}
\begin{center}
{\Large Supplemental Appendix to ``Bootstrap Likelihood Ratio Tests:\\ Quotient Geometry and Conditional Inference''}\\[2ex]
{\large Purevdorj Tuvaandorj}
\end{center}
\bigskip

\noindent This Supplemental Appendix contains the directional expansion (Lemma~\ref{lem:direxp}) and the proof of Proposition~\ref{prop:model} (Section~\ref{app:auxgeom}), proofs for the examples (Section~\ref{app:regularity}) and for Section~\ref{sec:apps} (Section~\ref{app:apps}), the implementation of the conditional bootstrap (Section~\ref{app:implement}) and additional simulations (Section~\ref{sec:practicalsim}). Unprefixed numbers refer to the main text.

\section{Auxiliary results}\label{app:auxgeom}

\paragraph{Directional expansion.}
In the notation of Proposition~\ref{prop:model}, write $\ell_n(P)\defeq L_n(\theta)$ for $P=P_\theta$, so that $\overline{L}_n(\eta_0)=\ell_n(Q_0)$ and $\ell_n(P)-\ell_n(Q_0)=\sum_i\log(1+\chi_Ps_P(X_i))$ when $P\ll Q_0$ and $\chi_P<\infty$. The proof of Proposition~\ref{prop:model} rests on the following expansion, which does not involve the product structure.

\begin{lemma}[Directional expansion]\label{lem:direxp}
Let $Q_0$ be a regular point of the null model, and let $\cP'\subset\cP$ be a model containing $Q_0$ that satisfies \ref{M1} and \ref{M2} with $\cP'$ in place of $\cP$. Let $\cD\subset L^2(Q_0)$ be a closed set of unit-norm functions that contains every limit point of $s_P$, $P\in\cP'$, as $\chi_P\to0$, and has a dense subset of derivatives of paths in $\cP'$ in the sense of \ref{M3}. Then, under $Q_0$,
\[
\sup_{P\in\cP'}\ell_n(P)-\ell_n(Q_0)=\tfrac12\sup_{d\in\cD}\plus{\G_nd}^2+o_p(1).
\]
\end{lemma}

\begin{proof}[Proof of Lemma~\ref{lem:direxp}]
Let $\mathcal R$ be the class in \ref{M2} for $\cP'$, $\overline{\mathcal R}$ its closure in $L^2(Q_0)$, and $x_-\defeq\min(x,0)$. Note that $d_{\mathrm{H}}(P,Q_0)\le\chi_P$, and that $\cD\subset\overline{\mathcal R}$: if $d$ is the derivative of a path $P_u$, then $\chi_{P_u}=u+o(u)$ and $s_{P_u}\to d$.

\paragraph*{Step 1: Bracketing.} By \ref{M2}, $\mathcal R$ is $Q_0$-Donsker with a square-integrable envelope $F$ \citep[Theorem~19.5]{vanderVaart1998}. An $L^2(Q_0)$-bracket $[l,u]$ within $[-F,F]$ gives brackets for $s^2$ and $s_-^2$, $s\in[l,u]$, of $L^1(Q_0)$-width at most $2\|u-l\|\,\|F\|$, so $\{s^2:s\in\mathcal R\}$ and $\{s_-^2:s\in\mathcal R\}$ are Glivenko--Cantelli \citep[Theorem~19.4]{vanderVaart1998}. Elements of $\overline{\mathcal R}$ have mean zero and unit norm, so $s_-\ne0$; as $\overline{\mathcal R}$ is compact, $\inf_{s\in\overline{\mathcal R}}\E_{Q_0}[s_-^2]>0$.

\paragraph*{Step 2: Rate.} Since $\log(1+x)\le x-x_-^2/2$ for $x>-1$, every $P$ with $0<d_{\mathrm{H}}(P,Q_0)\le\varepsilon$ and $\ell_n(P)\ge\ell_n(Q_0)$ satisfies, with $s=s_P$,
\[
0\le\sum_i\log(1+\chi_Ps(X_i))\le\chi_Pn^{1/2}\G_ns-\tfrac12\chi_P^2\sum_is_-(X_i)^2,
\]
so $\chi_P\le2n^{-1/2}\G_ns/(n^{-1}\sum_is_-(X_i)^2)$ \citep[Inequality~1.1]{Gassiat2002}. By Step~1 and \ref{M1} with $\varepsilon'=\varepsilon$, $\sup\{n^{1/2}\chi_P:P\in\cP',\ \ell_n(P)\ge\ell_n(Q_0)\}=O_p(1)$.

\paragraph*{Step 3: Quadratic expansion.} For fixed $M$, uniformly over $P$ with $0<\chi_P\le Mn^{-1/2}$ and $t\defeq n^{1/2}\chi_P$,
\[
\ell_n(P)-\ell_n(Q_0)=t\,\G_ns_P-\tfrac12t^2+o_p(1).
\]
Indeed, $\max_i\chi_P|s_P(X_i)|\le Mn^{-1/2}\max_iF(X_i)=o_p(1)$ \citep[Theorem~9.7]{Hansen2022}, $|\log(1+x)-x+x^2/2|\le|x|^3$ for $|x|\le1/2$, and $n^{-1}\sum_is_P(X_i)^2=1+o_p(1)$ uniformly by Step~1, so the remainder is at most $\max_i\chi_P|s_P(X_i)|\,t^2n^{-1}\sum_is_P(X_i)^2=o_p(1)$.

\paragraph*{Step 4: Limit directions.} The left side of the display in the lemma is nonnegative, so by Steps~2--3 and $tg-t^2/2\le\plus{g}^2/2$, with probability at least $1-\epsilon$ it is at most $\frac12\sup\{\plus{\G_ns_P}^2:0<\chi_P\le Mn^{-1/2}\}+o_p(1)$ for some $M=M(\epsilon)$. By compactness of $\overline{\mathcal R}$, for every $\rho>0$ there is $\varepsilon'>0$ such that every $s_P$ with $\chi_P\le\varepsilon'$ is within $\rho$ of $\cD$, and asymptotic equicontinuity of $\G_n$ on $\overline{\mathcal R}$ gives the upper bound. Conversely, if $d$ is the derivative of a path $P_u$, $t>0$ and $u_n\defeq tn^{-1/2}$, then $n^{1/2}\chi_{P_{u_n}}\to t$ and $s_{P_{u_n}}\to d$, so $\ell_n(P_{u_n})-\ell_n(Q_0)=t\,\G_nd-\tfrac12t^2+o_p(1)$ by Step~3 and equicontinuity. Maximizing over finite sets of such $d$ and of $t$ gives the lower bound, because $\G_n$ is asymptotically equicontinuous on the compact $\cD$ and $\sup_{d\in\cD}\plus{\G_nd}^2$ is tight.
\end{proof}

\begin{proof}[Proof of Proposition~\ref{prop:model}]
For the full model, \ref{M3} gives Lemma~\ref{lem:direxp} with $\cD=\cl\,\Sph(L+K_0)$, which is \eqref{eq:directional}, and with the Donsker condition in \ref{M3}, Assumption~\ref{ass:directional} holds. The null model $\cP_0\subset\cP$ satisfies \ref{M1} and \ref{M2}. If $P\in\cP_0$ and $\chi_P\to0$, then $d_{\mathrm{H}}(P,Q_0)\to0$, so $P=Q_\eta$ with $\eta\to\eta_0$, and since $\E_{Q_0}[(dQ_\eta/dQ_0-1-(\eta-\eta_0)'s_0)^2]=o(\|\eta-\eta_0\|^2)$, $s_P$ approaches the unit vector $(\eta-\eta_0)'s_0/\|I_0^{1/2}(\eta-\eta_0)\|$ of $L$; each unit vector $c's_0$ is the derivative of $u\mapsto Q_{\eta_0+uc}$. Lemma~\ref{lem:direxp} with $\cP'=\cP_0$ and $\cD=\Sph(L)$ gives
\[
\sup_{\theta\in\Theta_0}L_n(\theta)-\overline{L}_n(\eta_0)=\tfrac12\sup_{c\ne0}\frac{\plus{c'\Delta_n}^2}{c'I_0c}+o_p(1)=\tfrac12\Delta_n'I_0^{-1}\Delta_n+o_p(1).
\]
Since $L_n(\tilde\theta)\ge\overline{L}_n(\eta_0)$, Step~2 of the proof of Lemma~\ref{lem:direxp} gives $\chi_{P_{\tilde\theta}}=O_p(n^{-1/2})$, so $d_{\mathrm{H}}(P_{\tilde\theta},Q_0)\conp0$, which gives \ref{A0} and $\tilde\eta\conp\eta_0$. As $\chi_{Q_\eta}\ge\frac12\|I_0^{1/2}(\eta-\eta_0)\|$ near $\eta_0$ by the same expansion, $n^{1/2}(\tilde\eta-\eta_0)=O_p(1)$, which completes \ref{A1}.
\end{proof}

\section{Proofs for the examples}\label{app:regularity}

\paragraph{One hidden unit.} We refer to the conditions of Proposition~\ref{prop:nn} as \inpart{N1}{N1}, the bounds on $\E[\|x\|^4]$ and $|\psi(x;\gamma)|$; \inpart{N2}{N2}, the Donsker and Glivenko--Cantelli conditions and continuity of $J$; and \inpart{N3}{N3}, the bound on $\lambda_{\min}(J(\gamma))$.
For $\psi(x;\gamma)=\varphi(\gamma'\tilde x)$ with $\varphi$ a sigmoid, $\max(t,0)$ or $1\{t>0\}$, slopes of $\gamma$ bounded away from zero, and $x$ with a density positive on an open set $\mathcal O$ met by the hyperplanes $\{\gamma'\tilde x=0\}$, $\gamma\in\Gamma$, \ref{N2} holds because the first two are Lipschitz in $\gamma$ and indicators of half-spaces form a Vapnik--Chervonenkis class \citep[Sections~2.6 and~2.7]{vanderVaartWellner1996}, and \ref{N3} from linear independence of $1,x,\psi(x;\gamma)$ on $\mathcal O$ and compactness of $\Gamma$.

\begin{proof}[Proof of Proposition~\ref{prop:nn}]
Write $\vartheta\defeq(\beta',a)'\in\R^{d_x+2}$ and
\[
L_n(\vartheta;\gamma)\defeq\sum_i\bigl(y_i\vartheta'z(x_i;\gamma)-\log(1+e^{\vartheta'z(x_i;\gamma)})\bigr).
\]
Let $\vartheta_0\defeq(\beta_0',0)'$. The null log-likelihood is $L_n((\beta',0)';\gamma)$ for any $\gamma$, and the full one satisfies $\sup_{\theta\in\Theta}L_n(\theta)=\sup_{\gamma\in\Gamma}\sup_{\vartheta}L_n(\vartheta;\gamma)$.

\paragraph*{Step 1: Non-degeneracy.} By \ref{N3}, $\lambda_{\min}(J(\gamma))\ge\lambda>0$ on $\Gamma$. Since $\omega(\gamma)^2=J_{aa}(\gamma)-J_{a\beta}(\gamma)I_0^{-1}J_{\beta a}(\gamma)$ is the reciprocal of the last diagonal element of $J(\gamma)^{-1}$, $\inf_\Gamma \omega(\gamma)^2\ge\lambda$. By \ref{N1}, $c(\gamma)$ and $\omega(\gamma)$ are bounded on $\Gamma$.

\paragraph*{Step 2: Uniform local quadratic approximation.} For $h\in\R^{d_x+2}$, a third-order Taylor expansion of $t\mapsto\log(1+e^t)$, whose third derivative is bounded by $1$, gives
\[
L_n(\vartheta_0+hn^{-1/2};\gamma)-L_n(\vartheta_0;\gamma)=h'S_n(\gamma)-\tfrac12h'J_n(\gamma)h+r_n(h;\gamma),
\]
with $S_n(\gamma)\defeq\G_n[z(x;\gamma)(y-p_0)]$, $J_n(\gamma)\defeq n^{-1}\sum_ip_0(x_i)(1-p_0(x_i))z(x_i;\gamma)z(x_i;\gamma)'$, and
$|r_n(h;\gamma)|\le\|h\|^3(6\,n^{1/2})^{-1}n^{-1}\sum_i\|z(x_i;\gamma)\|^3$. By \ref{N1}, $\|z(x;\gamma)\|\le C(1+\|x\|)$ uniformly in $\gamma$, so $\sup_{\gamma}\sup_{\|h\|\le M}|r_n|=O_p(n^{-1/2})$ for each $M$. By \ref{N2}, $\sup_\gamma\|S_n(\gamma)\|=O_p(1)$ and $\sup_\gamma\|J_n(\gamma)-J(\gamma)\|\conp0$.

\paragraph*{Step 3: Localization by concavity.} For fixed $\gamma$, $h\mapsto f_n(h;\gamma)\defeq L_n(\vartheta_0+hn^{-1/2};\gamma)-L_n(\vartheta_0;\gamma)$ is concave. Let $f(h;\gamma)\defeq h'S_n(\gamma)-\tfrac12h'J(\gamma)h$, with maximizer $h^*(\gamma)\defeq J(\gamma)^{-1}S_n(\gamma)$ and maximum $\tfrac12S_n'J^{-1}S_n$. Given $\epsilon>0$, choose $M$ with $\Pr(\sup_\gamma\|h^*(\gamma)\|>M)<\epsilon$, and define $\delta_n\defeq\sup_\gamma\sup_{\|h-h^*(\gamma)\|\le1}|f_n(h;\gamma)-f(h;\gamma)|$, which is $o_p(1)$ by Step~2 on the ball of radius $M+1$. By the convexity lemma of \citet[Lemma~2]{HjortPollard1993}, applied to $-f_n(\cdot;\gamma)$, whenever $\delta_n<\lambda/4$, $f_n(\cdot;\gamma)$ attains its maximum at some $\hat h(\gamma)$ with $\|\hat h(\gamma)-h^*(\gamma)\|<1$, simultaneously for all $\gamma$, because $f(h^*;\gamma)-f(h;\gamma)\ge\lambda/2$ when $\|h-h^*(\gamma)\|=1$ and $f_n(\cdot;\gamma)$ is concave. On that event, with $\gamma$ suppressed, $f(h^*)-\delta_n\le f_n(h^*)\le f_n(\hat h)\le f(\hat h)+\delta_n\le f(h^*)+\delta_n$. Hence,
\[
\sup_\gamma\Bigl|\sup_hf_n(h;\gamma)-\tfrac12S_n(\gamma)'J(\gamma)^{-1}S_n(\gamma)\Bigr|=o_p(1).
\]
Because $\vartheta_0$ is interior to $B\times[-A,A]$ and $\sup_\gamma\|\hat h(\gamma)\|=O_p(1)$, the parameter constraints are inactive w.p.a.1. The same argument without the last regressor gives \ref{A1} for the null logit, with $\Delta_n=\G_n[\tilde x(y-p_0)]$; the logit family is DQM with information $I_0$, nonsingular by \ref{N3}, so $\beta_0$ is a regular point of the null (Definition~\ref{def:nullpar}), and \ref{A0} holds with $\mathcal H$ the interior of $B$, by consistency of the logit MLE.

\paragraph*{Step 4: Partitioned inverse.} Write $S_n(\gamma)=(\Delta_n',\G_n[\psi(x;\gamma)(y-p_0)])'$. The partitioned-inverse formula gives
\[
S_n'J^{-1}S_n-\Delta_n'I_0^{-1}\Delta_n=\frac{(\G_n[\psi(x;\gamma)(y-p_0)]-J_{a\beta}(\gamma)I_0^{-1}\Delta_n)^2}{\omega(\gamma)^2}=(\G_ne(\cdot;\gamma))^2,
\]
because $J_{a\beta}(\gamma)I_0^{-1}=c(\gamma)'$ and $\G_n$ is linear. Combining with Step~3, $\mathrm{LR}_n=\sup_\gamma(\G_ne(\cdot;\gamma))^2+o_p(1)=\sup_{e\in\cE}\plus{\G_ne}^2+o_p(1)$.

\paragraph*{Step 5: The class $\cE$.} Each $e(\cdot;\gamma)$ has mean zero and unit variance, and $\ip{e(\cdot;\gamma)}{v'\tilde x(y-p_0)}=v'\E[p_0(1-p_0)\tilde x(\psi(x;\gamma)-c(\gamma)'\tilde x)]/\omega(\gamma)=0$ by the definition of $c(\gamma)$, so $\cE\perp L$. Moreover, $e(\cdot;\gamma)=\omega(\gamma)^{-1}\psi(x;\gamma)(y-p_0)-\omega(\gamma)^{-1}c(\gamma)'\tilde x(y-p_0)$ is a combination, with coefficients bounded uniformly in $\gamma$ by Step~1, of an element of the Donsker class in \ref{N2} and the finitely many square-integrable functions $\tilde x(y-p_0)$. Such classes are Donsker \citep[Section~2.10]{vanderVaartWellner1996}, so $\cE$ together with the components of $\tilde x(y-p_0)$ is Donsker, which verifies Assumption~\ref{ass:expansion}. Theorems~\ref{thm:drift} and~\ref{thm:boot} give the remaining claims.
\end{proof}

\paragraph{Instrument relevance.}
In the IV example of Section~\ref{sec:setup}, let $Y\in\R^{n\times2}$ and $Z\in\R^{n\times k}$ stack the observations and let $\hat\rho^2$ be the largest eigenvalue of $(Y'Y)^{-1}Y'Z(Z'Z)^{-1}Z'Y$. Under $H_0:\pi=0$, $\mathrm{LR}_n=-n\log(1-\hat\rho^2)$ is the reduced-rank regression statistic of \citet{Anderson1951} for rank zero against rank one.

\begin{proposition}[Instrument relevance]\label{prop:iv}
For $H_0:\pi=0$ in this model, \inpart{prop:iv:a}{a}~conditionally on $Z$ with $Z'Z$ nonsingular and $n\ge k+2$, the null distribution of $\mathrm{LR}_n$ does not depend on $\Sigma$, so the bootstrap from $\mathcal N(0,Y'Y/n)$ is exact; \inpart{prop:iv:b}{b}~$\mathrm{LR}_n=\sigma_1(\Xi_n)^2+o_p(1)$ under $Q_0$ and $\cE$ is $Q_0$-Donsker, so Assumption~\ref{ass:expansion} holds.
\end{proposition}

\begin{proof}[Proof of Proposition~\ref{prop:iv}]
(a) Under $H_0$, $Y=U\Sigma^{1/2}$ with $U$ having i.i.d.\ $\mathcal N(0,I_2)$ rows independent of $Z$. Replacing $Y$ by $YA$ for a nonsingular $A$ does not change the eigenvalues of $(Y'Y)^{-1}Y'Z(Z'Z)^{-1}Z'Y$, so $\mathrm{LR}_n$ is the same function of $(U,Z)$ for every $\Sigma$, and its bootstrap counterpart, computed from $Y^*=U^*(Y'Y/n)^{1/2}$, is the same function of $(U^*,Z)$.

(b) Let $\hat\Xi_n\defeq(Z'Z)^{-1/2}Z'U$, so that $\hat\rho^2$ is the largest eigenvalue of $(U'U)^{-1}\hat\Xi_n'\hat\Xi_n$. Under $H_0$, $\Xi_n=n^{-1/2}\Sigma_{zz}^{-1/2}Z'U$, and as $n^{-1}U'U\conas I_2$ and $n^{-1}Z'Z\conas\Sigma_{zz}$, $\hat\Xi_n-\Xi_n=o_p(1)$ and $\mathrm{LR}_n=n\hat\rho^2+o_p(1)=\sigma_1(\Xi_n)^2+o_p(1)$. The elements $e_{a,b}\defeq(a'\Sigma_{zz}^{-1/2}z)(b'\Sigma_0^{-1/2}r)$ of $\cE$ satisfy $|e_{a,b}-e_{a',b'}|\le\|\Sigma_{zz}^{-1/2}z\|\,\|\Sigma_0^{-1/2}r\|\,(\|a-a'\|+\|b-b'\|)$ with a square-integrable constant, so $\cE$ is Donsker \citep[Example~19.7]{vanderVaart1998}. As $\G_ne_{a,b}=a'\Xi_nb$ and $\cE=-\cE$, $\sigma_1(\Xi_n)^2=\sup_\cE\plus{\G_ne}^2$.
\end{proof}

\paragraph{Structural null.} The next result gives the facts about $H_0:\gamma=0$ used in Sections~\ref{sec:exact} and~\ref{sec:entangled}, with $\mathrm{CLR}$ as defined there; $S_n$ and $T_n$ are defined in the proof.

\begin{proposition}[Structural null in the IV model]\label{prop:iv2}
In the Gaussian IV model with $k\ge2$ and $H_0:\gamma=0$, let $\Pi_n\defeq[0\ \,h/n^{1/2}]$, $h\in\R^k$.
\begin{enumerate}[label=(\alph*)]
\item\label{prop:iv2:a} $\mathrm{LR}_n=\mathrm{CLR}(S_n,T_n)+o_p(1)$ and $[S_n\ \,T_n]\dto[S\ \,T_h]$, with $S\sim\mathcal N(0,I_k)$ independent of $T_h\sim\mathcal N(\Sigma_{zz}^{1/2}h/\sigma_{x\cdot y},I_k)$, where $\sigma_{x\cdot y}^2$ is the variance of $x$ given $(y,z)$.
\item\label{prop:iv2:b} The limit $\mathrm{CLR}(S,T_h)$ is strictly stochastically larger than $\chi^2_1$ and tends to $\chi^2_1$ as $\|h\|\to\infty$; at $\pi=0$, the restricted-fit bootstrap is inconsistent.
\end{enumerate}
\end{proposition}

\begin{proof}[Proof of Proposition~\ref{prop:iv2}]
Under $H_0$, $\beta=0$ or $\pi=0$, so $\Theta_0$ contains every $\beta$ at $\pi=0$ and is saturated; its image is $\mathcal V_0\defeq\{[0\ \,\pi]:\pi\in\R^k\}$, and the null is regular (Section~\ref{sec:assumptions}). The restricted MLE factorizes as $y\sim \mathcal N(0,\sigma_{yy})$ and $x\,\vert\, y,z\sim \mathcal N(z'\pi+(\sigma_{xy}/\sigma_{yy})y,\sigma^2_{x\cdot y})$, so $\tilde\pi$ is the coefficient on $Z$ in the regression of $x$ on $(Z,y)$.

(a) Let $\hat\Pi\defeq(Z'Z)^{-1}Z'Y$, $\hat\Sigma\defeq(Y-Z\hat\Pi)'(Y-Z\hat\Pi)/n$, $b_0\defeq(1,0)'$, $a_0\defeq(0,1)'$, $S_n\defeq(Z'Z)^{-1/2}Z'Yb_0/(b_0'\hat\Sigma b_0)^{1/2}$ and $T_n\defeq(Z'Z)^{-1/2}Z'Y\hat\Sigma^{-1}a_0/(a_0'\hat\Sigma^{-1}a_0)^{1/2}$. With $D(\Pi)\defeq\hat\Sigma^{-1/2}(\hat\Pi-\Pi)'Z'Z(\hat\Pi-\Pi)\hat\Sigma^{-1/2}$, the LR statistic with known variance $\hat\Sigma$ is $\inf_{\mathcal V_0}\operatorname{tr}D-\inf_{\mathcal V_1}\operatorname{tr}D=\mathrm{CLR}(S_n,T_n)$ \citep{Moreira2003}, and $\mathrm{LR}_n=\inf_{\mathcal V_0}m_n-\inf_{\mathcal V_1}m_n$ with $m_n\defeq n\log\det(I_2+D/n)$. As $\operatorname{tr}D-(\operatorname{tr}D)^2/(2n)\le m_n\le\operatorname{tr}D$ and $m_n\ge n\log(1+\operatorname{tr}D/n)$, and $\operatorname{tr}D(\Pi_n)=O_p(1)$ with $\Pi_n\in\mathcal V_0\subset\mathcal V_1$, near-minimizers over either set have $\operatorname{tr}D=O_p(1)$, so $\mathrm{LR}_n=\mathrm{CLR}(S_n,T_n)+O_p(n^{-1})$. Given $Z$, $(Z'Z)^{-1/2}Z'Y$ is matrix normal with mean $(Z'Z)^{1/2}\Pi_n\to\Sigma_{zz}^{1/2}[0\ \,h]$, row covariance $I_k$ and column covariance $\Sigma$, and $\hat\Sigma\conp\Sigma$; since $b_0'\Sigma\Sigma^{-1}a_0=0$, $[S_n\ \,T_n]\dto[S\ \,T_h]$ with $S\sim\mathcal N(0,I_k)$ independent of $T_h\sim\mathcal N(\Sigma_{zz}^{1/2}h/\sigma_{x\cdot y},I_k)$, and $\mathrm{LR}_n\dto\Lambda(h)\defeq\mathrm{CLR}(S,T_h)$, with distribution $\mu(h)$.

(b) As $\mathrm{CLR}(s,t)=\lambda_{\max}(ss'+tt')-\|t\|^2$, $\mathrm{CLR}(s,t)\ge(u's)^2$ for $u\defeq t/\|t\|$ and $\mathrm{CLR}(s,t)\ge\|s\|^2-\|t\|^2$. With $u\defeq T_h/\|T_h\|$, independent of $S$, $(u'S)^2\sim\chi^2_1$, $\Lambda(h)\ge(u'S)^2$, and $\Lambda(h)>c\ge(u'S)^2$ with positive probability for every $c>0$ because $k\ge2$, so $\Lambda(h)$ is strictly stochastically larger than $\chi^2_1$. From the closed form $\mathrm{CLR}(s,t)=\tfrac12\bigl(s's-t't+((s's-t't)^2+4(s't)^2)^{1/2}\bigr)$, $\mathrm{CLR}(s,t)-(u's)^2=O(\|t\|^{-2})$ as $\|t\|\to\infty$ for fixed $s$, so $\mu(h)\to\chi^2_1$ as $\|h\|\to\infty$. At $\pi_0=0$, $n^{1/2}\tilde\pi\dto\mathcal N(0,\sigma^2_{x\cdot y}\Sigma_{zz}^{-1})$, so $\hat H_n$ has a limit $H$ whose first-stage component is normal with full support, and the limit of $\mathrm{LR}_n$ depends on the local null parameter only through that component $h$, the covariance-score directions being harmless. Since $\mu$ is continuous (Proposition~\ref{prop:dichotomy}\ref{prop:dichotomy:a}) and $\mu(h)\to\chi^2_1\ne\mu(0)$, $\mu(h)\ne\mu(0)$ on a nonempty open set of $h$, so $\Pr(\mu(H)\ne\mu(0))>0$, and Proposition~\ref{prop:dichotomy}\ref{prop:dichotomy:b} gives inconsistency.
\end{proof}

\section{Proofs for the applications}\label{app:apps}

\paragraph{Endogenous binary treatment.} Let $\phi_0\defeq1-\phi_1$. With the cells ordered as in Section~\ref{sec:weakiv}, the cell probabilities given $Z_i=0$ are
\[
q(\pi,\zeta)=\bigl(C(\pi_2,\zeta;\pi_3),\ \pi_1-C(\pi_1,\zeta;\pi_3),\ \zeta-C(\pi_2,\zeta;\pi_3)\bigr)',
\]
those given $Z_i=1$ are $q(\pi,\zeta+\beta)$, and $L_n(\theta)$, $\theta=(\pi_1,\pi_2,\pi_3,\zeta,\beta)'$, is the multinomial log-likelihood of the cells given the instruments. The parameter space bounds $|\pi_3|\le\bar\rho<1$ and keeps $\pi_1$, $\pi_2$, $\zeta$ and $\zeta+\beta$ in $[\varepsilon_0,1-\varepsilon_0]$ for some $\varepsilon_0\in(0,1/2)$. Write $\partial_1C,\partial_2C,\partial_3C$ for the partial derivatives of $C$ with respect to its three arguments, and
\[
a(\pi,\zeta)\defeq\frac{\partial q(\pi,\zeta)}{\partial \zeta}=\bigl(\partial_2C(\pi_2,\zeta;\pi_3),\ -\partial_2C(\pi_1,\zeta;\pi_3),\ 1-\partial_2C(\pi_2,\zeta;\pi_3)\bigr)'.
\]
Let $n_z$ be the number of observations with $Z_i=z$, $q^0\defeq q(\pi^0,\zeta^0)$ the common null cell vector at $\beta=0$ and $V_q\defeq\operatorname{diag}(q^0)-q^0q^{0\prime}$. Then, $\Omega=V_q/(\phi_0\phi_1)$ and, as in Section~\ref{sec:weakiv}, $g(\pi_2)=\Omega^{-1/2}a(\pi(\pi_2),\zeta^0)$, with $\pi(\pi_2)$ as in the proof of Proposition~\ref{prop:hm}.

\begin{proof}[Proof of Lemma~\ref{lem:hm-null}]
Let $M$ be the $3\times3$ Jacobian of $(\pi_1,\pi_3,\zeta)'\mapsto q(\pi,\zeta)$. Its columns are
\[
(0,1-\partial_1C(\pi_1,\zeta),0)',\qquad(\partial_3C(\pi_2,\zeta),-\partial_3C(\pi_1,\zeta),-\partial_3C(\pi_2,\zeta))',\qquad a(\pi,\zeta),
\]
and expanding along the first column gives $\det M=-(1-\partial_1C(\pi_1,\zeta))\,\partial_3C(\pi_2,\zeta)$. For the Gaussian copula, $\partial_3C(\pi_2,\zeta)$ is the bivariate normal density at $(\Phi^{-1}(\pi_2),\Phi^{-1}(\zeta))$, which is positive, and $\partial_1C(\pi_1,\zeta)=\Phi\bigl((\Phi^{-1}(\zeta)-\pi_3\Phi^{-1}(\pi_1))/(1-\pi_3^2)^{1/2}\bigr)<1$, so $\det M\ne0$ whenever $|\pi_3|<1$. The same holds for the Ali--Mikhail--Haq copula $C(u,v;\pi_3)=uv/(1-\pi_3(1-u)(1-v))$ used by \citet{HanMcCloskey2019}: $\partial_3C(u,v;\pi_3)=uv(1-u)(1-v)/(1-\pi_3(1-u)(1-v))^2>0$ on the interior, and $\partial_1C(u,v;\pi_3)$, a conditional distribution function in $v$, is less than one for $v<1$. The Jacobian of the null model has columns $(m_1',m_1(\zeta+\beta)')'$, $(m_2',m_2(\zeta+\beta)')'$, $(a(\pi,\zeta)',a(\pi,\zeta+\beta)')'$ and $(0',a(\pi,\zeta+\beta)')'$, where $m_1,m_2$ are the first two columns of $M$ and $m_j(\zeta+\beta)$ their values at $\zeta+\beta$. If a linear combination with coefficients $c_1,\dots,c_4$ vanishes, its first block is $M(c_1,c_2,c_3)'=0$, so $c_1=c_2=c_3=0$, and then $c_4a(\pi,\zeta+\beta)=0$, so $c_4=0$, as the first and third components of $a$ sum to one.
\end{proof}

\begin{proof}[Proof of Proposition~\ref{prop:hm}]
At $\beta=0$, the first and third cell probabilities sum to $\zeta$, so $\zeta=\zeta^0$ for every parameter observationally equivalent to the truth. Given $\pi_2$, the equations $C(\pi_2,\zeta^0;\pi_3)=q^0_1$ and $\pi_1-C(\pi_1,\zeta^0;\pi_3)=q^0_2$ have at most one solution $(\pi_1(\pi_2),\pi_3(\pi_2))'$, because $\partial_3C>0$ and $1-\partial_1C>0$ (proof of Lemma~\ref{lem:hm-null}). As $C$ increases strictly in its first argument and in $\pi_3$, $\pi_3(\pi_2)$ decreases strictly in $\pi_2$, and so does $\pi_1(\pi_2)$, because $\pi_1-C(\pi_1,\zeta^0;\pi_3)$ increases in $\pi_1$ and decreases in $\pi_3$. Hence, the set $\mathcal I_2$ of $\pi_2$ for which $\pi(\pi_2)$ lies in the parameter space is a compact interval, no bound binds at $\pi(\pi_2)$ for $\pi_2$ in its interior, and at least one binds at each end point. By the implicit function theorem and $\det M\ne0$, $\pi(\pi_2)$, $g$ and $u$ are continuously differentiable on an open interval containing $\mathcal I_2$.

\paragraph*{Step 1: Localization.} The null parametrization is one-to-one: given $Z_i=0$, the first and third cell probabilities sum to $\zeta$, the first then determines $\pi_3$ and the second $\pi_1$, as $\partial_3C>0$ and $1-\partial_1C>0$, and given $Z_i=1$ the sum is $\zeta+\beta$. With compactness, the inverse is continuous, and Lemma~\ref{lem:hm-null} gives Assumption~\ref{ass:null}. Under $\beta=h/n^{1/2}$, both $\hat q_0$ and $\hat q_1$ converge to $q^0$, and by the usual argument for multinomial likelihoods every $\theta$ at which $L_n(\theta)$ is within $O_p(1)$ of the saturated maximum has both conditional cell vectors within $O_p(n^{-1/2})$ of $q^0$. Since the first and third components of $q(\pi,\zeta)$ sum to $\zeta$, such parameters have $\beta=O_p(n^{-1/2})$ and $(\pi',\zeta)'$ within $o_p(1)$ of the identified curve. Parameters whose cell vectors are at distance at least $\epsilon>0$ from $q^0$ have a log-likelihood deficit of order $n$ and do not contribute.

\paragraph*{Step 2: Quadratic expansion.} Write $q_z$ for the model cell vectors, $x_z\defeq\hat q_z-q_z$ and $\bar x\defeq\phi_0x_0+\phi_1x_1$. Uniformly over parameters with $q_z-q^0=O(n^{-1/2})$, and using $n_z/n\to\phi_z$,
\[
2\bigl(L_n(\theta)-L_n^{\mathrm{sat}}\bigr)=-n\sum_{z}\phi_zx_z'V_q^{-1}x_z+o_p(1).
\]
The identity $\sum_z\phi_zx_zx_z'=\bar x\bar x'+\phi_0\phi_1(x_1-x_0)(x_1-x_0)'$ splits the right side into a level term $-n\bar x'V_q^{-1}\bar x$ and the difference term $-\|X_n-n^{1/2}\,\Omega^{-1/2}(q_1-q_0)\|^2$.

\paragraph*{Step 3: Fixed $\pi_2$ in the interior.} Fix $\pi_2$ in the interior of $\mathcal I_2$. The map $(\pi_1,\pi_3,\zeta)'\mapsto\phi_0q(\pi,\zeta)+\phi_1q(\pi,\zeta+\beta)$ has Jacobian $M$ at $\beta=0$, so by the inverse function theorem every cell vector near $q^0$ is its value at a unique $(\pi_1,\pi_3,\zeta)'$ near $(\pi_1(\pi_2),\pi_3(\pi_2),\zeta^0)'$ when $\beta$ is small; hence the level term can be made zero by a perturbation of order $n^{-1/2}$, and with $\beta=b/n^{1/2}$ the difference $n^{1/2}\,\Omega^{-1/2}(q_1-q_0)$ equals $bg(\pi_2)+o(1)$. Minimizing over $b$,
\[
\mathrm{PL}_n(\pi_2)\defeq2\,\Bigl(\sup_{\theta:\,\pi_2\ \mathrm{fixed}}L_n(\theta)-L_n^{\mathrm{sat}}\Bigr)=\mathrm{PL}(\pi_2,X_n)+o_p(1),
\]
with $\mathrm{PL}(\pi_2,x)\defeq-\|x\|^2+(u(\pi_2)'x)^2$,
uniformly over $\pi_2$ in every compact interval $J$ in the interior of $\mathcal I_2$, by continuity of the inverse map and of $g$.

\paragraph*{Step 4: End points.} At the end points of $\mathcal I_2$, a bound binds and Step~3 does not apply directly. Fix $\bar\rho'\in(\bar\rho,1)$ and $\varepsilon_0'\in(0,\varepsilon_0)$, and let $\mathrm{PL}'_n$ be defined as $\mathrm{PL}_n$ with these bounds. Enlarging the parameter set can only increase the supremum, so $\mathrm{PL}_n\le \mathrm{PL}'_n$, and Step~3 applies to $\mathrm{PL}'_n$ uniformly on $\mathcal I_2^\delta\defeq\{\pi_2:d(\pi_2,\mathcal I_2)\le\delta\}$ for small $\delta>0$, because $\mathcal I_2^\delta$ lies in the interior of the identified interval of the enlarged model. For $\pi_2\notin\mathcal I_2^\delta$, no parameter with the original bounds and that value of $\pi_2$ has cell vectors within some $\epsilon(\delta)>0$ of $q^0$, by compactness, so these values do not contribute by Step~1. Hence, for every compact $J$ in the interior of $\mathcal I_2$ and every small $\delta$,
\[
\sup_{\pi_2\in J}\mathrm{PL}(\pi_2,X_n)+o_p(1)\ \le\ \sup_{\pi_2}\mathrm{PL}_n(\pi_2)\ \le\ \sup_{\pi_2\in\mathcal I_2^\delta}\mathrm{PL}(\pi_2,X_n)+o_p(1).
\]
As $|\mathrm{PL}(\pi_2,x)-\mathrm{PL}(\pi_2',x)|\le2\|u(\pi_2)-u(\pi_2')\|\,\|x\|^2$, $u$ is uniformly continuous near $\mathcal I_2$ and $X_n$ is tight, for all $\epsilon,\gamma>0$ there are $J$ and $\delta$ such that, for large $n$, both bounds are within $\epsilon$ of $\sup_{\mathcal I_2}\mathrm{PL}(\cdot,X_n)$ with probability at least $1-\gamma$. Therefore, $\sup_{\pi_2}\mathrm{PL}_n(\pi_2)=\sup_{\pi_2\in\mathcal I_2}\mathrm{PL}(\pi_2,X_n)+o_p(1)$.

\paragraph*{Step 5: Limits.} The true $\pi_{2,0}$ is interior to $\mathcal I_2$, so $\mathrm{LR}_n=\sup_{\pi_2}\mathrm{PL}_n(\pi_2)-\mathrm{PL}_n(\pi_{2,0})=\sup_{\mathcal I_2}(u'X_n)^2-(u_0'X_n)^2+o_p(1)$ and $\mathrm{AR}_n=-\mathrm{PL}_n(\pi_{2,0})=\|X_n\|^2-(u_0'X_n)^2+o_p(1)$. Conditionally on the instruments, the CLT and $\E[\hat q_1-\hat q_0]=q(\pi^0,\zeta^0+h/n^{1/2})-q(\pi^0,\zeta^0)=a(\pi^0,\zeta^0)h/n^{1/2}+O(n^{-1})$ give $X_n\dto\mathcal N(hg_0,I_3)$, and the CMT gives the two limits; $\|X\|^2-(u_0'X)^2=\|(I-u_0u_0')X\|^2\sim\chi^2_2$ because $(I-u_0u_0')g_0=0$. The jointly Gaussian $t=u_0'X$ and $X-tu_0$ are uncorrelated, hence independent, and the distribution of the second is free of $h$.

\paragraph*{Step 6: Geometry and the bootstrap.} Since $d(x,\operatorname{span}u)^2=\|x\|^2-(u'x)^2$, the limit is $T(X)=d(X,L)^2-d(X,C_1)^2$ with $L=\operatorname{span}\{u_0\}$ and $C_1=\bigcup_{\pi_2\in\mathcal I_2}\operatorname{span}\{u(\pi_2)\}$; the level scores are asymptotically independent of $X_n$, enter neither statistic, and lie in the harmless part $L_s$ of Theorem~\ref{thm:condlimit}\ref{thm:condlimit:d}. If $C_1+L=C_1$ and $v=u(\pi_2^*)\ne\pm u_0$, then $v+su_0\in C_1$ for every $s$, and $s=-u_0'v$ gives a nonzero element of $C_1$ orthogonal to $u_0$; so if no $u(\pi_2)$ is orthogonal to $u_0$, \ref{G2} fails. If $\dot u(\pi_{2,0})\ne0$, such a $v$ exists. The local position of the restricted fit converges to $t$, whose distribution has full support on $L$; by Theorems~\ref{thm:diagnostic}\ref{thm:diagnostic:b} and~\ref{thm:condlimit}\ref{thm:condlimit:c}, the restricted-fit bootstrap is consistent at $\beta=0$ if and only if the distribution $F(\cdot\,\vert\, t)$ of $T(tu_0+W)$, $W\sim\mathcal N(0,I_3-u_0u_0')$, which is continuous in $t$, does not depend on $t$. Let $\nu<1$ be the maximum of $\|(I_3-u_0u_0')u(\pi_2)\|$ on $\mathcal I_2$. Then, $T(W)\le\nu^2\|W\|^2$, so $1-F(c\,\vert\,0)\le e^{-c/(2\nu^2)}$. With $e_0\defeq\dot u(\pi_{2,0})/\|\dot u(\pi_{2,0})\|$, evaluating at $\pi_2=\pi_{2,0}+r/t$ and using $u'u_0=1-\tfrac12\|u-u_0\|^2$ gives $\liminf_{t\to\infty}T(tu_0+W)\ge\sup_r(2r\dot u(\pi_{2,0})'W-r^2\|\dot u(\pi_{2,0})\|^2)=(e_0'W)^2$, so by Fatou's lemma $\liminf_t(1-F(c\,\vert\, t))\ge\Pr(\chi^2_1>c)$, which exceeds $e^{-c/(2\nu^2)}$ for large $c$, so $F(\cdot\,\vert\, t)$ depends on $t$. Finally, fix $t$ and let $f(w)\defeq T(tu_0+w)$, $w\perp u_0$. As $\dot u(\pi_{2,0})'u_0=0$, $(u(\pi_2)'(tu_0+w))^2$ has derivative $2t\dot u(\pi_{2,0})'w$ at $\pi_{2,0}$ and, for $t=0$, equals $(\pi_2-\pi_{2,0})^2(\dot u(\pi_{2,0})'w)^2+o((\pi_2-\pi_{2,0})^2)$, so $f(w)>0$ unless $e_0'w=0$. As $f$ is continuous, convex and unbounded with $f(0)=0$, each set $\{f=c\}$, $c>0$, is the boundary of a convex set with interior, hence Lebesgue-null, and $\{a<f<b\}$ is open and nonempty for $0\le a<b$; as $W$ has a positive density on $u_0^\perp$, $F(\cdot\,\vert\, t)$ is continuous and strictly increasing on $(0,\infty)$, which contains $c_\alpha(t)$, as Theorem~\ref{thm:fitcond} requires.
\end{proof}

\begin{proof}[Proof of Proposition~\ref{prop:nonlin}]
Under $H_0$, $y\sim \mathcal N(X_0(\zeta',\beta')',\sigma^2I_n)$, and $P_0y$ and $\|\hat e_0\|$ are jointly complete and sufficient for $(\zeta',\beta',\sigma)'$. The direction $\hat e_0/\|\hat e_0\|$ is uniformly distributed on the unit sphere of the range of $M_0$ whatever the parameters, so it is independent of the sufficient statistic by Basu's theorem \citep[Proposition~3]{Dufour1990}; $M_0\xi/\|M_0\xi\|$ has the same distribution. Hence, $y=P_0y+\|\hat e_0\|\,\hat e_0/\|\hat e_0\|$ has the stated conditional distribution. Given the sufficient statistic, $S(y),S(y^*_1),\dots,S(y^*_B)$ are exchangeable under $H_0$, so the rank of $S(y)$ among them is uniform up to ties, and conservative or randomly tie-broken ranks have the stated size by \citet{Dufour2006}.
\end{proof}

\section{Implementing the conditional bootstrap}\label{app:implement}

We describe three ways to compute the conditional critical value: exact simulation where it is available, a fast projection shortcut and the kernel reweighting covered by Theorem~\ref{thm:fitcond}. We illustrate the last two in the model of Section~\ref{sec:weakiv}.

\paragraph{What to condition on.}
Section~\ref{sec:fitcond} conditions on components $\eta_e$ such that the scores of the remaining components lie in the harmless part $L_s$ (Theorem~\ref{thm:condlimit}\ref{thm:condlimit:d}); in the examples, level, scale and covariance parameters lie in $L_s$. Conditioning on further components is asymptotically harmless but wastes draws.

\paragraph{(A) Exact conditional simulation.}
When the conditional distribution of the data given a sufficient statistic for the nuisance parameter is known, drawing from it gives an exact test: in Gaussian linear models, the residual direction given the fit and residual norm is uniform on a sphere (Proposition~\ref{prop:nonlin}), and in discrete exponential families conditional Monte Carlo applies \citep{LindqvistTaraldsen2005}.

\paragraph{(B) Projection draws.}
Suppose the likelihood depends on the data through a statistic $\hat q$ that is asymptotically normal with mean $q(\eta)$ under $Q_\eta$, such as the cell frequencies $(\hat q_0',\hat q_1')'$ of Section~\ref{sec:weakiv}. Let $\tilde V_q$ be the variance of $\hat q$ under $Q_{\tilde\eta}$, $\tilde V_q^{1/2}$ its symmetric square root, and $\tilde P^\perp$ the orthogonal projection onto the orthogonal complement of the column space of $\tilde V_q^{-1/2}\partial q(\tilde\eta)/\partial\eta'$. Draw
\[
q^*\defeq q(\tilde\eta)+\tilde V_q^{1/2}\tilde P^\perp\xi^*,\qquad \xi^*\sim\mathcal N(0,I),
\]
and recompute the LR statistic from $q^*$ in place of $\hat q$, with the same optimizer and starting values as for the data. After whitening by $\tilde V_q^{-1/2}$, the perturbation is orthogonal to the null Jacobian at the fit, so the restricted score at $\tilde\eta$ vanishes for every draw, exactly for multinomial cells and to first order in general, and the bootstrap restricted fit equals $\tilde\eta$ to first order. In the limit experiment, the draws are $t+Z^*$ with $Z^*_L=0$ and $Z^*_\perp\sim\mathcal N(0,I)$, whose LR has the conditional distribution $F(\cdot\,\vert\, t)$ of Theorem~\ref{thm:condlimit}\ref{thm:condlimit:b}. We do not prove that the distribution of $\mathrm{LR}_n^*$ approximates $F(\cdot\,\vert\,\hat H_n)$, which would need the expansion of the LR statistic as a function of $\hat q$ to hold uniformly over $n^{-1/2}$-neighborhoods of $q(\tilde\eta)$ rather than in probability under the model, so (B) is a shortcut, to be checked against (C), which Theorem~\ref{thm:fitcond} covers. In the IV model, (B) is a parametric analogue of the conditional bootstrap of \citet{MoreiraPorterSuarez2004} that fixes the first-stage statistic at its observed value.

\paragraph{(C) Kernel reweighting.}
This is the procedure of Section~\ref{sec:fitcond}: draw from $Q_{\tilde\eta}$ as in the ordinary bootstrap and weight draw $j$ by $w_j\defeq K(D_j^*/b)$. The truncated Gaussian kernel used below is not Lipschitz, but it lies between Lipschitz kernels whose expectations under the Gaussian limit distribution of $D_n^*/b$ differ arbitrarily little, so the proof of Theorem~\ref{thm:fitcond}\ref{thm:fitcond:a} applies to it. In place of $\hat\Sigma_e^{-1/2}n^{1/2}$, one may standardize by the bootstrap covariance of $\tilde\eta^*_e$, when its normalized version converges to $\Sigma_e$, or, when $d_e=1$, by a robust spread such as the interquartile range divided by $1.349$; this is asymptotically equivalent when $B$ also increases. With a Gaussian kernel truncated at three bandwidths and $D_j^*$ approximately $\mathcal N(0,I_{d_e})$, the effective number of draws $(\sum_jw_j)^2/\sum_jw_j^2$ is approximately
\[
B\,\frac{(b^2(2+b^2))^{d_e/2}}{(1+b^2)^{d_e}},
\]
about $0.45B$ for $d_e=1$ and $b=0.35$, and $0.2B$ for $d_e=2$. For a symmetric kernel, such as the truncated Gaussian used below, the smoothing bias of the critical value is of order $b^2$ when $F(\cdot\,\vert\, t)$ is smooth in $t$ and has a positive density at $c_\alpha(t)$. We recommend a few hundred effective draws and $p$-values at two bandwidths. Counting the observed sample as a draw with weight $K(0)$ (step~4 of Section~\ref{sec:fitcond}) is an optional finite-$B$ adjustment borrowed from the add-one rule of Monte Carlo tests: the weighted draws are not exchangeable with the data, so it does not make the test exact, and its effect is an empirical matter. The method is practical for $d_e\le3$; beyond that, (B) is preferable where applicable.

\paragraph{Illustration.}
In the endogenous-treatment model, $\eta_e=\beta$, the take-up effect. Table~\ref{tab:hm-implement} compares the ordinary bootstrap, projection draws (B) and kernel reweighting (C) of $800$ ordinary draws at three bandwidths. All rejection frequencies lie between $0.040$ and $0.063$; with $3{,}000$ replications one Monte Carlo standard error is about $0.004$. The largest deviations are at $\beta=0$, where the kernel versions reject in $0.061$ to $0.063$ of replications, so they are mildly oversized, and the projection draws in $0.058$. Reweighting only $199$ draws gives $54$, $90$ and $119$ effective draws and changes the frequencies by at most $0.008$. Replication by replication, projection and kernel $p$-values have correlation at least $0.989$ and agree at level $0.05$ in at least $98\%$ of replications; ordinary bootstrap $p$-values are nearly as close, as its inconsistency is mild here. The table uses the basic rule of step~4. In a small check on $300$ replications at $\beta=0$, the optional adjustment together with the data's optimizer for the draws (Section~\ref{sec:practicalsim}) lowers the kernel frequencies by $0.010$ to $0.013$.

\begin{table}[ht]\begin{center}
\caption{Null rejection frequencies at level $0.05$ of the conditional bootstrap in the endogenous-treatment model}\label{tab:hm-implement}
\setlength{\tabcolsep}{2.5pt}
\begin{tabular}{cccccc}\toprule
& & & \multicolumn{3}{c}{Kernel reweighting}\\\cmidrule(lr){4-6}
$\beta$ & Ordinary & Projection & $b=0.2$ (218) & $b=0.35$ (362) & $b=0.5$ (479)\\\midrule
0 & 0.050 & 0.058 & 0.063 & 0.063 & 0.061\\
0.025 & 0.040 & 0.043 & 0.044 & 0.044 & 0.043\\
0.05 & 0.057 & 0.057 & 0.057 & 0.058 & 0.058\\
0.1 & 0.050 & 0.046 & 0.051 & 0.050 & 0.051\\
\bottomrule\end{tabular}\end{center}
\floatnotes{LR test with $n=1000$ in the design of Table~\ref{tab:hm}, with $3{,}000$ replications independent of those used there; ordinary restricted-fit bootstrap and projection draws with $B=199$, and kernel reweighting of $800$ ordinary draws with a Gaussian kernel truncated at three bandwidths and bandwidth $b$ in units of the bootstrap spread of $\tilde\beta^*$; average effective number of draws in parentheses.}
\end{table}

\section{Additional simulations}\label{sec:practicalsim}

\paragraph{Endogenous binary treatment: conditional draws.}
The projection draws (B) use the multinomial variance of the two conditional cell vectors at the restricted fit, scaled by the observed $n_z$; here, no draw has a negative cell probability. The finite-sample LR statistics of Tables~\ref{tab:hm} and~\ref{tab:hm-wald} profile $\pi_2$ numerically on a $0.04$ grid with parabolic refinement, which approximates the continuous profile statistic of Proposition~\ref{prop:hm}, and each draw starts the optimizer from the data's profile. In $300$ replications of Table~\ref{tab:hm} at each of $\beta=0$ and $0.025$, starting the draws as for the data changes none of the $1{,}200$ resampling decisions. In $100$ replications at $\beta=0$, a $0.01$ grid changes $2$ of $200$ resampling decisions, one each way, and in the three with the largest changes a $0.0025$ grid moves the statistic by at most $0.08$, although their $p$-values move by up to $0.20$ between the $0.04$ and $0.01$ grids; on a $0.005$ grid, the $\chi^2_1$ decision changes in at most $0.8\%$ of $400$ samples at each of three values of $\beta$.

\paragraph{Design.}
Limit-experiment computations use $100{,}000$ replications, critical values interpolated at $33$ values of $|t|$ in $[0,8]$, nearly constant beyond, and a grid of spacing $0.0025$ on $\mathcal I_2$. Spacing $0.0005$ with the end points of $\mathcal I_2$ resolved to $10^{-5}$ changes rejection probabilities by at most $0.003$ and raises the Gaussian critical values at $t=0$ by about $0.08$; for the design of Table~\ref{tab:hm} we report the refined values. The curve $u(\mathcal I_2)$ lies close to a plane through the origin: the root-mean-square singular values of the matrix of its points are $0.95$, $0.30$ and $0.04$. Projected on the plane orthogonal to $u_0$, the lines $\operatorname{span}\{u(\pi_2)\}$ have directions within an arc of $32^\circ$ out of $180^\circ$; the arc is $45^\circ$ in the Ali--Mikhail--Haq design below and at most $62^\circ$ over the $32$ designs below, so in each design the completed cone is not the whole space. The mean drift of $t$ is $n^{1/2}\,\beta\,\|g_0\|=1.29\,n^{1/2}\,\beta$, so $\beta=0.2$ at $n=1000$ gives about $8$. In the limit experiment of Proposition~\ref{prop:hm}, the conditional critical value rises from $2.5$ at $t=0$, where the restricted-fit bootstrap uses $3.1$, to between $3.8$ and $4.0$ for $t\ge3$. For mean drifts up to $5$, the null rejection probability is $0.049$ to $0.051$ for the conditional test, $0.042$ to $0.052$ for the restricted-fit bootstrap and $0.028$ to $0.052$ for $\chi^2_1$, and over $32$ designs with $\pi_1\in\{0.3,0.6\}$, $\pi_{2,0}\in\{0.4,0.7\}$, $\pi_3\in\{\pm0.4,\pm0.8\}$ and $\zeta\in\{0.2,0.5\}$, it is $0.023$ to $0.054$ for the restricted-fit bootstrap at drifts $0$ to $3$.

\paragraph{Comparison with the robust Wald test.}
The simulations of \citet{HanMcCloskey2019} use the Ali--Mikhail--Haq copula, for which the critical values of the robust Wald test are tabulated. We implement it as in their Section~7.3: the Wald statistic for $\pi_2$, with the largest tabulated $0.95$ quantile over a $50\%$ confidence interval for $n^{1/2}\beta$ plus the size-correction factor. At their alternatives $\pi_2=0.2$ and $0.7$, no null parameter with $|\pi_3|<1$ reproduces the cell probabilities given $Z_i=0$, so the levels alone reveal the alternative and power there says nothing about weak identification; these points are reported separately. The robust Wald test is conservative under weak identification, and at the interior alternatives the conditional LR test is up to $31$ percentage points more powerful at $\beta=0.2$ and $21$ at $\beta=0.4$; size-adjusted at the true $\beta$, infeasibly and with the conditional test at $0.045$ since its $p$-values are discrete, the gaps are up to $27$ and $5$ points, so at $\beta=0.4$ the gap mostly reflects this conservatism. Here, $\mathcal I_2=[0.17,0.64]$. In the limit experiment, the restricted-fit bootstrap rejects with probability $0.019$ at zero drift and the $\chi^2_1$ test with $0.001$, against $0.049$ for the conditional test, which explains why the ordinary bootstrap under-rejects at $\beta=0$ and, at interior alternatives, is up to $9$ points less powerful than the conditional test. The AR test, valid because the null is regular, profiles out the nuisance parameters.

\begin{table}[ht]\begin{center}
\caption{Rejection frequencies at level $0.05$ in the Ali--Mikhail--Haq design}\label{tab:hm-wald}
\begin{tabular}{ccccccc}\toprule
& & \multicolumn{3}{c}{$\mathrm{LR}$} & $\mathrm{AR}$ & Robust\\\cmidrule(lr){3-5}\cmidrule(lr){6-6}
$\beta$ & True $\pi_2$ & $\chi^2_1$ & Restricted fit & Conditional & $\chi^2_2$ & Wald\\\midrule
0 & 0.4 & 0.000 & 0.020 & 0.052 & 0.049 & 0.002\\
0.1 & 0.4 & 0.024 & 0.056 & 0.050 & 0.046 & 0.008\\
0.2 & 0.4 & 0.047 & 0.051 & 0.045 & 0.053 & 0.040\\
0.4 & 0.4 & 0.053 & 0.050 & 0.052 & 0.046 & 0.028\\\midrule
0.2 & 0.3 & 0.160 & 0.194 & 0.189 & 0.120 & 0.181\\
0.2 & 0.5 & 0.152 & 0.166 & 0.161 & 0.118 & 0.096\\
0.2 & 0.6 & 0.530 & 0.532 & 0.619 & 0.409 & 0.308\\
0.4 & 0.3 & 0.564 & 0.568 & 0.561 & 0.463 & 0.436\\
0.4 & 0.5 & 0.545 & 0.544 & 0.539 & 0.452 & 0.331\\
0.4 & 0.6 & 0.992 & 0.990 & 0.994 & 0.982 & 0.955\\
\midrule
0.2 & 0.2 & 0.679 & 0.587 & 0.763 & 0.556 & 0.513\\
0.2 & 0.7 & 0.984 & 0.958 & 0.991 & 0.955 & 0.653\\
0.4 & 0.2 & 0.997 & 0.994 & 0.999 & 0.994 & 0.989\\
0.4 & 0.7 & 1.000 & 1.000 & 1.000 & 1.000 & 1.000\\
\bottomrule\end{tabular}\end{center}
\floatnotes{Ali--Mikhail--Haq copula, $n=1000$, $H_0:\pi_2=0.4$, $\pi_1=0.6$, $\pi_3=0.4$, $\zeta=0.2$, $\phi_1=1/2$, alternatives changing $\pi_2$ only; bottom: restricted fit on the bound $|\pi_3|=0.99$. $3{,}000$ replications; $B=199$ draws for both resampling tests, projection draws as in Table~\ref{tab:hm}.}
\end{table}

\paragraph{The first decay parameter.}
For the Nelson--Siegel decay $\lambda$ itself, without the second curvature term and restricted to place the peak of the curvature loading between $3$ and $120$ months, the monthly conditional sets cover a median of $40\%$ of that range on a log scale and contain the value $0.0609$ of \citet{DieboldLi2006} in $70\%$ of the months, against $54\%$ for the $\chi^2_1$ sets. Here, the fitted slope and curvature terms are large relative to the residual variation, with a median of $34$ times the residual standard deviation, and in simulations calibrated to each month, as in Table~\ref{tab:svensson}, the conditional sets and the restricted-fit bootstrap cover with probability $0.950$ and $0.949$, and the $\chi^2_1$ and Wald sets with $0.912$ and $0.814$; resampled residuals change these coverages by at most $0.001$, and Gaussian errors with the cross-maturity covariance used for $\lambda_2$ give $0.991$, $0.991$, $0.980$ and $0.889$, so the conditional sets are conservative.

\paragraph{Yield-curve computations.}
We use the dataset of \citet{DieboldLi2006}, whose 1985:01--2000:12 subsample matches their Table~1 to within $0.002$. For $\lambda_2$, the admissible set is the union of two intervals, with the peak of $f_2(\cdot;\lambda_2)$ between $3.0$ and $19.6$ months or between $44.2$ and $120$ months. The statistic minimizes $SSR$ over $314$ values of $\lambda_2$, log-spaced within each interval, and the tests are evaluated at $95$ such values in the data and at the true value in the simulations, so the reported sets are grid-based inversions of exact tests. The Wald interval is $\hat\lambda_2\pm1.96\,\widehat{\mathrm{se}}$, with $\hat\lambda_2$ minimizing the residual sum of squares over the same grid and refined within a grid cell, by bounded minimization in the data and parabolic interpolation in $\log\lambda_2$ in the simulations, and $\widehat{\mathrm{se}}=\hat\sigma/|R_{55}|$, with $\hat\sigma^2=SSR/(n-5)$ and $R$ the triangular factor of the $17\times5$ Jacobian $[X(\hat\lambda_2)\ \,\hat\beta_4\,\partial f_2(\cdot;\hat\lambda_2)/\partial\lambda]$. Each simulated sample draws from the month's least-squares fit, with $\sigma_t^2=SSR/(n-4)$, and Gaussian errors, the month's residuals resampled with replacement and rescaled by $(n/(n-5))^{1/2}$, which also counts $\hat\lambda_2$ among the fitted parameters, or Gaussian errors with covariance $c_tV_\varepsilon$. For $\lambda_1$, the corresponding choices are $SSR/(n-3)$ and $(n/(n-4))^{1/2}$, with the same $V_\varepsilon$: the Nelson--Siegel residuals also contain the second curvature term in the months where it is significant. Here, $V_\varepsilon$ is the minimum-norm least-squares solution of $\sum_tM_tV_\varepsilon M_t=\sum_tr_tr_t'$, projected on the positive semidefinite cone, with $M_t$ the residual-maker of month $t$'s fitted regressors and $r_t$ its residuals, and $c_t$ makes the expected residual variance equal to $\sigma_t^2$. The correction matters because fitted residuals are orthogonal to the fitted regressors: $V_\varepsilon$ is not identified, and is set to zero, only along the level, slope and first-curvature loadings, which every Svensson regression absorbs. As $M_t$ depends on the fitted $\lambda_2$, the equation holds only approximately, and $V_\varepsilon$ calibrates the design rather than estimating the error covariance without bias. The restricted-fit bootstrap draws from the fitted null with the maximum likelihood variance. The completed statistic depends on the data only through the direction of the null residuals, so, unlike the other tests, its coverage does not depend on the error scale; resampled-residual coverages are within $0.007$ of the Gaussian ones.

\paragraph{Weighted average power for a break date.}
The break-date limit experiment of \citet[Section~5.2]{ElliottMullerWatson2015} illustrates Theorem~\ref{thm:condWAP}. The observation is the path $G(s)\defeq\mathbb W(s)-s\mathbb W(1)-\delta(\min(\pi,s)-\pi s)$, $s\in[0,1]$, with $\mathbb W$ a standard Wiener process. The likelihood ratio of $(\pi,\delta)'$ against $\delta=0$ is $\exp(-\delta G(\pi)-\tfrac12\delta^2\pi(1-\pi))$, so under $H_0:\pi=\pi_0$ the model is a Gaussian exponential family in $\delta$ with complete sufficient statistic $G(\pi_0)$, the restricted fit: similar tests have Neyman structure, and the argument of Theorem~\ref{thm:condWAP}\ref{thm:condWAP:b} applies conditionally on $G(\pi_0)$. The approximately least favorable test of \citet{ElliottMullerWatson2015} and the best similar test of Theorem~\ref{thm:condWAP}\ref{thm:condWAP:b} therefore both reject when the averaged likelihood ratio exceeds a function of $G(\pi_0)$.

With their weights, $\pi$ uniform on $[0.15,0.85]$ and $\delta\sim\mathcal N(0,100)$ truncated to $[-20,20]$, the weighted average power of the best similar test is within about half a percentage point of the estimated bound over all valid tests from their approximately least favorable distribution, and that of the best invariant test is $4$ to $5$ points lower (Table~\ref{tab:emw}), so in these designs nearly all of their test's advantage over invariant tests comes from dropping invariance, not similarity. The $0.95$ quantile of the null distribution of the LR statistic, from $100{,}000$ replications at each $\delta\in\{-20,-17.5,\dots,20\}$, lies in $[6.64,6.87]$ at $\pi_0=0.2$ and in $[6.59,7.39]$ at $\pi_0=0.4$, with Monte Carlo error of about $0.05$, so the ordinary bootstrap, which uses the quantile at the fitted $\delta$, is inconsistent, but only mildly: its null rejection frequencies lie in $[0.046,0.053]$ on that grid, and those of all tests in Table~\ref{tab:emw} in $[0.041,0.054]$.

\begin{table}[ht]\begin{center}
\caption{Weighted average power in the break-date experiment}\label{tab:emw}
\setlength{\tabcolsep}{4pt}
\begin{tabular}{lcccccc}\toprule
$\pi_0$ & Bound & EMW & Best similar & Conditional LR & Best invariant & Ordinary LR \\ \midrule
$0.2$ & 0.522 & 0.513 & 0.517 & 0.513 & 0.476 & 0.513 \\
$0.4$ & 0.446 & 0.441 & 0.444 & 0.441 & 0.395 & 0.440 \\
\bottomrule
\end{tabular}\end{center}
\floatnotes{Level $0.05$; weights of \citet{ElliottMullerWatson2015}; Brownian paths on $1{,}000$ steps; conditional critical values interpolated at $33$ values of $|G(\pi_0)|$; $100{,}000$ replications under the weighted alternative.}
\end{table}

\end{document}